\documentclass[aps,prb, %
graphicx,
amsmath, amssymb,
twocolumn, %
reprint %
]{revtex4-2}
\usepackage[utf8]{inputenc}
\usepackage[T1]{fontenc}
\usepackage[british]{babel}
\usepackage{dcolumn}%
\usepackage{bm}%
\usepackage{linop, braket} %
\usepackage{xparse} %
\usepackage[makeroom]{cancel}

\usepackage[caption=false]{subfig}

\usepackage[normalem]{ulem}

\usepackage{tikz}
\usetikzlibrary{shapes.geometric, arrows}

\tikzstyle{startstop} = [rectangle, rounded corners, minimum width=8cm, minimum height=1.3cm,text centered, draw=black]
\tikzstyle{process} = [rectangle, minimum width=8.0cm, minimum height=1cm, text centered, draw=black]
\tikzstyle{decision} = [diamond, aspect=2, text badly centered, minimum width=9.0cm, draw=black]
\tikzstyle{io} = [trapezium, trapezium left angle=70, trapezium right angle=110, minimum width=4.0cm, minimum height=1cm, text centered, draw=black]

\tikzstyle{blank} = [draw=none, fill=none]
\tikzstyle{arrow} = [thick,->,>=stealth]
\tikzstyle{line} = [thick,-,>=stealth]

\def\nodespacing{-0.4cm}

\renewcommand{\vec}[1]{\mathbf{#1}} %
\newcommand{\intd}[1]{\mathrm{d}\vec{#1}} %

\newcommand{\fdv}[2]{\frac{\delta #1}{\delta #2 }}
\newcommand{\LXCgrad}[1]{\delta v^{#1}(\vec{r})}

\NewDocumentCommand{\vorb}{o}{ %
\IfNoValueTF{#1}
    {\phi_{i}^\mathrm{LXC}(\vec{r})} %
    {\phi_{i}^\mathrm{LXC}(\vec{#1})}
}
\NewDocumentCommand{\tarden}{o}{ %
\IfNoValueTF{#1}
    {\rho_\mathrm{t}(\vec{r})} %
    {\rho_\mathrm{t}(\vec{#1})}
}

\NewDocumentCommand{\vden}{o}{ %
\IfNoValueTF{#1}
    {\rho_v(\vec{r})} %
    {\rho_v(\vec{#1})}
}

\NewDocumentCommand{\grndden}{o}{ %
\IfNoValueTF{#1}
    {\rho_0(\vec{r})} %
    {\rho_0(\vec{#1})}
}

\NewDocumentCommand{\vlfx}{o}{ %
\IfNoValueTF{#1}
    {v_\mathrm{LFX}(\vec{r})} %
    {v_\mathrm{LFX}(\vec{#1})}
}

\NewDocumentCommand{\vexx}{o}{ %
\IfNoValueTF{#1}
    {v_\mathrm{xOEP}(\vec{r})} %
    {v_\mathrm{xOEP}(\vec{#1})}
}

\NewDocumentCommand{\vlxcmin}{o}{ %
\IfNoValueTF{#1}
    {v_{0,\mathrm{t}}(\vec{r})} %
    {v_{0,\mathrm{t}}(\vec{#1})}
}

\NewDocumentCommand{\vlxc}{o}{ %
\IfNoValueTF{#1}
    {v(\vec{r})} %
    {v(\vec{#1})}
}

\NewDocumentCommand{\vlxcdfa}{mo}{ %
	\IfNoValueTF{#2}
	{v_\mathrm{LXC}^\mathrm{#1}(\vec{r})} %
	{v_\mathrm{LXC}^\mathrm{#1}(\vec{#2})}
}

\newcommand{\Uobj}{U_{\rho_\mathrm{t}}}

\newcommand{\xcpot}{v_\mathrm{xc}} %

\newcommand{\deltaxc}{\Delta_\mathrm{xc}}

\NewDocumentCommand{\ksxc}{mo}{ %
	\IfNoValueTF{#2}
	{v_\mathrm{xc}^\mathrm{#1}(\vec{r})} %
	{v_\mathrm{xc}^\mathrm{#1}(\vec{#2})}
}

\NewDocumentCommand{\gksx}{o}{ %
	\IfNoValueTF{#1}
	{\op{v}{\mathrm{x}}{\mathrm{GKS}}} %
	{\op{v}{\mathrm{x}}{\mathrm{#1}}}
}

\newcommand{\BHO}{$\mathrm{BaHfO}_3$}
\newcommand{\BTO}{$\mathrm{BaTiO}_3$}
\newcommand{\STO}{$\mathrm{SrTiO}_3$}
\newcommand{\BZO}{$\mathrm{BaZrO}_3$}
\newcommand{\KMF}{$\mathrm{KMgF}_3$}

\usepackage{subfiles}

\begin{document}

\title{Local Exchange-Correlation Potentials by Density Inversion in Solids}
\author{Visagan Ravindran}
\author{Nikitas I. Gidopoulos}
\author{Stewart J. Clark}
\affiliation{Department of Physics, Durham University, South Road, Durham, DH1 3LE, UK}
\email[Corresponding Author: ]{s.j.clark@durham.ac.uk}
\date{7th August 2025}

\begin{abstract}
    Following Hollins \textit{et al.} [J. Phys.: Condens. Matter \textbf{29}, 04LT01 (2017)], we invert the electronic ground state densities for various semiconducting and insulating solids calculated using
    several density functional approximations within the generalised Kohn-Sham (GKS) scheme, which includes Hartree-Fock (HF) theory, hybrid schemes and the LDA+$U$ method.
    To appraise the role of locality vs non-locality in the effective KS/GKS potential, the band structures from the resulting {\em local exchange-correlation} (LXC) Kohn-Sham potential %
    are then compared with the band structures of the original GKS method.
    We find the LXC potential obtained from the HF density systematically predicts band gaps in good agreement with experiment, including for strongly correlated transition metal monoxides (TMOs).
    Furthermore, we find that the HSE06 and PBE0 hybrid functionals yield similar densities and LXC potentials to each other. In weakly correlated systems, these potentials are also similar to PBE.
    For LDA+$U$ densities, the LXC potential partly reverses the excessive flattening of bands caused by too-large Hubbard $U$ values.
    For meta-GGAs, we find only small differences between the GKS and LXC results, demonstrating that the non-locality of meta-GGAs is weak.
\end{abstract}
\maketitle

\section{Introduction}
Density functional theory (DFT) has enjoyed considerable success as a framework for modelling and obtaining theoretical insight into the behaviour of materials at a relatively low computational cost. Within the Kohn-Sham(KS)\cite{Kohn-Sham_DFT} formulation, this is realised through a fictitious auxiliary non-interacting system of electrons constructed to yield an identical ground state density $\grndden$ to the interacting system.
Although DFT is a formally exact theory in principle, in practice the exchange-correlation (XC) $E_\mathrm{xc} [\rho] $ contribution to the
total energy density functional needs to be approximated.
Consequently, the resulting XC potential of any density functional approximation (DFA),
\begin{equation}
    \xcpot ^{\rm DFA} [\rho] (\vec{r}) = \fdv{E_\mathrm{xc}^{\rm DFA}[\rho]}{\rho(\vec{r})},
    \label{eq:xc_pot_defn}
\end{equation}
will differ from the corresponding $\xcpot [\rho] (\vec{r})$ of the exact theory.

The development and refinement of approximations for $E_\mathrm{xc} [\rho]$ remains challenging, not least because the exact $E_\mathrm{xc} [\rho]$ or $\xcpot [\rho] (\vec{r})$ are not experimentally observable and thus, one lacks a benchmark to which one can compare.

Although previous work using many-body perturbation theory\cite{delta_xc_Sham_Schuelter_1983,delta_xc_Sham_Schuelter_1986_Si,Sham_Schuelter_1987_arsenides,Sham_Schuelter_1987_trends_xc,delta_xc_Sham_Schuelter_1988} has been carried out in simple semiconductors to gain some insight and study the behaviour of the exact $\xcpot [\rho] (\vec{r})$, an alternative, albeit indirect approach to construct the exact $\xcpot [\rho] (\vec{r})$ is through the inverse KS or \textit{density inversion} problem.
Because of the one-to-one correspondence between the KS potential $v_s(\vec{r})$ and $\grndden$ established by the first Hohenberg-Kohn theorem\cite{Hohenberg-Kohn_theorems,Nikitas_Interface_WFT_DFT}, one can  seek the \textit{local exchange-correlation} (LXC) potential $v_\mathrm{LXC}(\vec{r})$,
that will adopt a target density $\tarden$ as its ground state.
If, for a given number of electrons $N$, the target density is obtained by sufficiently accurate means, for instance from quantum Monte Carlo (QMC) calculations\cite{den_inversion_Lucia_QMC_2023} such that it may be regarded as ``exact'', the technique can be used to study the corresponding ``exact'' KS potential, which is determined up to a constant, for fixed $N$.

Moreover, density inversion can also be used to study and analyse the quality of approximate densities and their corresponding local XC potentials.
Considerable work has been carried out in recent years in understanding errors within DFT calculations\cite{LDA_CASTEP,Self_Interaction_Corrected_TMOs_1990_Svane,Localisation_Delocalisation_Error_Yang_2008,delocalisation_error_perspective_2023,band_gaps_in_GKS_2017,Uncertainty_in_DFT_energy_Corrections_2021,DFT_Errors_2012_Wang,Perspective_Burke_2012,Geometry_Errors_Vuckovic2020,Geometry_Errors_Vuckovic2022,Errors_for_Surfaces_2016,systematic_exchange_errors_2009,Gidopoulos_Lathiotakis_2012_Screening_Charge,Gidopoulos_Lathiotakis_SC_Charge_2015,Pitts_screening_charge_2018,Callow_Gidopoulos_Faraday_Discussion}.

According to an insightful analysis by Burke and coworkers, the error
in the approximate total energy of a DFT calculation may be partitioned into the so-called ``functional error'' and the ``density-driven error''\cite{density_driven_errors_Burke_2013,density_driven_errors_Burke_2014,density_driven_errors_Burke_2019,density_driven_errors_Burke_2022,DC_DFT_explained_Burke_2022}. The former is the most common and refers to the quality of the approximation for the XC energy density functional $E_\mathrm{xc}[\rho]$ for the exact density.
The significance of the latter can be investigated using density inversion\cite{density_driven_errors_inversion_Burke_2020}. It arises in ``abnormal'' systems \cite{density_driven_errors_Burke_2013}, where the density is particularly sensitive to changes in $v_\mathrm{xc}(\vec{r})$, giving rise to large errors in the density, which are further exacerbated by the self-consistent cycle.
In such instances, evaluating the total energy in a DFA such as PBE using the Hartree-Fock (HF) density in a non-self-consistent manner can give improved total energies, such as in water clusters\cite{DC_SCAN_Water_Clusters_Perdew_2021,DC_SCAN_Perdew_2022,DC_DFT_Reaction_Barriers_2023_Perdew,DC_SCAN_Water_Clusters_Burke_2023,DC_DFT_Burke_2023_spin_contamination}.

Despite the appeal of density inversion as a benchmarking and functional development tool for the aforementioned reasons, the inversion of the density remains a somewhat formidable task due to the sensitivity of the potential to small changes in the density\cite{den_inversion_numerical_issues_Jensen_Wasserman_2018,den_inversion_numerical_issues_Shi_Wasserman_2021} and thus the problem remains an active area of research with various inversion schemes proposed
\cite{den_inversion_Werden_Davidson_1984,den_inversion_Tozer_1995,den_inversion_Görling_1992,den_inversion_Leeuwen_Baerends_1994,den_inversion_Zhao_Morrison_Parr_1994,den_inversion_Savin_Umrigar_Gonze_1998,den_inversion_Wu_Yang_2003,den_inversion_Peirs_Van_Neck_Waroquier_2003,den_inversion_Kadantsev_Stott_2004,den_inversion_Ryabinkin_Staroverov_2012,den_inversion_Kumar_2019,den_inversion_Tim_Callow_2020,den_inversion_Sofia_Extension_to_Tim_Callow_2022,HF_inv_Görling_1995,HF_inv_Staroverov_2013,HF_inv_Staroverov_2014,HF_inv_as_density_func_problem_Holas_1993,HF_inv_as_density_func_problem_Nagy_1997,Stott_HF_inv_1994_small_atoms}.
In this work, we use the scheme developed by Hollins \textit{et al.}\cite{Hollins_LFX_2017} to compare the LXC potential obtained via inversion of densities produced by approximations with increasing degree of non-locality in their single-particle Hamiltonians.
We then compare the shape of the calculated band structures (not just the band gaps) from the LXC potential and from the self-consistent non-local potential. We also compare the total energy differences, when the total energy functional is evaluated using the self-consistent orbitals and using the orbitals obtained via the LXC potential.

This paper is structured as follows. In section \ref{section:theory}, we overview our method used to invert densities to find the LXC potential and its implementation in a plane-wave pseudopotential code.
We discuss he exchange-correlation derivative discontinuity for the
optimised effective potential (OEP) and LXC potentials in a separate subsection \ref{section:delta_xc}.
Section \ref{section:comp_details} describes the various computational parameters we have used and also introduces the systems used in this study.
In section \ref{section:RESULTS}, we present computed band structures from the LXC potentials and from self-consistent KS and generalised Kohn-Sham (GKS) schemes.
Section \ref{section:local_func_dens} presents the inversion of local/semi-local LDA and PBE densities to quantify the degree of numerical error in the inversion algorithm.
We then discuss the inversion of HF densities and hybrid functionals in sections \ref{section:LFX} and \ref{section:hybrids} where we also discuss the difference in total energy when a given DFA is evaluated using self-consistent GKS orbitals and the LXC orbitals as a quantitative indicator of the non-locality.
Results from the inversion of LDA+$U$ densities in transition metal monoxides (TMOs) are presented in section \ref{section:LDA+U}. We then discuss the inversion of densities from a meta-GGA functional (rSCAN) when treated in a GKS scheme.
Finally, we draw conclusions and discuss plans for future work in section \ref{section:CONCLUSIONS}.

\section{Theory}\label{section:theory}

We follow the density inversion method of references \cite{Hollins_LFX_2017,den_inversion_Tim_Callow_2020,den_inversion_Sofia_Extension_to_Tim_Callow_2022}.
We consider a non-interacting system of particles moving within a local effective potential $\vlxc$ with a ground state density $\vden$.
The many-body Hamiltonian of this non-interacting system can be expressed as a sum of single-particle Hamiltonians $\op{h}(\vec{r}_i, \vec{p}_i)$ with single-particle energies $\varepsilon_{v,i}$,
\begin{equation}
	\op{H}{v} = \sum_{i=1}^N \op{h}(\vec{r}_i, \vec{p}_i) = \sum_{i=1}^N \left(-\frac{\nabla_i^2}{2} + v(\vec{r}_i)\right).
	\label{eq:KS_like_Hamiltonian}
\end{equation}

The goal is to invert a ``target'' density, $\tarden$, obtained from some electronic structure or quantum chemical method, to find the local effective potential $\vlxc$ with ground state density equal to the target density, $\vden = \tarden$.

Consider the Coulomb energy $\Uobj[v]$ associated with the difference between the target density and the ground state density of the local potential $\vlxc$, $\tarden - \vden$:
\begin{equation}
    \Uobj[v] = \frac{1}{2}\iint\intd{r} \intd{r'}
    \frac{[\tarden-\vden][\rho_\mathrm{t}(\vec{r'})-\rho_v(\vec{r'})]}
    {|\vec{r}-\vec{r}'|} \geq 0.
    \label{eq:Coulomb_diff}
\end{equation}
When the two densities are equal, $\tarden=\vden$, the Coulomb energy vanishes, $\Uobj[v]=0$, and thus the KS potential is found and the density inversion problem solved.
The KS potential is customarily partitioned into an external $v_\mathrm{ext}(\vec{r})$, Hartree $v_\mathrm{H}(\vec{r})$ and exchange-correlation $v_\mathrm{xc}(\vec{r})$ potentials
\begin{equation}
	v_s(\vec{r}) = v_\mathrm{ext}(\vec{r})
	+ v_\mathrm{H}(\vec{r}) + v_\mathrm{xc}(\vec{r}).
	\label{defn:KS_pot}
\end{equation}
In this work, we will refer to the exchange-correlation contribution to the KS potential obtained via density inversion of a target density $\tarden$ as the \textit{local exchange-correlation} (LXC) potential $v_\mathrm{LXC}(\vec{r})$ (emphasising that it is a local multiplicative potential).

 Since $\Uobj[v]\geq 0$, it is possible to invert a given target density $\tarden$ through the minimisation of $\Uobj[v]$.
A change in the potential, $\vlxc \to \vlxc + \epsilon \; \LXCgrad{}$ with $\epsilon>0$ causes a change in $\Uobj$
\begin{equation}
    \Uobj[v+\epsilon \; \delta v] - \Uobj[v] = \
    \epsilon \int\intd{r}\; \LXCgrad{}
    \frac{\delta \Uobj[v]}{\delta v(\vec{r})}
    .
    \label{eq:derivation_inter_1}
\end{equation}
We apply the chain rule, noting the change in the density $\delta \rho_v(\vec{r})$ is related to the density-density response function,
\begin{equation}
	\delta \rho_v(\vec{r}) = \int\intd{r'} \; \chi_v(\vec{r},\vec{r'}) \delta v(\vec{r}').
\end{equation}
Thus, we express the change in the Hartree energy $\Uobj$ given in Eq. \eqref{eq:derivation_inter_1} by:
 \begin{equation} \label{eq:derivation_inter_2}
 	\begin{split}
 		&\Uobj[v + \epsilon \; \delta v] - \Uobj[v]
 		\\
 		=
 		\epsilon &\int\intd{r}\; \LXCgrad{}
 		\int\intd{r'}
 		\frac{\delta \vden[r']}{\delta v(\vec{r})}
 		\frac{\delta \Uobj}{\delta \vden[r']}
 		 \\
 		=
 		-\epsilon &\int \intd{r}\;  \delta v(\vec{r})
 		\int\intd{r'} \chi_v(\vec{r}',\vec{r})
 		\int\intd{r''}\frac{\tarden[r'']-\vden[r'']}{|\vec{r'}-\vec{r''}|}
 		.
 	\end{split}
 \end{equation}
We now choose for Eq. \eqref{eq:derivation_inter_2}:
\begin{equation}
    \delta \vlxc =
    -\int \intd{r'} \frac{
    \tarden[r']-\vden[r']
    }{|\vec{r}-\vec{r'}|} .
    \label{eq:LXC_Grad}
\end{equation}

This choice of $\delta v(\vec{r})$ ensures that the value of $\Uobj$ is reduced, $\Uobj[v+\epsilon \; \delta v ] < \Uobj[v] $,
for sufficiently small $\epsilon>0$ since  the response function $\chi_v(\vec{r},\vec{r'})$ is a negative semi-definite operator.
Moreover, the choice of $\delta v(\vec{r})$ in Eq. \eqref{eq:LXC_Grad}
leads to faster convergence compared to simply using the density difference, $\vden - \tarden$, particularly in regions of low density\cite{Hollins_LFX_2017}.
For these reasons, $\delta \vlxc$ can be regarded as an \textit{effective} gradient of $\Uobj$.

\begin{figure}
	\centering
	\begin{tikzpicture}[node distance=1.75cm,every node/.style={scale=1.0}]
		\node (SCF) [align=center, startstop] {
			Calculate the target density $\tarden$
			\\
			(in this work self-consistently via KS/GKS equations)
		};
		\node (Pot_Init) [align=center, startstop, below of= SCF, yshift=\nodespacing]
		{Initialise LXC potential to LDA potential of $\tarden$ \\
			$v^{n=1}(\vec{r})
			=
			v_\mathrm{Hxc}^\mathrm{LDA}[\rho_t{}](\vec{r})+v_\mathrm{ext}(\vec{r})$.
		};
		\node (den_init)
		[align=center, startstop, below of=Pot_Init, yshift=\nodespacing]
		{Calculate initial KS orbitals $\{\phi_{i, \mathrm{LXC}}^{n=1} (\vec{r})\}$
			\\
			and density $\rho_v^{n=1}(\vec{r})$
			for $v^{n=1}(\vec{r})$.
		};

		\node (LFX_grad)
		[align=center, process, below of= den_init, yshift=\nodespacing]
		{
			Calculate effective gradient of potential for $n$-th iteration\\
			$\displaystyle \LXCgrad{n}
			=
			-
			\int\intd\vec{r}'
			\frac{\rho_\mathrm{t}(\vec{r}') -\rho_v^n(\vec{r}')}{|\vec{r}-\vec{r}'|}.
			$
		};

		\node (LFX_line_search)
		[align=center, process, below of= LFX_grad, yshift=\nodespacing]
		{
			Determine optimal value of $\epsilon$ via\\  parabolic line search
			in $U_{\rho_\mathrm{t}}^n[v; \epsilon]$.
		};

		\node (LFX_pot_calc)
		[align=center, process, below of= LFX_line_search, yshift=\nodespacing]
		{
			Update total potential using optimised $\epsilon$\\
			$\displaystyle v^{n+1}(\vec{r})
			=
			v^{n}(\vec{r})
			+
			\epsilon
			\delta v^n(\vec{r})
			$
		};
		\node (LFX_den_calc)
		[align=center, process, below of= LFX_pot_calc, yshift=\nodespacing]
		{
			Find $\rho_v^{n+1}(\vec{r})$ and $\{\phi_{i, \mathrm{LXC}}^{n+1} (\vec{r})\}$   for  $\displaystyle v^{n+1}_\mathrm{LXC}(\vec{r})$
		};
		\node (check)
		[align=center, decision, below of= LFX_den_calc, yshift=-1.4cm]
		{
			Does $\Uobj[v]$ satisfy \\  convergence criterion?
		};
		\node (repeat)
		[blank, right of= check, xshift=4cm]
		{};
		\node (have_LFX)
		[align=center, io, below of= check, yshift=-1.5cm]
		{
			Compute band structure  using  \\ $v_\mathrm{LXC}(\vec{r})$ corresponding to $\rho_t(\vec{r})$.
		};
		\draw [arrow] (SCF) -- (Pot_Init);
		\draw [arrow] (Pot_Init) -- (den_init);
		\draw [arrow] (Pot_Init) -- (den_init);
		\draw [arrow] (den_init) -- (LFX_grad);
		\draw [arrow] (LFX_grad) -- (LFX_line_search);
		\draw [arrow] (LFX_line_search) -- (LFX_pot_calc);
		\draw [arrow] (LFX_pot_calc) -- (LFX_den_calc);
		\draw [arrow] (LFX_den_calc) -- (check);
		\draw [arrow] (check.east)  node[anchor=east, yshift=4.25cm]{no} |- (LFX_grad);
		\draw[arrow] (check) -- node[anchor=east, xshift=-0.5cm, yshift=0.25cm]{yes}(have_LFX) ;
	\end{tikzpicture}
	\caption{Steepest descent algorithm to calculate the local exchange-correlation (LXC) potential $v_\mathrm{LXC}(\vec{r})$.
		The total potential $v(\vec{r})$ is the sum of $v_\mathrm{LXC}(\vec{r})$, Hartree $v_\mathrm{H}(\vec{r})$ and external (electron-nuclear)  $v_\mathrm{ext}(\vec{r})$ potentials such that in the $n$-th iteration
		$v_\mathrm{LXC}^n(\vec{r})=v^n(\vec{r}) - v_\mathrm{H}[\rho_v^n](\vec{r}) - v_\mathrm{ext}(\vec{r})$.
	}
	\label{fig:LXC_algorithm}
\end{figure}
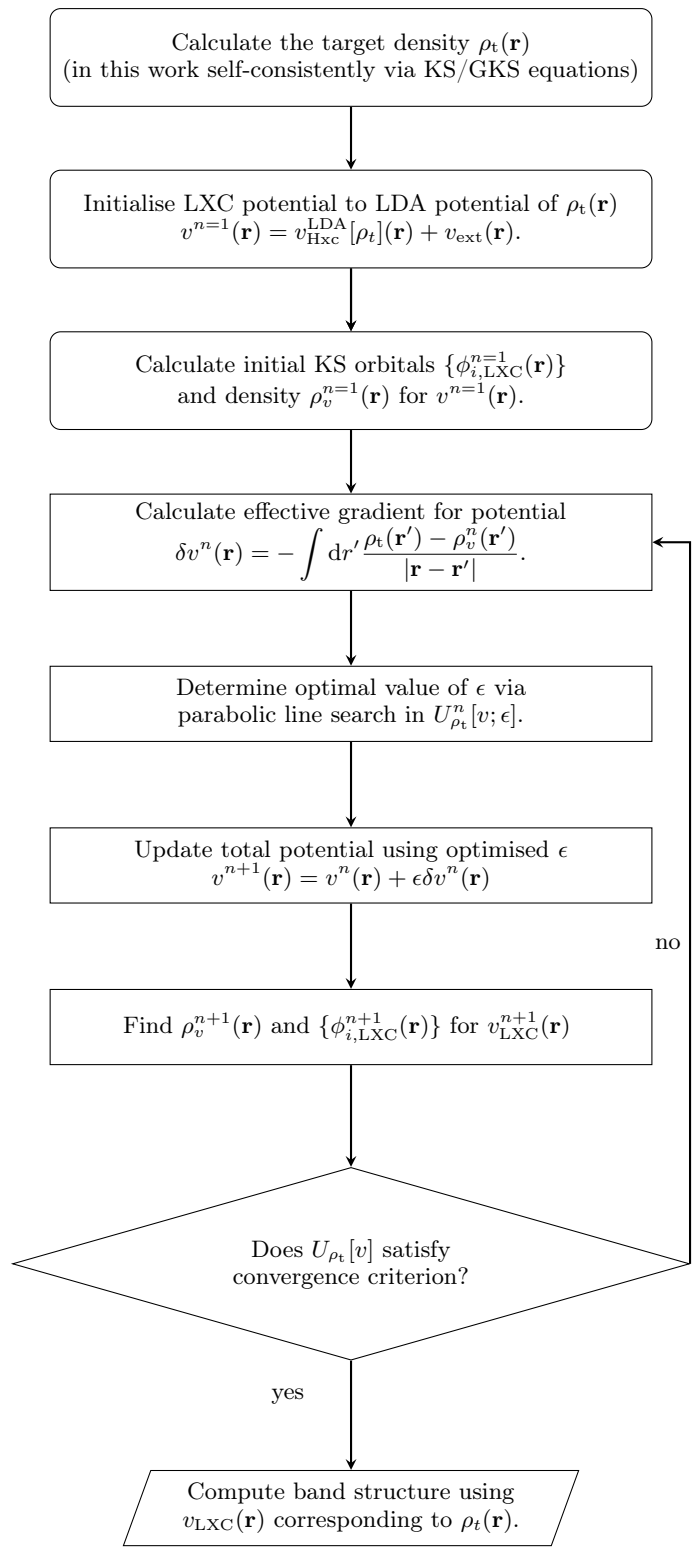

In a steepest descent algorithm as outlined in Fig. \ref{fig:LXC_algorithm}, the potential is thus updated by taking a step in the direction of the effective gradient
\begin{equation}
    \vlxc \to \vlxc + \epsilon \; \delta \vlxc,
    \label{eq:steepest_descents}
\end{equation}
where $\epsilon$ is the step size calculated using a line search, as discussed further in section \ref{section:comp_details}.
The density $\vden$ %
is then updated by solving the KS equations
\begin{equation}
    \left(-\frac{\nabla^2}{2} + \vlxc \right)\vorb = \varepsilon_{i} \vorb,
    \label{eq:pot_step_grad}
\end{equation}
to find the orbitals $\vorb$ of
the updated local effective potential $\vlxc$.
We then calculate the Hartree energy $\Uobj[v]$, see Eq. \eqref{eq:Coulomb_diff} of the density difference $\tarden - \vden$.
This iterative procedure is repeated until the target density $\tarden$ and the density of the local effective potential, $\vden$, are equal within a small tolerance such that $\Uobj[v]$ is effectively zero.
In practice, we use a more efficient conjugate gradient\cite{Fletcher_Reeves_CG} algorithm.

As an aside, we note the inverse KS problem is strictly defined for densities which are ground state $v$-representable, that is to say, ground state densities of local multiplicative potentials $v(\vec{r})$.

\subsection{KS $\{ \vorb \} $ and GKS $\{ \phi_i^\mathrm{GKS} \}$ Orbitals and their Total Energies}\label{subsection:total_energies_theory}
Following Kohn\cite{Kohn_Nobel_Lecture_1999},
the orbitals that yield the density $\vden$ of the non-interacting system bound by a local potential $\vlxc$ can be regarded as
density optimal in the sense that they yield an identical density to the target density $\tarden$.
In standard KS theory, these are the KS orbitals, which are functionals of $\vden$ and $\vlxc$ is the KS potential of $\vden$.

Suppose now that the target density $\tarden$ is obtained via a density functional approximation (DFA) whose total energy has an explicit dependence on the single-particle orbitals $\phi_i(\vec{r})$ but not on the density $\rho(\vec{r})$.
The total energy of this DFA is still an \textit{implicit} functional of the density, since the KS orbitals are determined by the KS potential which in turn depends on the density.
Such DFAs can be treated within the standard KS scheme, using single-particle equations that contain a local, multiplicative potential $v(\vec{r})$ via the OEP method\cite{Kuemmel_Orbital_Dependent_Review_2008,GKS_OEP_SCAN_Perdew_2016}.
Alternatively, one can instead consider them in a generalised Kohn-Sham (GKS) scheme with non-local single-particle equations, obtaining a lower (in general) total energy minimum.

The KS orbitals $\vorb$ obtained from the inversion of the GKS
density, which we refer to hereafter as the LXC orbitals, are not energy optimal as they do not fully minimise the total energy functional of the DFA, $E^\mathrm{DFA}[\{ \phi_i \} ]$,
since in the case of such DFAs, the GKS scheme yields a lower energy in general.
Equivalently, the effective potential $\vlxc$ with an LXC contribution potential $\vlxcdfa{DFA}$ obtained by inverting
this GKS DFA density $\rho_t^\mathrm{DFA}(\vec{r})$ is not the minimising potentials of $E^\mathrm{DFA}$;
instead the minimising (GKS) potential has a non-local XC $\op{v}{\mathrm{xc}}{\mathrm{DFA}}$, where in general, $v^\mathrm{DFA}_\mathrm{LXC}(\vec{r})\neq \op{v}{\mathrm{xc}}{\mathrm{DFA}}$.
Therefore, if one evaluates $E^\mathrm{DFA}$ in a \textit{non-self-consistent} manner using the orbitals $\vorb$ from the inversion, then
\begin{equation}
    E^\mathrm{DFA}[\{\phi_i^\mathrm{LXC}\}] - E^\mathrm{DFA}[\{\phi_i^\mathrm{GKS}\}] \geq 0,
    \label{eq:inv_eng_diff}
\end{equation}
where $\phi_i^\mathrm{GKS}(\vec{r})$ are the self-consistent set of GKS orbitals associated with the minimising non-local GKS potential. The magnitude of the energy difference in Eq. \eqref{eq:inv_eng_diff} will depend on the degree of non-locality in the Hamiltonian $\op{H}{\mathrm{GKS}}{\mathrm{DFA}}$.

In the case of HF densities, the potential obtained via inversion of the HF density is the local Fock exchange (LFX) potential \cite{Stott_HF_inv_1994_small_atoms,HF_inv_Görling_1995,HF_inv_Staroverov_2013,HF_inv_Staroverov_2014,Hollins_LFX_2017,LFX_Metals_2017,den_inversion_Tim_Callow_2020,den_inversion_Sofia_Extension_to_Tim_Callow_2022},
$\vlxcdfa{HF} \equiv \vlfx$.
The inversion of HF densities yields a local potential similar to the exact-exchange-only (EXX) potential, or exchange-only OEP (xOEP) (without correlation).
The similarity is particularly evident in weakly correlated
systems\cite{den_inversion_Ryabinkin_Staroverov_2012,HF_inv_as_density_func_problem_Holas_1993,HF_inv_as_density_func_problem_Nagy_1997,EXX_Semiconductors_Görling_1997,EXX_Semiconductors_Görling_1999,EXX_Delta_X_Grüning_Rubio_2006,HF_inv_Staroverov_2013,HF_inv_Staroverov_2014,EXX_RPA_Kresse_2014,Hollins_Hylleraas_OEP,Hollins_LFX_2017,Trushin_Görling_2019_EXX_perovskites}.
This is because the xOEP potential and the LFX potential minimise physically equivalent energy differences\cite{Hollins_LFX_2017,Nikitas_Interface_WFT_DFT}.

\subsection{Exchange-Correlation Derivative Discontinuity and Band Gap}\label{section:delta_xc}

In exact KS theory, the KS band gap, $E_{\mathrm{g},s} (N) = \varepsilon_{N+1} (N) - \varepsilon_{N} (N)$, of an $N$-electron system between the
lowest unoccupied orbital, $\varepsilon_{N+1} (N)$, (conduction band minimum, CBM) and the highest occupied orbital, $\varepsilon_N (N) $, (valence band maximum, VBM)
is not equal to the fundamental gap, %
\begin{equation}
E_\mathrm{g} (N) = I(N) - A(N) ,
\end{equation}
where $E_\mathrm{g} (N)$ is equal to the difference between the ionization potential $I(N) = E(N-1)-E(N)$ and the electron affinity $A(N) = E(N) - E(N+1)$.
(Here, $E(M)$ denotes the ground state energy of the $M$-electron system, $M=N,N\pm1$.)
This is because the KS potential is a non-analytic function of electron number.
When the number of electrons increases by an infinitesimal fraction  through the integer value $N$, the exact
KS potential changes by a constant known as the exchange and correlation disconinuity, $\deltaxc (N) >0$.

The discontinuity of the exact KS potential is not captured in local (LDA) and semi-local (GGAs) XC DFAs. We remark that using an ensemble formulation, it is possible to recover a non-zero approximation to the exact $\deltaxc (N)$ even for local and semi-local approximations, but nevertheless the approximate LDA/GGA discontinuity vanishes in the thermodynamic limit \cite{Kraisler_Kronik_2013_Delta_XC,Kraisler_Kronik_2014_Delta_XC,HollinsPhDThesis}.
Hence, in these approximations, the prediction for the fundamental gap of solids, as given by the approximate KS gap alone, systematically underestimates the value of $E_\mathrm{g} (N)$.

More advanced XC approximations such as meta-GGAs, DFT+$U$, and hybrid functionals yield a non-local XC potential when incorporated in a GKS calculation. The approximate non-local GKS potential is continuous as a function of electron number and the prediction for $E_\mathrm{g} (N)$
is given by the GKS gap without any correction,
$$
E_\mathrm{g} = I(N) - A(N) = \varepsilon_{N+1}^\mathrm{GKS} (N) - \varepsilon_{N}^\mathrm{GKS} (N) .
$$
The GKS gap value for these approximations tends to predict the fundamental gap more accurately than LDA and GGAs.

The same XC approximations can also yield a local multiplicative potential when the DFA total energy is minimised with the OEP method.
In that case, the approximate OEP potential has a finite XC discontinuity $\deltaxc^\mathrm{OEP} (N)$ which approximates the discontinuity $\deltaxc (N)$ of the exact KS scheme. $\deltaxc^\mathrm{OEP} (N)$ is given by the
difference between the GKS and OEP gaps,
\begin{multline} \label{dxc_lxc}
    \deltaxc^\mathrm{OEP} (N) = \left( \varepsilon_{N+1}^\mathrm{GKS} (N) - \varepsilon_{N}^\mathrm{GKS} (N) \right) \\
    - \left( \varepsilon_{N+1}^\mathrm{OEP} (N) - \varepsilon_{N}^\mathrm{OEP} (N) \right).
\end{multline}
In this work, together with standard hybrid functionals we have also employed ad-hoc generalisations of these functionals (Eq. \eqref{eq:PBE0_like_hybrid} for instance), by introducing a weight $\alpha$ in front of the non-local Fock exchange energy and potential terms and scanned over $\alpha$, $0 \le \alpha \le 1$ in our calculations to assess the effect of non-locality.
Obviously, the GKS gaps of these ad-hoc hybrid schemes do not provide predictions for the fundamental gap for any random value of $\alpha$, but only for the optimal value of $\alpha$, which is about $\alpha \sim 0.25$ (see Refs. \cite{B3LYP_1,PBE0,HSE03} and references therein). The same is true for our scans of DFT+$U$ calculations using a range of Hubbard-$U$ values. The prediction of the fundamental gap is only given by an optimal value for the Hubbard-$U$, which can be obtained
independently with linear response\cite{Hubbard_U_linear_response_2005,Hubbard_U_linear_response_2006}.

Our inversions of the GKS target density for the various XC approximations yield local KS potentials (dubbed LXC potentials) that are very close to the corresponding OEP XC potentials, with the important difference that the inverted LXC potentials do not strictly have an XC discontinuity, because they are not functional derivatives of an XC energy functional.
We note it is possible to incorporate in the inversion algorithm a constraint for the so-called screening charge, which changes discontinuously from $Q_\mathrm{scr} (N^-)=N-1$ to $Q_\mathrm{scr} (N^+) = N$,
when the number of electrons increases infinitesimally past the integer value $N$ with $N^- = N-\delta$, $N^+ = N+\delta$ and $\delta>0$ as $\delta \rightarrow 0$\cite{Gidopoulos_Lathiotakis_2012_Screening_Charge,Gidopoulos_Lathiotakis_SC_Charge_2015,Callow_Gidopoulos_Faraday_Discussion,den_inversion_Tim_Callow_2020}.
However, the extra screening charge $\Delta Q_\mathrm{scr} = 1$ delocalises in a solid and this discontinuity vanishes in the thermodynamic limit.

Despite the lack of XC discontinuity in a strict sense, due to the similarity between the inverted LXC potentials for hybrid (including HF), DFT+$U$ and meta-GGA densities
with the corresponding OEP potentials, we still loosely refer to the XC discontinuity of the LXC potentials, having in mind that the discontinuity actually refers to the OEP potential which LXC approximates.

Finally, we address the question in the literature whether it is justified or misleading to report uncorrected OEP-KS band-gaps (and by extension here LXC-KS bandgaps) for materials without including $\deltaxc^\mathrm{OEP}$.
In our work, there is no reason to omit the discontinuity %
in the case of LXC potentials for densities obtained by meta-GGAs and hybrid schemes, since we can obtain an estimate of $\deltaxc^\mathrm{OEP}$ from \eqref{dxc_lxc}.
In a future publication we will obtain an estimate of $\deltaxc^\mathrm{OEP}$ for DFT+$U$ calculations, once the linear response method of Ref. \cite{Hubbard_U_linear_response_2005} is implemented in our code.
For our work, the question refers to the exchange-only OEP potential and the LFX potential obtained from the inversion of HF densities.

The exchange and correlation discontinuity $\Delta_{xc}$ of the exact KS potential is the sum of two large contributions, $\Delta_\mathrm{x}>0 $ and $\Delta_\mathrm{c} < 0$.
The order of magnitude for each of these energies is several electronvolts but when added together, the two discontinuities almost cancel each other out giving a net correction $\deltaxc = \Delta_\mathrm{x} + \Delta_\mathrm{c}$, which is typically positive, $\deltaxc > 0$, with an order of magnitude for weakly correlated systems of a few tenths of an electronvolt.

One could argue that since both xOEP and LFX methods omit electronic correlation, one should omit the correlation contribution to the discontinuity, $\Delta_\mathrm{c}$ altogether from the band gap prediction.
However, by Brillouin’s theorem, in M{\o}ller–Plesset (MP) perturbation theory, the first-order corrected state (MP1) has the same density as HF. Hence, the HF density is also the density of a weakly interacting system, and it makes sense to consider the correlation contribution to the total energy (post-SCF)
at the second order level (MP2), corresponding to the MP1 weakly interacting state. The same situation holds for the xOEP density (see Ref. \cite{Hollins_LFX_2017} for a discussion). Hence, it is meaningful to consider together $\Delta_{x}$ and $\Delta_{c}$ even in a HF/LFX or xOEP calculation that omits correlation in the self-consistent cycle.
To conclude, we argue that the inclusion of $\Delta_\mathrm{x}$ without $\Delta_\mathrm{c}$ to an xOEP band-gap result (or to an LFX band-gap result, based on the similarity between xOEP-LFX) introduces a large systematic error of several eV to the xOEP/LFX prediction. Obviously, the accurate prediction for the band gap would be to include the whole of $\deltaxc$, but it is difficult to obtain an accurate estimate of $\Delta_\mathrm{c}$.
A small number of post-SCF calculations for $\Delta_\mathrm{c}$ in the literature overestimate the magnitude of $\Delta_\mathrm{c}$, predicting falsely insulators to be metallic as shown in Refs. \cite{MP2_Solids_1983_Suhai,MP2_Kresse_2010,MP2_Berkelbach_2021}. In this work, on balance, between including $\Delta_\mathrm{x}$ without $\Delta_\mathrm{c}$ and omitting $\deltaxc$ altogether, we chose the latter since the latter error is about an order of magnitude smaller than the former error.

With regard to strongly correlated systems, namely Mott insulators, the hallmark of a Mott insulator is a large $\deltaxc$ correction relative to the KS band gap, or equivalently that the majority of the contribution to the fundamental gap is from $\deltaxc$.
In cases where the LFX gap greatly underestimates the experimental fundamental gap, we believe this to be confirmation of strong Mott physics at play in FeO and that the correlation energy not taken into account by both HF and LFX is significant.

\section{Computational Details}\label{section:comp_details}
The algorithm to carry out the inversion of the density to find the LXC potential has been implemented in the plane-wave DFT code CASTEP\cite{CASTEP}.
As is usual with plane-wave implementations, the density, orbitals and potentials are represented on rectilinear grids\cite{Car_Parrinello_1985}. The KS orbitals are described within a spherical region of reciprocal space with a radius equal to the cutoff wave vector $G_\mathrm{cut}$, while the density and potential are non-zero in a region of $2G_\mathrm{cut}$.
The density is inverted through minimisation of the Coloumb energy difference $\Uobj[v]$, see Eq. \eqref{eq:Coulomb_diff} and is performed in real space to enable direct variation of the potential, see Eq. \eqref{eq:LXC_Grad}.

\begin{figure}[t]
    \centering
    \includegraphics[width=\linewidth]{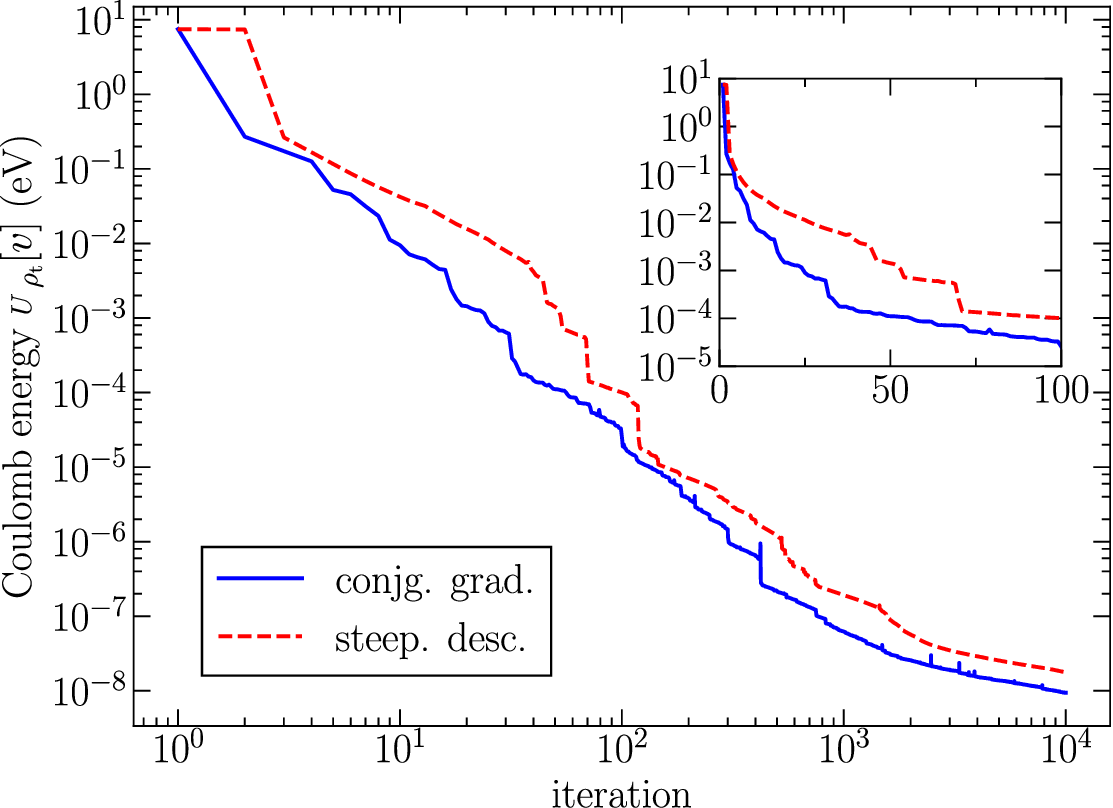}
    \caption{Convergence of the Coulomb energy difference $\Uobj[v]$, (refer to Eq. \eqref{eq:Coulomb_diff}), for inversion of the Hartree-Fock (HF) density of silicon using steepest descent (dotted red line) and Fletcher-Reeves-based\cite{Fletcher_Reeves_CG} conjugate gradient (solid blue line) algorithms. }
    \label{fig:SD_vs_CG_minimiser}
\end{figure}

A Fletcher-Reeves-based conjugate gradient algorithm\cite{Fletcher_Reeves_CG} is the algorithm of choice for the minimisation and is used to compute the search direction (in steepest descent this is simply Eq. \eqref{eq:LXC_Grad}).
The potential is then corrected along this search direction multiplied by a prefactor $\epsilon$, whose optimal value is determined via a line search that minimises $\Uobj[v]$ using a parabolic three-step fit.
Fig. \ref{fig:SD_vs_CG_minimiser} compares the convergence of $\Uobj[v]$ for the inversion of the HF density of silicon for steepest descent and conjugate gradient algorithms with the initial LXC potential set to the local density approximation (LDA) potential generated by the HF density.
Although the steepest descent algorithm exhibits a similarly shaped decrease in $\Uobj[v]$ to conjugate gradient, the rate of convergence starts to fall off rapidly after around 20 iterations only reaching a value of $4 \times 10^{-4} \text{ eV}$
after around 100 iterations, whereas the conjugate gradient algorithm reaches a value of $2\times 10^{-5} \text{ eV}$.
In both cases, the effective gradient term, see Eq. \eqref{eq:LXC_Grad},  tends to zero as expected from a variational method.

For our actual calculations, the differences in $\Uobj[v]$ were monitored over a window of $K$ iterations (we find $K=3$ works well), denoted $\{U_{\rho_\mathrm{t},K}[v]\}$.
The minimisation procedure was repeated until the \textit{difference} between the largest and smallest values of $\Uobj$ within the window is smaller than the threshold value of $\xi$, that is to say,
\begin{equation}
 \Delta \Uobj = \max \{U_{\rho_\mathrm{t},K}\} - \min \{U_{\rho_\mathrm{t},K}\}<\xi.
 \label{eq:convg_condition}
\end{equation}
We find that $\xi = 10^{-5} \text{ eV/atom}$ was sufficiently small to ensure that the calculation was well converged.
This ensured that the algorithm did not terminate prematurely due to a plateau in the energy landscape.

For all calculations presented here, the plane-wave cutoff energy $E_\mathrm{cut} = G_\mathrm{cut}^2/2$ and Monkhorst-Pack\cite{Mohkhorst_Pack_Grid} grid for Brillouin zone sampling were selected such that the self-consistent total energy was converged to less than $1 \text{ meV}$. The inversion algorithm was considered converged when the difference of the Hartree energy within a window of three iterations was less than $10^{-5}$ eV/atom.

With regard to the choice of pseudopotentials, we used norm-conserving pseudopotentials (NCPs)
from CASTEP's \textup{NCP19}\cite{CASTEP_NCP19_Reference} on-the-fly pseudopotential library at the same level of theory as the self-consistent calculation used to generate the target density.
The exception to this was HF and B3LYP\cite{B3LYP_1,B3LYP_2,B3LYP_3} hybrid functionals where LDA\cite{Kohn-Sham_DFT,LDA_CASTEP} NCPs were used, while in the case of PBE0\cite{PBE0} and HSE06\cite{HSE06,HSE03} functionals, PBE\cite{PBE} NCPs were used since these functionals contain predominantly PBE exchange and entirely PBE correlation.

For each solid, the experimental lattice parameters were taken from Ref. \cite{Madelung_Basic_Data} unless stated otherwise. The diamond cubic structure (space group: $Fd\bar{3}m$) was used for silicon, diamond and germanium; the zincblende structure (space group: $F\bar{4}3m$) for BAs, BP, CdS (for which the wurtzite structure [space group: $P6_3 mc$] was also computed), CdSe, GaAs, GaP, InP, SiC and ZnS;
the rocksalt or halite structure (space group: $Fm\bar{3}m$) for CaO, LiF, MgO and NaCl.
For the perovskites materials $\textrm{BaTiO}_3$, $\textrm{SrTiO}_3$, $\textrm{BaHfO}_3$, $\textrm{BaZrO}_3$ and $\textrm{KMgF}_3$, the ideal cubic phase (space group: $Pm\bar{3}m$) was used.
For the transition metal monoxide (TMOs), the rocksalt structure was used with a primitive rhombohedral computational cell of the AFM II magnetic structure commensurate with the antiferromagnetic ordering between alternating cubic (111) planes.

\section{Results and Discussion}\label{section:RESULTS}
\subsection{LDA and GGA Densities}\label{section:local_func_dens}
We begin with the inversion of target densities, $\tarden$, generated by the self-consistent solution of the KS equations for the two lowest rungs of Jacob's ladder\cite{Jacobs_ladder},
namely the local density approximation\cite{Kohn-Sham_DFT,LDA_CASTEP} (LDA) and generalised gradient approximations (GGAs).
For a purely local DFA like the LDA, the potential within the single-particle KS Hamiltonian is a local, multiplicative potential.
By construction, the LXC potential obtained by inversion of the LDA target density, $\vlxcdfa{LDA}$ is expected to be identical (up to a constant) to the LDA potential for that system $\ksxc{LDA}$
\begin{equation}
	\vlxcdfa{LDA} = \ksxc{LDA},
	\label{eq:LXC_LDA}
\end{equation}
by the Hohenberg-Kohn theorems.

In the case of a GGA such as the PBE\cite{PBE} functional,
although the XC energy functional $E_\mathrm{xc}[\rho,\nabla \rho]$ is semi-local as a result of the use of gradient expansions of the density,
the potential $\ksxc{GGA}$ is a local, multiplicative potential and can still be treated within the KS scheme without the use of the OEP method. For the same reason as target densities from LDA, we expect
\begin{equation}
	\vlxcdfa{GGA} = \ksxc{GGA}.
	\label{eq:LXC_GGA}
\end{equation}

\begin{figure}[t]
    \centering
    \includegraphics[width=0.8\columnwidth]{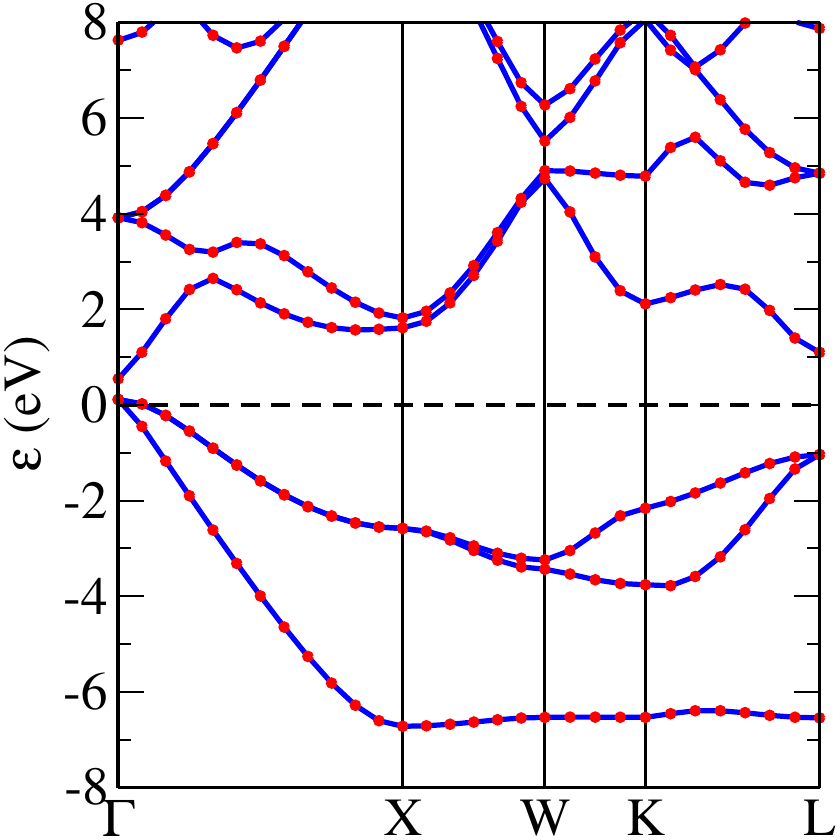}
    \caption{
        Computed band structures of GaAs using PBE (solid blue) and LXC-PBE (dotted red), the latter obtained via inversion of the PBE density.
        The Fermi energy has been set to 0 eV. Note that the two band structures are indistinguishable.}
    \label{fig:locpot_BS}
\end{figure}

Despite the statements contained within Eqs.\eqref{eq:LXC_LDA} and \eqref{eq:LXC_GGA} being %
obviously true, the inversion of the target densities from LDA and GGAs remains nonetheless insightful as a
benchmark to quantify the numerical error from the minimisation algorithm outlined in section \ref{section:RESULTS}, since the LXC potential from the DFA (or DFA-LXC potential for short), $\vlxcdfa{DFA}$ is known \textit{a priori} to the inversion.
We first performed a self-consistent calculation of the target density $\tarden$ and KS potential $v_s(\vec{r})$ (see Eq. \eqref{defn:KS_pot}) by self-consistently solving the KS equations for the LDA and PBE functionals and obtained the band structure for these functionals (using $v_s(\vec{r})$).
The LDA and PBE band gaps are given in Table \ref{table:non_TMO_gaps}.
We found that the LXC-LDA\footnote{
    For LDA target densities, the LXC potential was initialised to the PBE potential calculated from the target density. If the LXC potential was instead initialised to the LDA potential calculated from the target density like elsewhere in this work, the inversion is already converged without any further iteration, and more importantly remains converged.
}
and LXC-PBE band structures were indistinguishable
from the LDA and PBE band structures for all systems considered, with the mean pairwise absolute difference (MAD) between the LXC-DFA and DFA band gap being around 3 meV for both LDA and PBE. An example of this is shown in Fig. \ref{fig:locpot_BS} for GaAs using the PBE target density where the LXC-PBE band structure is indistinguishable from the PBE band structure.
An example of the inversion of the LDA density in diamond is given in the supplemental material\cite{Supplementary_Material}.

\begin{table*}[t]
	\caption{
		Computed band gaps (in eV) for various semiconductors and insulators.
		Experimental (expt.) band gaps and lattice parameters are from Ref. \cite{Madelung_Basic_Data} unless indicated otherwise. Zero-point renormalisation (ZPR) values to experimental gaps are obtained from Ref. \cite{Cardona_Thewalt_ZPR_values} and references therein unless stated otherwise.
		The LXC band gaps via density inversion are given in brackets  next to the corresponding DFA band gaps except for LDA and PBE
		which differed by around 3 meV or less from the self-consistent gap.
		Mean absolute errors (MAE) and mean absolute relative error (MARE) for each DFA and LXC (in brackets) are quoted with respect to the experimental gaps, accounting for ZPR,  where available.
		MAD and MARD refer to the mean pairwise absolute difference and mean pairwise absolute relative difference between pairs of LXC and GKS band gaps.
	}
	\label{table:non_TMO_gaps}
	\begin{ruledtabular}
		\begin{tabular}{c@{\hskip -0.75em}ddd@{\hskip -0.75em}dd@{\hskip -0.75em}dd@{\hskip -0.75em}dd@{\hskip -0.75em}dd@{\hskip -0.75em}ddd}
			&     \multicolumn{1}{r}{LDA}&     \multicolumn{1}{r}{PBE}&
			\multicolumn{2}{c}{rSCAN}&
			\multicolumn{2}{c}{B3LYP}&
			\multicolumn{2}{c}{PBE0}&
			\multicolumn{2}{c}{HSE06}&
			\multicolumn{2}{c}{HF}&
			\multicolumn{1}{c}{ZPR}& %
			\multicolumn{1}{r}{expt.}\\\hline
			BAs&	1.17&	 1.24&	1.43& (1.37)&	2.51& (1.54)&	2.53& (1.35)&	1.97& (1.36)&	7.75& (1.97)&
			-& 1.82\footnote{See Ref.\cite{BAs_band_gap}}\\
			BP&	1.19&	 1.26&	1.43& (1.38)&	2.57& (1.59)&	2.53& (1.36)&	1.94& (1.36)&	7.40& (1.79)&
			-& 2.1\\
			C\footnote{diamond}&	4.12&	 4.18&	4.39& (4.37)&	5.97& (4.59)&	6.06& (4.36)&	5.42& (4.37)&	12.21& (4.75)&
			-0.370& 5.5\\
			CaO&	3.51&	 3.67&	4.18& (4.01)&	5.64& (4.19)&	6.00& (4.18)&	5.29& (4.20)&	13.81& (5.63)&
			-0.357\footnotemark[11]& 6.93\\
			CdS\footnote{wurtzite structure}&	0.92&	 1.20&	1.62& (1.44)&	2.43& (1.42)&	2.83& (1.54)&	2.23& (1.53)&	8.67& (2.39)&
			-& 2.48\\
			CdS\footnote{zincblende structure}&	0.87&	 1.16&	1.57& (1.39)&	2.39& (1.36)&	2.78& (1.49)&	2.19& (1.48)&	8.58& (2.34)&
			-0.062& 2.5\\
			CdSe&	0.34&	 0.61&	1.07& (0.75)&	1.72& (0.83)&	2.11& (0.94)&	1.57& (0.92)&	7.97& (1.85)&
			-0.034& 1.9\\
			GaAs&	0.20&	 0.44&	0.89& (0.43)&	1.23& (0.47)&	1.70& (0.68)&	1.27& (0.66)&	6.78& (0.93)&
			-0.045& 1.52\\
			GaP&	1.44&	 1.60&	1.85& (1.70)&	2.70& (1.77)&	2.81& (1.67)&	2.23& (1.67)&	7.33& (1.95)&
			-0.086& 2.35\\
			Ge&	0.00&	 0.00&	0.31& (0.00)&	0.54& (0.00)&	1.09& (0.20)&	0.72& (0.18)&	5.91& (0.46)&
			-0.052& 0.79\\
			InP&	0.38&	 0.61&	1.01& (0.58)&	1.38& (0.65)&	1.75& (0.79)&	1.27& (0.77)&	6.36& (1.08)&
			-0.048& 1.42\\
			LiF&	8.78&	 9.01&	9.95& (9.60)&	11.76& (9.75)&	12.18& (9.69)&	11.43& (9.71)&	21.80& (10.91)&
			-1.231\footnotemark[11]& 13.6\\
			MgO&	4.57&	 4.64&	5.50& (5.25)&	7.02& (5.46)&	7.25& (5.31)&	6.57& (5.33)&	15.75& (6.87)&
			-0.533\footnotemark[11]& 7.9\\
			NaCl&	4.66&	 5.07&	5.79& (5.34)&	6.77& (5.35)&	7.21& (5.43)&	6.50& (5.43)&	13.81& (6.20)&
			-& 8.97\\
			Si&	0.48&	 0.58&	0.77& (0.71)&	1.70& (0.87)&	1.69& (0.69)&	1.14& (0.69)&	6.12& (1.17)&
			-0.062& 1.17\\
			SiC&	1.31&	 1.36&	1.73& (1.69)&	2.90& (1.85)&	2.85& (1.59)&	2.23& (1.60)&	8.27& (2.30)&
			-0.175\footnotemark[11]& 2.42\\
			ZnS&	1.80&	 2.04&	2.57& (2.33)&	3.32& (2.16)&	3.67& (2.22)&	3.06& (2.21)&	9.03& (2.19)&
			-0.105& 3.78\\
			$\textrm{BaHfO}_3$\footnote{Lattice parameters from Ref.\cite{BaHfO3_Lat_Param}, expt. band gap from Ref.\cite{BaHfO3_band_gap}}&	3.46&	 3.63&	4.14& (3.98)&	5.27& (4.18)&	5.62& (4.21)&	5.11& (4.19)&	12.66& (5.78)&
			-& 6.1\\
			$\textrm{BaTiO}_3$\footnote{Lattice parameters from Refs.\cite{BaTiO3_Lat_Param_1,BaTiO3_Lat_Param_2}, expt. band gap from Ref.\cite{BaTiO3_band_gap}}&	1.75&	 1.83&	2.18& (2.09)&	3.30& (2.26)&	3.69& (2.32)&	2.98& (2.31)&	11.67& (4.07)&
			-& 3.2\\
			$\textrm{BaZrO}_3$\footnote{Lattice parameters from Refs.\cite{BaZrO3_Lat_Param_1,BaZrO3_Lat_Param_2}, expt. band gap from Ref.\cite{BaZrO3_band_gap}}&	3.09&	 3.20&	3.78& (3.68)&	4.81& (3.85)&	5.12& (3.85)&	4.62& (3.83)&	12.13& (5.40)&
			-& 5.3\\
			$\textrm{KMgF}_3$\footnote{Lattice parameters from Refs.\cite{KMgF3_Lat_Param_1_and_gap,KMgF3_Lat_Param_2}, expt. band gap from Ref.\cite{KMgF3_Lat_Param_1_and_gap}}&	6.88&	 7.24&	8.14& (7.81)&	9.88& (8.00)&	10.38& (8.03)&	9.63& (8.02)&	19.86& (9.72)&
			-& 12.4\\
			$\textrm{SrTiO}_3$\footnote{Lattice parameters from Refs.\cite{SrTiO3_Lat_Param}, expt. band gap from Ref.\cite{SrTiO3_band_gap}}	&	1.78&	 1.88&	2.33& (2.22)&	3.39& (2.42)&	3.80& (2.48)&	3.08& (2.47)&	11.90& (4.25)&
			-0.44\footnote{See Ref.\cite{ZPR_SrTiO3_Bhandari_Schlifgaarde}}& 3.25\\
			\hline
			\text{MAE}& 1.87& 1.70& 1.24& (1.44)& 0.61& (1.33)& 0.57& (1.34)& 0.58& (1.34)& 6.45& (0.69)&
			{}&  {}\\
			\text{MARE}&
			51.1\%& 45.4\%& 30.4\%& (39.0\%)&
			16.3\%& (35.3\%)& 19.2\%& (34.6\%)& 10.7\%& (35.0\%)&
			241.0\%& (16.8\%)&
			{}& {}\\
			\text{MAD}&    \multicolumn{1}{c}{$10^{-3}$}&  \multicolumn{1}{c}{$10^{-3}$}&
			\multicolumn{2}{c}{0.20}&   \multicolumn{2}{c}{1.12}&   \multicolumn{2}{c}{1.42}&   \multicolumn{2}{c}{0.83}&   \multicolumn{2}{c}{6.90}&
			{}& {}\\
			\text{MARD}&   0.18\% &  0.15\% &  \multicolumn{2}{c}{14.6\%}& \multicolumn{2}{c}{36.9\%}& \multicolumn{2}{c}{40.7\%}& \multicolumn{2}{c}{28.9\%}& \multicolumn{2}{c}{68.6\%}&
			{}& {}\\
		\end{tabular}
	\end{ruledtabular}
	\footnotetext[11]{See Ref.\cite{Kresse_ZPR_values}}
\end{table*}

\subsection{Local Fock Exchange(LFX)}\label{section:LFX}
Per the discussion in section \ref{subsection:total_energies_theory}, the KS orbitals are density optimal\cite{Kohn_Nobel_Lecture_1999} in contrast to the HF orbitals, which are energy optimal giving the minimising
Slater determinant $\Phi_\mathrm{HF}$ of the HF energy functional.
The (unconstrained) minimisation of the HF functional yields single-particle equations which contain a non-local potential in the form of the Fock exchange operator
\begin{equation}
	\gksx[HF]\phi_j^\sigma(\vec{r})
	=
	-\int\intd{r'}
	\sum_i
	\frac{\phi_i^\sigma(\vec{r})[\phi_i^\sigma(\vec{r'})]^*}
	{|\vec{r}-\vec{r'}|}
	\phi_j^\sigma(\vec{r'})
	.
\end{equation}
In a wider sense HF together with hybrid DFAs can be considered within a GKS scheme wherein the GKS equations are single-particle equations with a non-local exchange potential as part of the full XC potential $\op{v}{\mathrm{xc}}{\mathrm{GKS}}$.
Alternatively, these DFAs can be considered within a KS scheme with a local XC potential $\ksxc{}$ but due to the orbital dependence, one cannot evaluate Eq. \eqref{eq:xc_pot_defn} directly and instead the OEP method must be used.
A similar LXC (more accurately, local exchange-only) potential can be obtained by inversion of the HF density, which, following the terminology of Hollins \textit{et al.}, we will refer to as the local Fock exchange\cite{Hollins_LFX_2017} (LFX) potential $\vlfx\equiv\vlxcdfa{HF}$.

\begin{figure}[t]
	\centering
	\includegraphics[width=0.80\columnwidth]{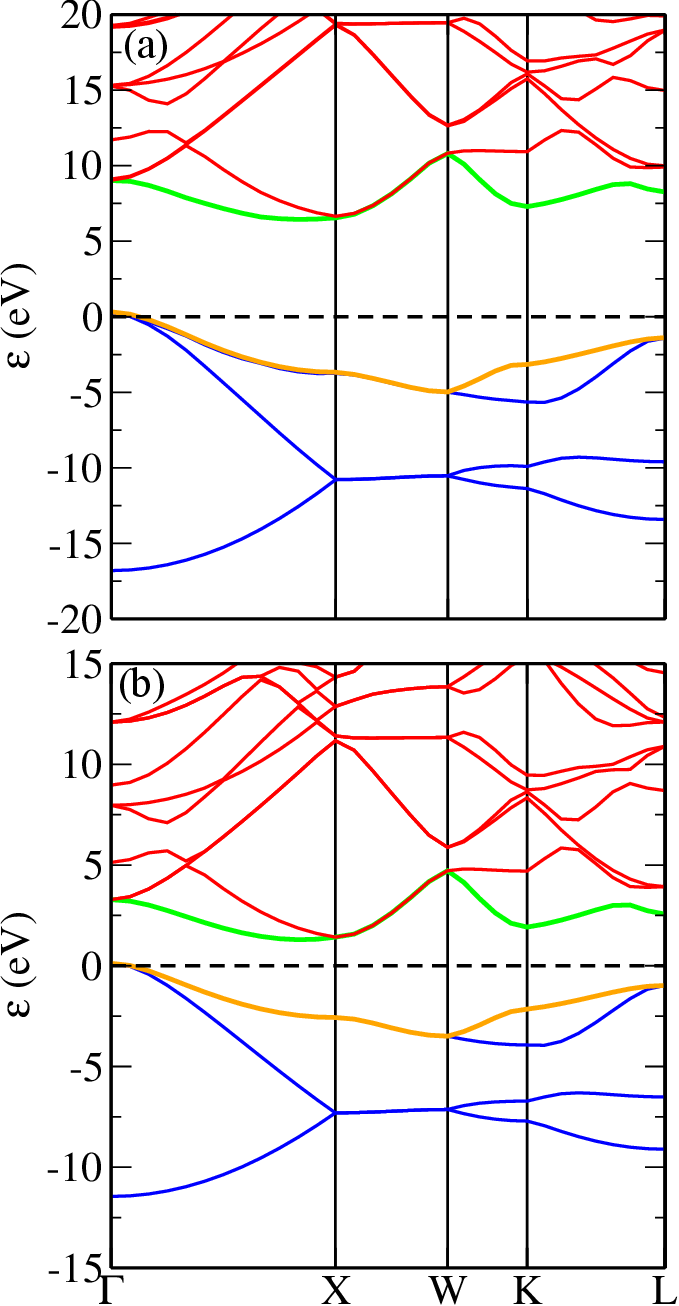}
	\caption{Computed band structures of Si for (a) HF and (b) LFX obtained by inversion of the HF density. The Fermi energy has been arbitrarily set to 0 eV in both band structures. Occupied bands are in blue, valence band in orange, conduction band in green and unoccupied bands in red.}
	\label{fig:Si_HF_den}
\end{figure}

Table \ref{table:non_TMO_gaps} gives the HF and LFX band gaps for various systems. As expected, the HF band gap is typically a massive overestimate of the experimental band gaps. In the literature, this effect is often attributed to the lack of correlation \cite{Ashcroft_Mermin}. However, see the discussion in Ref. \cite{Nikitas_hyper_HF} that suggests this is instead an artefact of the non-locality in the Fock exchange operator.
As a result of the strongly non-local nature of the HF exchange potential $\gksx[HF]$, the band structures obtained from $\vlfx$ will naturally be quite different from the original HF band structure obtained via $\gksx[HF]$,
unlike the LXC-LDA and LXC-PBE band structures where the potentials obtained by inversion, $\vlxcdfa{LDA}=\ksxc{LDA}$ and $\vlxcdfa{PBE}=\ksxc{PBE}$.
In contrast to HF, the LFX band gaps are generally a systematic improvement over the unphysically large HF band gaps, when compared to experiment.
As discussed in Refs.\cite{Hollins_LFX_2017,LFX_Metals_2017}, the LFX potential $\vlfx$ is expected to be similar to the xOEP potential $v_\mathrm{xOEP}(\vec{r})$ in weakly interacting systems whose exchange energy is much larger than the correlation energy. Therefore, the LFX band structures and band gaps obtained here and in Refs.\cite{Hollins_LFX_2017,LFX_Metals_2017} will be similar to those calculated using xOEP\cite{EXX_Semiconductors_Görling_1997,EXX_Semiconductors_Görling_1999,EXX_Delta_X_Grüning_Rubio_2006,Hollins_Hylleraas_OEP,EXX_RPA_Kresse_2014}.

In particular for Si, shown in Fig. \ref{fig:Si_HF_den} the LFX band gap happens to be equal to the experimental band gap of $1.17 \text{ eV}$ while the HF band gap is $6.12 \text{ eV}$.
In the case of Ge, local and semi-local DFAs\cite{Bachelet_1985_Ge_metallic,Ekuma_2013_Ge_metallic} like LDA and PBE
incorrectly predict it to be metallic rather than insulating.
On the other hand, the LFX potential correctly describes Ge qualitatively, with a gap $0.46 \text{ eV}$, suggesting that the failure of the LDA and PBE potentials is probably due to an inaccurate description of exchange.
Alternatively, these results can be improved by resorting to many-body perturbation theory techniques such as the $GW$ method\cite{GW_original_paper_Hedin_1965,GW_Ge_band_gap_correct_1_Rohlfing_1993,GW_Ge_band_gap_correct_2_Fleszar_2001}.
We note that large relative errors were obtained for LFX in GaAs (39\%) and ZnS (42\%).

In the case of the cubic perovskites \BTO, \STO{} and \BZO, we found that the computed LFX band gap was an overestimate in contrast to the other semiconductors and insulators in Table \ref{table:non_TMO_gaps}, although the same qualitative improvement over HF was observed.
Comparable overestimates in these cubic perovskites were obtained in the xOEP calculations by Betzinger \textit{et al.}\cite{Betzinger_Görling_2012_EXX_perovskites} and Trushin \textit{et al.}\cite{Trushin_Görling_2019_EXX_perovskites}.
As discussed in the latter, the error is not necessarily due to an inherent shortcoming of the xOEP/LFX potentials but rather to additional physical effects not considered here such as zero-point motion of the lattice that leads to band gap renormalisation. In transition metal perovskites such as \STO{}, this can be as large as $\sim 0.4 \text{ eV}$ as calculated in Ref. \cite{ZPR_SrTiO3_Bhandari_Schlifgaarde}, although the effect is largely insignificant in other semiconductors and insulators (see Ref. \cite{ZPR_Kresse_Semiconductors}).

\begin{figure}[t]
	\centering
	\includegraphics[width=\columnwidth]{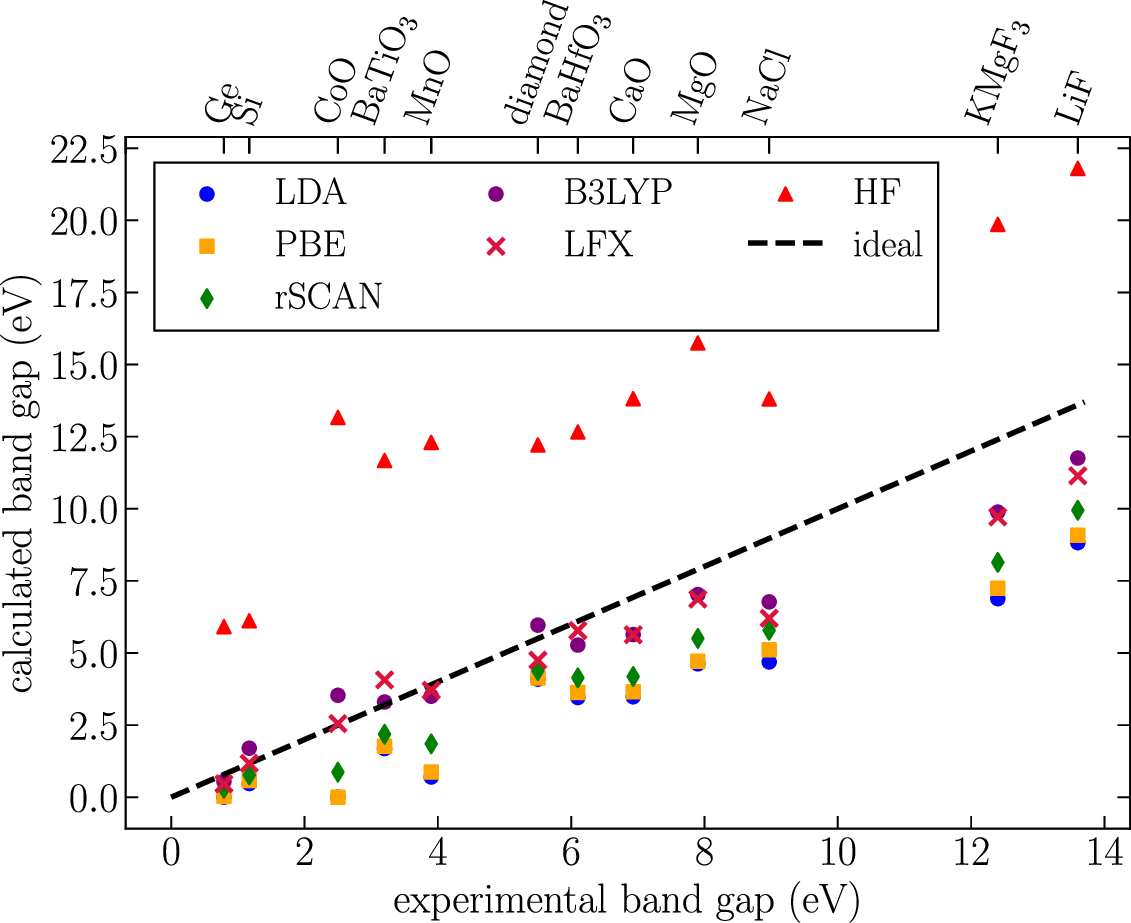}
	\caption{Comparison of calculated band gaps with experimental band gaps. Local Fock exchange (LFX) denotes the band gap calculated using the LXC potential obtained from the inversion of the Hartree-Fock (HF) density.
		The straight line indicates perfect agreement between theory and experiment; points above the line indicate that the band gap is overestimated while points below the line indicate the band gap is underestimated.
	}
	\label{fig:experi_vs_calc_gaps}
\end{figure}

Finally, as shown in Fig. \ref{fig:experi_vs_calc_gaps}, the degree to which the LFX band gap underestimates the fundamental band gap increases in systems with large fundamental gaps.

\subsection{Reducing non-locality: Hybrid Functionals}\label{section:hybrids}
We have seen in section \ref{section:local_func_dens} that the LXC potential obtained from the inversion of LDA and PBE densities is identical to the original potential LDA and PBE KS potentials, $\vlxcdfa{LDA}=\ksxc{LDA}$ and $\vlxcdfa{PBE}=\ksxc{PBE}$.
At the other end of the non-locality spectrum, in the previous section, we have shown that a target density generated by the non-local HF (exchange-only) potential results in a LFX potential $\vlfx$ with improved band structures and particularly band gaps compared to the HF results.
The natural question is what happens when the two are combined.

In hybrid functionals, the exchange energy $E_\mathrm{x}^\mathrm{DFA}[\rho,\phi_i]$ is a mixture of non-local Fock exchange and local/semi-local DFA exchange energy. Considered in a GKS scheme, this gives rise to a non-local exchange potential that is not as strongly non-local as the full HF exchange potential due to a smaller weighting on the Fock exchange energy $E_\mathrm{x}^\mathrm{HF}[\phi]$ in the total energy functional.
We considered target densities from three hybrid functionals: B3LYP\cite{B3LYP_1,B3LYP_2,B3LYP_3}, PBE0\cite{PBE0} and HSE\cite{HSE03} (we specifically used the HSE06\cite{HSE06} parametrization).

In all hybrid functionals considered, the LXC band gap obtained by the inversion of the respective hybrid target densities was lower than the calculated GKS band gap. This behaviour of the LXC potential $\vlxcdfa{DFA}$ is analogous to the behaviour of the LFX potential $\vlfx$, although the effect is reduced due to the smaller weighting of HF exchange $E_\mathrm{x}^\mathrm{HF}$ in the full exchange energy $E_\mathrm{x}^\mathrm{DFA}$.

The B3LYP functional mixes several exchange and correlation functionals with HF exchange $E_\mathrm{x}^\mathrm{HF}$,
\begin{equation}
	\begin{split}
		E_\mathrm{xc}^\mathrm{B3LYP}[\rho, \nabla \rho, \phi_i]
		=
		\alpha E_\mathrm{x}^\mathrm{HF}[\phi_i] + (1-\alpha) E_\mathrm{x}^\mathrm{LDA}[\rho]
		\\+ \beta \Delta E_\mathrm{x}^\mathrm{B88}[\rho, \nabla \rho]
		+ \gamma E_\mathrm{c}^\mathrm{LYP}[\rho,\nabla\rho]
		\\+ (1-\gamma) E_\mathrm{c}^\mathrm{LDA}[\rho].
	\end{split}
	\label{eq:B3LYP_defn}
\end{equation}
$\Delta E_\mathrm{x}^\mathrm{B88}$ is the Becke-88\cite{B88} (B88) gradient-correction (to the LSDA) and $E_\mathrm{c}^\mathrm{LYP}$ is the correlation functional of Lee, Yang and Parr\cite{B3LYP_2} (LYP)
with\cite{B3LYP_1} $\alpha=0.2, \beta = 0.72$ and $\gamma = 0.81$.
The only term in Eq. \eqref{eq:B3LYP_defn} that gives rise to a non-local potential in a GKS scheme is the HF exchange energy $E_\mathrm{x}^\mathrm{HF}[\phi_i]$ while all other terms are all explicitly local/semi-local density functionals which yield a local, multiplicative potential.
The LXC-B3LYP gaps had a smaller deviation from GKS-B3LYP than LFX and HF, highlighting the weaker non-locality in B3LYP compared to HF.
Nevertheless, we
note that the LXC-B3LYP band gaps are still typically larger than the LDA and PBE band gaps.
The reason for this is two-fold; the first is the aforementioned analogous behaviour of $\vlxcdfa{DFA}$ to $\vlfx$.
The second is the semi-local B88 functional $E_\mathrm{x}^\mathrm{B88}$ constructed as a correction to the L(S)DA which,
as noted by Becke\cite{B88}, obtains a greater portion of the full exchange energy due to recovering the correct asymptotic behaviour of the exchange energy density.

We stress that the non-linearity of the inversion process means that the LXC potential does not simply subtract off the non-local HF potential and replace it with the LFX potential.
Notably, in the case of Ge, the LXC-B3LYP potential $\vlxcdfa{B3LYP}$ gives a metallic band structure whereas the GKS-B3LYP band gap is 0.54 eV; the latter is also closer to the experimental value of 0.785 eV than the LFX band gap of 0.46 eV.
We also found that the LXC-B3LYP band gaps were similar to the GKS-rSCAN band gaps %
which we will explore further in section \ref{section:MGGAs}.

\begin{figure}[t]
	\centering
	\includegraphics[width=0.8\columnwidth]{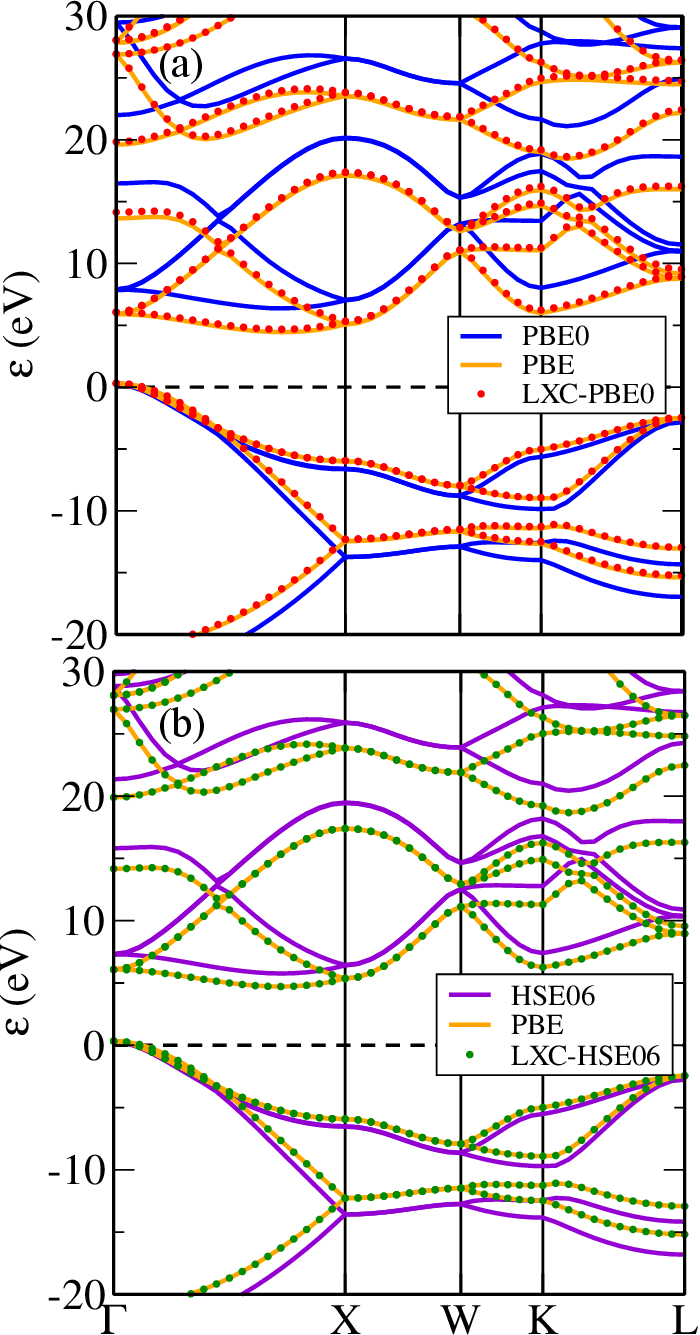}
	\caption{Computed DFA (solid) and LXC (dotted) band structures of diamond. In panels (a,b), the dotted red lines are LXC-PBE0 while in panels (c,d), the dotted green lines are LXC-HSE06. The solid lines are in (a) GKS-PBE0, (c) GKS-HSE06 and (b,d) PBE. The energy scale has been chosen such that the valence band maxima coincide and the Fermi energy is 0 eV.}
	\label{fig:diamond_hybrid_bs}
\end{figure}

In the case of PBE0, which contains $25\%$ non-local HF exchange with the remainder of the XC energy $E_\mathrm{xc}^\mathrm{PBE0}$ coming from semi-local PBE, one might expect the LXC band gap to be similar to the PBE band gap with the LXC potential similar to the PBE potential.
Indeed, this is the case as shown in Fig. \ref{fig:diamond_hybrid_bs}(a) for diamond where the LXC band structure obtained by inversion of the PBE0 target density is virtually identical to the PBE band structure.
Likewise, the remainder of Table~\ref{table:non_TMO_gaps} shows that the LXC-PBE0 band gap is close (but not necessarily identical) to the PBE gap.
Notably in Ge, LXC-PBE0 gave a small band gap of 0.20 eV in contrast to PBE and LXC-B3LYP, which predicted it to be metallic.

Turning our attention to a range-separated hybrid such as HSE06, the short-range exchange (a mixture of HF non-local exchange and PBE semi-local exchange) is screened by PBE exchange at long-range.
Since we have a different DFA from PBE0, it is not clear in advance how the densities and thus the LXC potentials from HSE06 and PBE0 will differ, in particular the degree to which screening in the former will affect the density and potential.
Indeed, the MAD between HSE06 and PBE0 GKS gaps is 0.60 eV, while the mean absolute pairwise relative difference MARD is 18.32\%  for the semiconductors and insulators in Table~\ref{table:non_TMO_gaps},  which is unsurprising due to the screened Fock operator in the former that can potentially lead to large differences in GKS eigenvalues.

However, it turns out that for the systems we considered, there is only a small difference between LXC-PBE0 and LXC-HSE06 band gaps with a MAD of $0.012 \text{ eV}$ (and a MARD 1.2\%).
The similarity of the two LXC potentials from PBE0 and HSE06, $\vlxcdfa{PBE0}$ and $\vlxcdfa{HSE06}$, as can be seen in the band structures in Fig. \ref{fig:diamond_hybrid_bs}, suggests that both DFAs yield similar densities. Furthermore, given the similarity of LXC-PBE0 to PBE, by extension, both PBE0 and HSE06 have similar densities to PBE and their respective LXC potentials are similar to the PBE potential $v_\mathrm{xc}^\mathrm{PBE}(\vec{r})$.
In particular, it appears that screening within the HSE06 DFA has a minimal effect on the density compared to unscreened PBE0.
Together with the fact that both HSE06 and PBE0 have the same weighting of Fock exchange (25\%), it is not too surprising that LXC-HSE06 is similar to LXC-PBE0.
The similarity of the LXC band structures suggests that the PBE0 and HSE06 density in diamond are similar.
The difference in band gaps obtained when these two DFAs are considered in a GKS scheme is thus due to the screening of HF exchange in HSE06 that is absent in PBE0. Therefore, despite having similar densities, the two DFAs yield different GKS band gaps.

\subsubsection{Total Energy Differences and Band Gaps}
The inversion of a DFA's density finds the local Kohn-Sham potential $v(\vec{r})$ that reproduces that density.
As discussed in section \ref{subsection:total_energies_theory},  the orbitals $\vorb$ obtained by solving the Kohn-Sham equations with $v_\mathrm{xc}(\vec{r}) = \vlxcdfa{DFA}$ do not fully minimise the total energy $E^\mathrm{DFA}$ for a non-local DFA due to the additional constraint that the potential obtained from the inversion must be local.
Therefore, the total energy difference between $E^\mathrm{DFA}[\{\phi_i^\mathrm{LXC}\}]$ and $E^\mathrm{DFA}[\{\phi_i^\mathrm{GKS}\}]$, defined in Eq. \eqref{eq:inv_eng_diff}, can thus be used as a metric to gauge the degree of non-locality in a particular GKS scheme.

\begin{table}[b]
  \begin{ruledtabular}
    \begin{tabular}{ccc}
      &HF(LFX)          &B3LYP \\ \hline
      BAs                       &390            &16\\
      BP                        &179            &5\\
      C-diamond                 &152            &6\\
      CaO                       &255            &14\\
      CdS (zincblende)          &347            &16\\
      CdSe                      &644            &23\\
      GaAs                      &536            &23\\
      GaP                       &348            &11\\
      Ge                        &519            &20\\
      InP                       &449            &12\\
      LiF                       &093            &5\\
      MgO                       &112            &5\\
      NaCl                      &109            &7\\
      Si                        &228            &9\\
      SiC                       &140            &5\\
      ZnS                       &371            &13\\
      \BHO                      &664            &23\\
      \BTO                      &840            &44\\
      \BZO                      &664            &38\\
      \KMF                      &290            &16\\
      \STO                      &777            &44
    \end{tabular}
    \caption{Total energy differences (in meV) between the evaluation of the total energy functional using the LFX/LXC orbitals and self-consistent HF/GKS orbitals as defined by Eq.\eqref{eq:inv_eng_diff}. }
    \label{table:tot_eng_diffs}
  \end{ruledtabular}
\end{table}

Table \ref{table:tot_eng_diffs} gives total energy differences for HF and B3LYP when evaluated using the GKS $\phi_i^\mathrm{GKS}(\vec{r})$ orbitals and LXC orbitals $\vorb$ obtained via the KS equations with the XC potential equal to the LXC potential $\vlxcdfa{DFA}$.
As one might expect, the total energy difference between LFX and HF is the largest since HF contains a strongly non-local exchange-only potential.
Moreover, these energy differences are similar but not identical to those between xOEP and HF\cite{Hollins_LFX_2017}.
We note that the xOEP potential satisfies the virial relation for exchange\cite{exchange_virial_Levy_Perdew_1985}
while LFX satisfies the virial relation only approximately to second order\cite{Hollins_LFX_2017}.
Turning now to B3LYP which contains only 20\% HF exchange, the constraint of a local potential is less ``severe'' and thus the orbitals $\vorb$ generated by $\vlxcdfa{B3LYP}$ are closer to the energy optimal GKS orbitals $\phi_i^\mathrm{GKS}(\vec{r})$ generated by the non-local GKS potential $\op{v}{\mathrm{xc}}{\mathrm{B3LYP}}$.
This can also be inferred from the smaller changes in the band structure between LXC-B3LYP and GKS-B3LYP compared to LFX and HF.

To better investigate the effects of non-locality on both the band structure and total energy difference, we considered a PBE0-like hybrid functional with parameter $\alpha$
\begin{equation}
	\begin{split}
	    E_\mathrm{xc}[\rho,\nabla\rho ,\phi_i] = \alpha E_\mathrm{x}^\mathrm{HF}[\phi_i]
	+
	(1-\alpha) E_\mathrm{x}^\mathrm{PBE}[\rho,\nabla\rho]
	\\+ E_\mathrm{c}^\mathrm{PBE}[\rho,\nabla\rho],
	\end{split}
	\label{eq:PBE0_like_hybrid}
\end{equation}
where $\alpha=25\%$ gives the standard PBE0 functional.
For various values of $\alpha$, the DFA defined by Eq. \eqref{eq:PBE0_like_hybrid} was treated in a GKS scheme from which the GKS band gap and target density were obtained before inversion to obtain the LXC potential and band gap.

\begin{figure*}[t]
	\centering
	\includegraphics[width=0.85\linewidth]{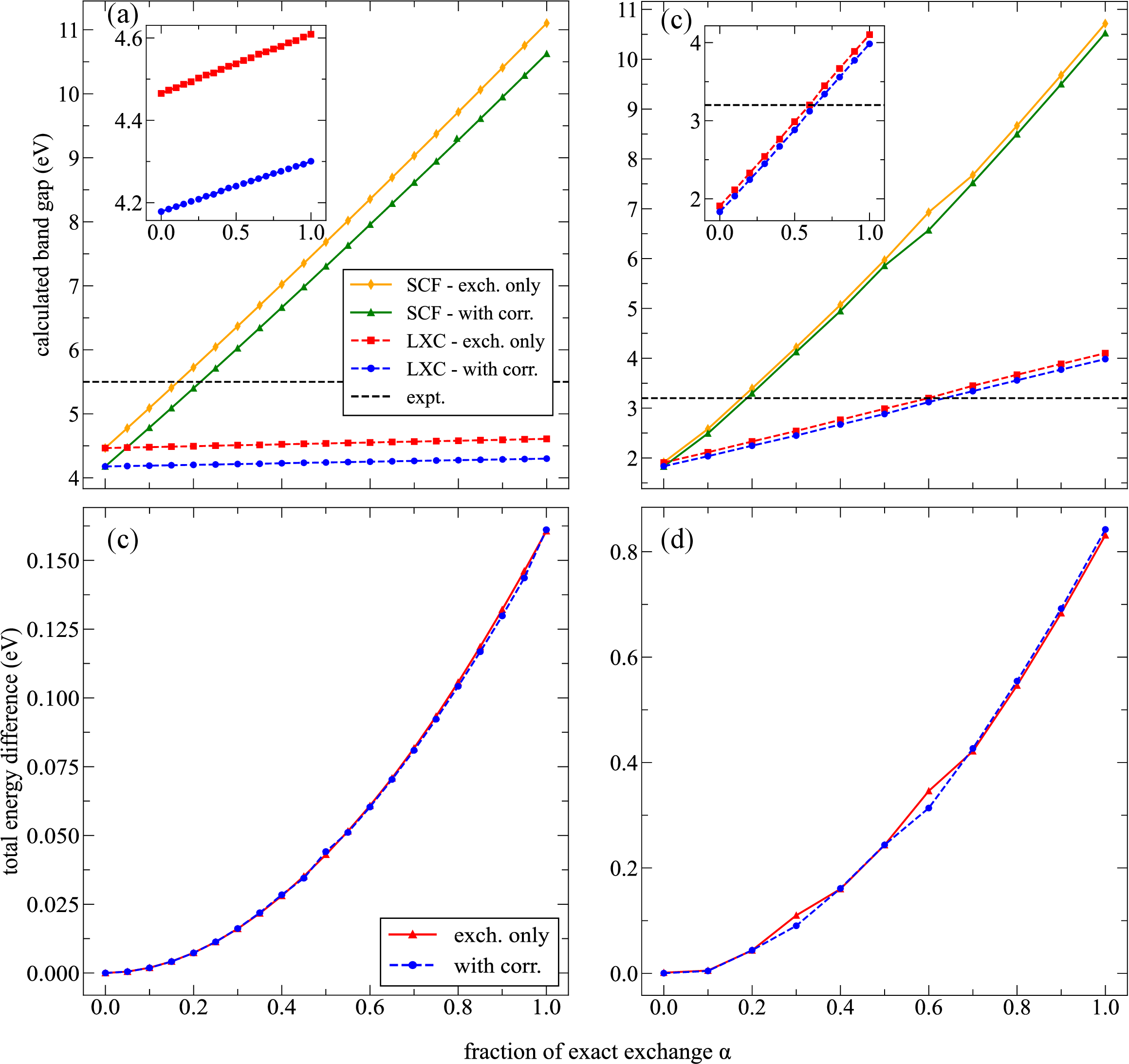}
	\caption{
		Calculated band gap using a PBE0-like hybrid functional (see Eq.\eqref{eq:PBE0_like_hybrid}) with varying $\alpha$ fraction of HF exchange for (a) diamond and (b) \BTO{}.
		Solid orange and green lines are obtained in GKS with and without PBE correlation respectively while the dashed red and blue are the LXC gaps obtained from inversion of the corresponding GKS density.
		Experimental (expt.) gaps are from Refs.\cite{Madelung_Basic_Data} and \cite{BaTiO3_band_gap} respectively.
		(c) and (d) show the difference in the total energy between evaluation using LXC and GKS orbitals according to Eq.\eqref{eq:inv_eng_diff} for diamond and \BTO{} respectively.
		}
        \label{fig:diamond_BTO_alpha}
\end{figure*}

The results are shown in Fig. \ref{fig:diamond_BTO_alpha} for diamond and \BTO{} with and without the inclusion of semi-local PBE correlation $E_\mathrm{c}^\mathrm{PBE}$.
At low $\alpha$, the total energy difference is small (zero at $\alpha=0$) but increases rapidly in a non-linear fashion as $\alpha\to 1$, i.e. full HF exchange. The GKS band gap increases faster than the LXC band gap with $\alpha$, where at $\alpha=1$ in the absence of correlation, one obtains the LFX gap.
We note that the inclusion of semi-local PBE correlation $E_\mathrm{c}^\mathrm{PBE}$ does not significantly change the difference in total energies although it does result in a reduction of both the GKS and LXC band gaps.

The difference in exchange-only band gaps and those with semi-local PBE correlation exhibits a weak dependence on $\alpha$ with the difference between the two for a given $\alpha$ remaining nearly constant; in diamond the difference in LXC band gaps between exchange-only results varied between 0.29--0.31 eV and 0.08-0.12 \text{ eV} in \BTO.
The connection between the band gap and total energy is not fully clear; in particular, a larger difference in total energies does not necessarily directly translate to a larger difference between GKS and LXC band gaps.
For \BTO (see Fig. \ref{fig:diamond_BTO_alpha} and Tables \ref{table:non_TMO_gaps} and \ref{table:tot_eng_diffs}) the HF gap is 11.67 eV and the LFX gap is $4.07 \text{ eV}$
while the total energy difference of the HF functional when evaluated using the LFX and HF orbitals is 0.840 eV.
By contrast, for diamond, the HF gap is $12.21 \text{ eV}$ and the LFX gap is $4.75 \text{ eV}$
yet the total energy difference is only $0.152 \text{ eV}$, more than a factor of 5 smaller than the energy difference for \BTO{} despite both systems having similar band gap differences between HF and LFX.

At high $\alpha$, one might expect that the inclusion of correlation would give a calculated LXC band gap closer to the experimental value. However, we note that while LFX (and xOEP) treat exchange exactly,
DFAs can still remain accurate in spite of the inexact treatment of exchange due to a cancellation of errors when the corresponding exchange and correlation DFAs are used in tandem, in this case, PBE exchange with PBE correlation.

\subsubsection{Transition Metal Monoxides (TMOs)}
\begin{table*}[t!]
  \caption{
    Same as table \ref{table:non_TMO_gaps} for transition metal monoxide band gaps (in eV). Calculations done with
    rock-salt structure using experimental lattice parameters from Ref.\cite{TMO_lat_params}.
  }
  \label{table:TMO_gaps}
  \begin{ruledtabular}
    \begin{tabular}{cddd@{\hskip -0.75em}dd@{\hskip -0.75em}dd@{\hskip -0.75em}dd@{\hskip -0.75em}dd@{\hskip -0.75em}ddd}
      &     \multicolumn{1}{r}{LDA}&     \multicolumn{1}{r}{PBE}&     \multicolumn{2}{c}{rSCAN}&     \multicolumn{2}{c}{B3LYP}&     \multicolumn{2}{c}{PBE0}&     \multicolumn{2}{c}{HSE06}&     \multicolumn{2}{c}{HF}&     \multicolumn{1}{r}{expt.}
      \\\hline
      CoO         &0.00         &0.00       &0.87 &(0.77)
                                                                                                                              &3.53 &(1.30)    &4.24 &(1.45)   &3.76 &(1.44) &13.17 &(2.55) &2.5\footnotemark[1]
      \\
      FeO         &0.00         &0.00       &0.45 &(0.46)
                                                                                                                              &\multicolumn{2}{c}{--}       &\multicolumn{2}{c}{--}
                                                                                                                                                                                           &\multicolumn{2}{c}{--}
                                                                                                                                                                                                                       &7.96 &(0.81)   &2.4\footnotemark[2]
      \\
      MnO         &0.81         &0.96       &1.85 &(1.81)
                                                                                                                              &3.50 &(2.02) &3.70 &(2.02)   &3.18 &(2.01)  &12.30 &(3.94) &3.9\footnotemark[3]
      \\
      NiO         &0.66         &1.23       &2.84 &(2.61)
                                                                                                                              &4.44 &(2.52) &5.41 &(2.72)   &4.67 &(2.71) &13.93 &(4.03) &4.0\footnotemark[4]
      \\
      \hline
      \text{MAE}&      2.83&  2.65&   1.70& (1.79)&   0.62& (1.52)&   1.12& (1.41)&   0.88& (1.41)&   8.64& (0.43)&
      \\
      \text{MARE}&     90.7\%&        86.1\%& 57.1\%& (59.6\%)&       20.8\%& (44.4\%)&       36.7\%& (40.8\%)&       28.6\%& (41.0\%)&       280.6\%& (17.6\%)&
      \\
      \text{MAD}&      \multicolumn{1}{c}{-}& \multicolumn{1}{c}{-}&  \multicolumn{2}{c}{0.10}&       \multicolumn{2}{c}{1.88}&       \multicolumn{2}{c}{2.39}&       \multicolumn{2}{c}{1.82}&   \multicolumn{2}{c}{9.01}&
      \\
      \text{MARD}&     \multicolumn{1}{c}{-}& \multicolumn{1}{c}{-}&
                                                                     \multicolumn{2}{c}{6.2\%}&        \multicolumn{2}{c}{49.6\%}&       \multicolumn{2}{c}{53.7\%}&       \multicolumn{2}{c}{46.8\%}&   \multicolumn{2}{c}{77.4}&
    \end{tabular}
    \footnotetext[1]{Value from Ref. \cite{CoO_band_gap_experimental}}
    \footnotetext[2]{Value from Refs.
      \cite{FeO_band_gap_experimental,TMOs_2006_Hybrid_Calculations}}
    \footnotetext[3]{Value from Ref. \cite{MnO_band_gap_experimental}}
    \footnotetext[4]{Value from Ref. \cite{NiO_band_gap_experimental}}
  \end{ruledtabular}
\end{table*}
The transition metal monoxides (TMOs) (Co,Fe,Mn,Ni)O, which adopt an AFM-II ([111]) insulating ground state present challenges for simple local and semi-local density functionals
due to strong inter-electron interactions, particularly for $d$-electrons.
These interactions lead to strong localisation that is often underestimated by these functionals as a result of their tendency to overly delocalise electrons due to the presence of self-interaction error (SIE)\cite{LDA_CASTEP,Self_Interaction_Corrected_TMOs_1990_Svane,Localisation_Delocalisation_Error_Yang_2008,DFT_U_Str_Correlation_Review_2014,delocalisation_error_perspective_2023}
leading to the prediction of both quantitatively as well as qualitatively incorrect material properties.
HF on the other hand does not have SIE and by extension, hybrid functionals which include a portion of HF exchange are able to partly correct the SIE\cite{TMOs_2006_Hybrid_Calculations,Gillen_Robertson_TMOs_Screened_Exchange_2013}.
These errors can also be corrected through DFT+$U$ (discussed in the next section), xOEP\cite{Engel_2009_TMOs_EXX,Hollins_Hylleraas_OEP} as well as many-body perturbation theory methods such as the $GW$ approximation\cite{TMOs_2013_Dynamical_Screen_Coulomb_Int_cRPA} and dynamical mean field theory\cite{DMFT_Review_Kotliar_2006,MnO_LDA_DMFT_calculation_Kuneš_2008,Mandal_Vanderbilt_TMOs_eDMFT_2019,Mandal_Vanderbilt_eDMFT_paramagnetic_TMOs_2019} (DMFT).

\begin{figure}[t]
	\centering
	\includegraphics[width=0.8\columnwidth]{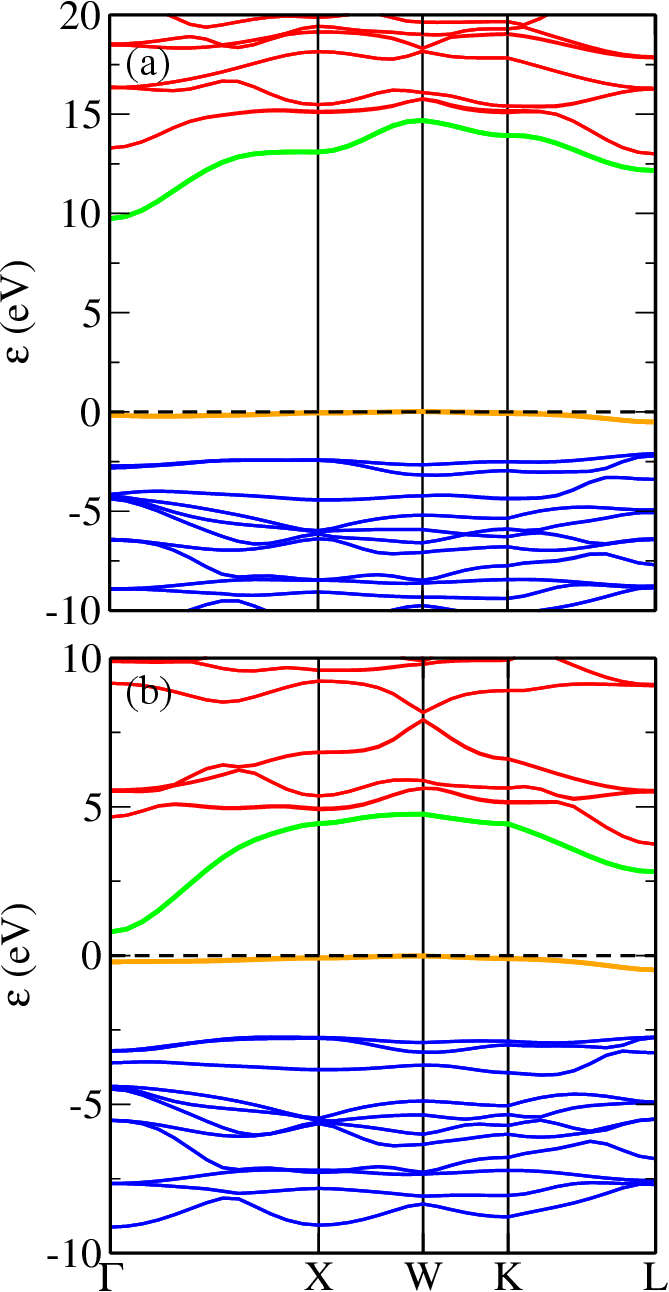}
	\caption{Computed band structure of FeO using (a) HF and (b) LFX. The Fermi energy has been arbitrarily set to 0 eV for both band structures. Colour scheme same as in Fig. \ref{fig:Si_HF_den}.}
	\label{fig:TMO_LFX_BS}
\end{figure}

The band gaps for various DFAs and the corresponding LXC gaps are given in Table \ref{table:TMO_gaps} for these TMOs in the rocksalt structure.
For MnO and NiO, the LSDA underestimates the band gap when compared to experiment
with a small improvement at the GGA level of theory using PBE.
However, in the case of CoO and FeO, the inaccuracy of the LSDA and PBE is more severe with both systems predicted to be falsely metallic rather than insulating.
In the case of a meta-generalised gradient approximation (meta-GGA) such as the regularised variant of the SCAN\cite{RSCAN_CASTEP} (rSCAN), all four TMOs were correctly predicted to be insulating which is somewhat unsurprising given that SCAN is expected to have a smaller SIE than GGAs\cite{TMOs_SCAN_Performance_1_Carter_2018,TMOs_SCAN_Performance_2_Carter_2020,SCAN}.

The LFX potential, by virtue of being obtained by inversion of the HF density, does not contain a SIE. On the other hand, the LFX potential contains a small systematic error due to over-localisation of the density from the omission of correlation in the self-consistent HF cycle that generates the density. This error tends to increase the value of the LFX band gap, typically for most materials by a few tenths of
an eV. In addition, by omitting the exchange and correlation discontinuity correction
$\deltaxc$ (see section \ref{section:delta_xc} for a discussion), the LFX result includes a small compensating systematic error of a similar order of magnitude as over-localisation, which tends to underestimate the value of the band gap.

As a result, the computed KS band gap using LFX had the lowest MAE of $0.43 \text{ eV}$ and MARE of 17.6\% for these systems compared to experiment, with accuracy comparable to hybrid functionals treated in a GKS scheme, in particular B3LYP and HSE06.

We note that a larger deviation between LFX and the experimental band gap was found in FeO, shown in Fig. \ref{fig:TMO_LFX_BS}, where LFX gave a 0.81 eV gap while the experimental gap is 2.4 eV.
As discussed in section \ref{section:delta_xc}, we believe this is due to strong Mott physics.
We point out that the qualitative picture is correct and in particular, we are not aware of another \textit{local} XC potential (besides LFX/xOEP) which predicts a finite band gap for FeO.

For FeO specifically, as the value of the predicted LFX band gap is small, we explored whether the systematic over-localisation error from the omission of correlation in the self-consistent HF cycle could change qualitatively the result. Hence, we carried out a HF + (PBE correlation) calculation. We found the LXC potential from the inverted density still predicts FeO to be insulating with a slightly smaller gap of 0.66~eV. This result is qualitatively correct, although quantitatively underestimates the true band gap of FeO by the omission of $\deltaxc$ which in this case is larger because of strong correlations in FeO.

In the other TMOs studied, the similarity of the LFX band gaps and experimental band gaps suggests weaker Mott physics and that the failure of local and semi-local DFAs is at least partially due to an inaccurate or incomplete description of exchange.

\begin{figure}[b]
	\centering
	\includegraphics[width=0.8\linewidth]   {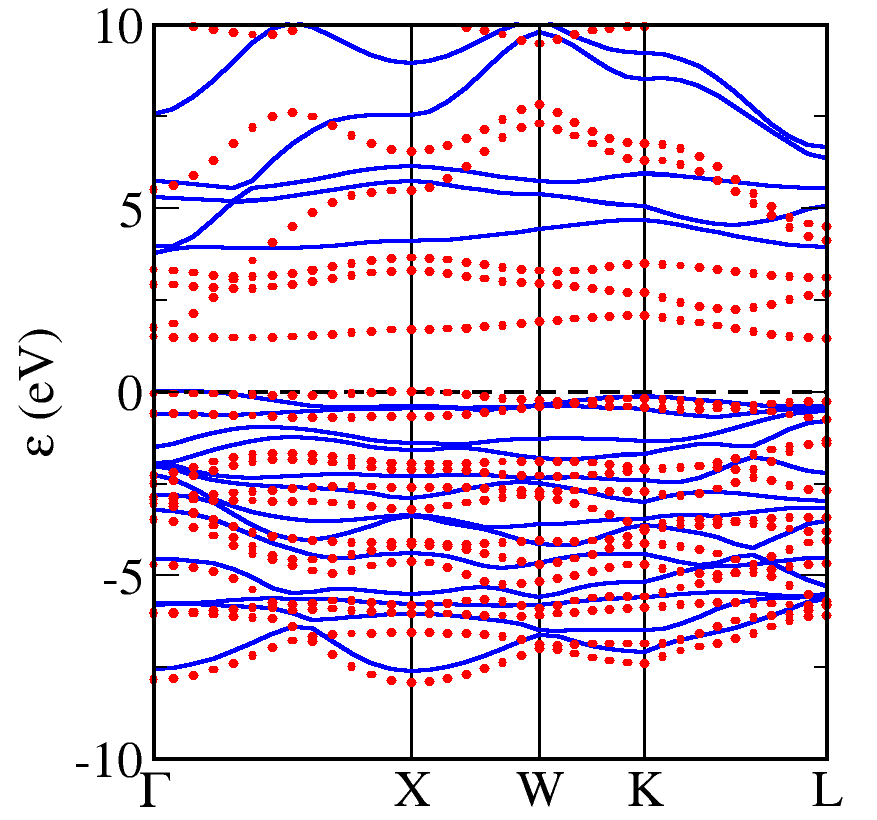}
	\caption{Computed LXC band structures for CoO using HSE06.
		Dotted red: LXC-HSE06, solid blue : HSE06
		The energy scale has been chosen such that the valence band maxima coincide and the Fermi energy is 0 eV.}
	\label{fig:CoO_HSE_inv_bs}
\end{figure}

The inversion of PBE0 and HSE06 target densities in these TMOs, however, yields an LXC potential that is quite different from the PBE potential.
Notably, in the case of CoO, both the LXC-PBE0 and LXC-HSE06  band structures (shown in Fig. \ref{fig:CoO_HSE_inv_bs}) are insulating with a band gap of $1.45 \text{ eV}$ and $1.44 \text{ eV}$ respectively while PBE predicts it to be metallic.
Likewise, the LXC-B3LYP potential $\vlxcdfa{B3LYP}$ gives a band gap of 1.30 eV which is perhaps not surprising since the weighting for HF exchange in the B3LYP, PBE0 and HSE06 functionals is similar with 20\% in B3LYP and 25\% in PBE0 and HSE06
(the non-locality is weaker in HSE06 since the HF exchange is screened by PBE exchange at long range).
Similar albeit less drastic differences were observed for MnO and NiO; in these systems, PBE gives the right qualitative behaviour but underestimates the band gaps giving $0.96 \text{ eV}$ and $1.23 \text{ eV}$ respectively.
However, we note that in these systems, the LXC-PBE0 and LXC-HSE06 gaps, $2.02 \text{ eV}$ and $2.01 \text{ eV}$ in MnO and $2.72 \text{ eV}$ and $2.71 \text{ eV}$ in NiO respectively, are much larger than the PBE gap.
The behaviour of the LXC potential for PBE0 and HSE06 densities for (Co,Ni,Mn)O is nonetheless consistent with the behaviour in the previously discussed systems in Table~\ref{table:non_TMO_gaps} insofar that PBE0 and HSE06 appear to give similar densities and therefore similar LXC potentials.
However, the large difference between these LXC potentials and the PBE potential suggests that the PBE0 and HSE06 densities differ greatly from the PBE density, especially in CoO where the SIE is large.

\begin{table}[b]
  \begin{ruledtabular}
    \begin{tabular}{ccc}
      Structure                 &HF(LFX)          &B3LYP \\
      \hline
      CoO &1.522  &0.126 \\
      FeO &0.496  &--    \\
      MnO &0.740  &0.047 \\
      NiO &1.516  &0.104 \\
    \end{tabular}
    \caption{
      Total energy differences (in eV) between the evaluation of the total energy functional using the LFX/LXC orbitals and self-consistent HF/GKS orbitals as defined in Eq.\eqref{eq:inv_eng_diff} for the transition metal monoxides.
      Rock-salt structure with experimental lattice parameters from ref\cite{TMO_lat_params} are used.}
    \label{table:tot_eng_diffs_TMOs}
  \end{ruledtabular}
\end{table}

Table \ref{table:tot_eng_diffs_TMOs} gives the differences in total energies as defined in Eq. \eqref{eq:inv_eng_diff} for HF and B3LYP when the respective total energy functional is evaluated using the LFX/LXC orbitals and HF/GKS orbitals.
In the TMOs studied in this work, we found that the total energies differences were typically larger than the other systems studied for both B3LYP and HF, cf. Table \ref{table:tot_eng_diffs}.
We found that the energy differences in TMOs were on average larger than the other systems studied in Table \ref{table:tot_eng_diffs} suggesting that the degree of non-locality for HF and B3LYP is stronger in the TMOs.
For instance, \STO{} had the largest energy difference of $0.777 \text{ eV}$ and 0.044 eV for LFX and B3LYP respectively while the smallest for LFX in TMOs was FeO with $0.496 \text{ eV}$ and MnO with $0.740 \text{ eV}$.

\subsection{Reduced Non-Locality: LDA+$U$}\label{section:LDA+U}
The Hubbard model\cite{Hubbard_Model_OG,Tasaki_1998_Review_Hubbard_intro,Tasaki_1998_Review_Hubbard_rigorous} is the simplest model Hamiltonian capable of capturing a Mott transition. In the so-called DFT+$U$ method, the `Hubbard-$U$' term gives an energy contribution to a subset of orbitals penalising double occupancy and thus results in greater localisation of the electrons occupying these orbitals.
This partially corrects the delocalisation error and SIE, particularly in local and semi-local DFAs, with the role of the Hubbard-$U$ term in the total energy expression analogous to that of HF in hybrid functionals.
More recently, Koopmans-compliant\cite{Koopmans_compliant_1_2010,Koopmans_compliant_2_2018,Koopmans_compliant_3_2022,Koopmans_Compliant_Review_2019} functionals have been developed to tackle the issue of SIE within standard DFAs through a restoration of the piecewise-linearity condition known from exact DFT as obtained by Perdew, Parr, Levy and Balduz\cite{delta_xc_Perdew_Parr_Levy_Balduz_1982}.
Since the Hubbard-$U$ correction in DFT+$U$ is applied to a subset of all orbitals (typically strongly-correlated $d$ and $f$ electrons in most calculations), it is by definition a non-local potential and thus the LXC potential
$v_\mathrm{LXC}^{\mathrm{DFT}+U}(\vec{r})$ derived from a DFT+$U$ target density $\rho_\mathrm{t}^{\mathrm{DFT}+U}(\vec{r})$ will naturally differ from the original, self-consistent non-local potential $\op{v}{\mathrm{LXC}}{{\mathrm{DFT}+U}}$ for DFT+$U$ or GKS potential $\op{v}{\mathrm{xc}}{\mathrm{GKS}}$.

\begin{figure}[t]
	\centering
	\includegraphics[width=0.9\linewidth]{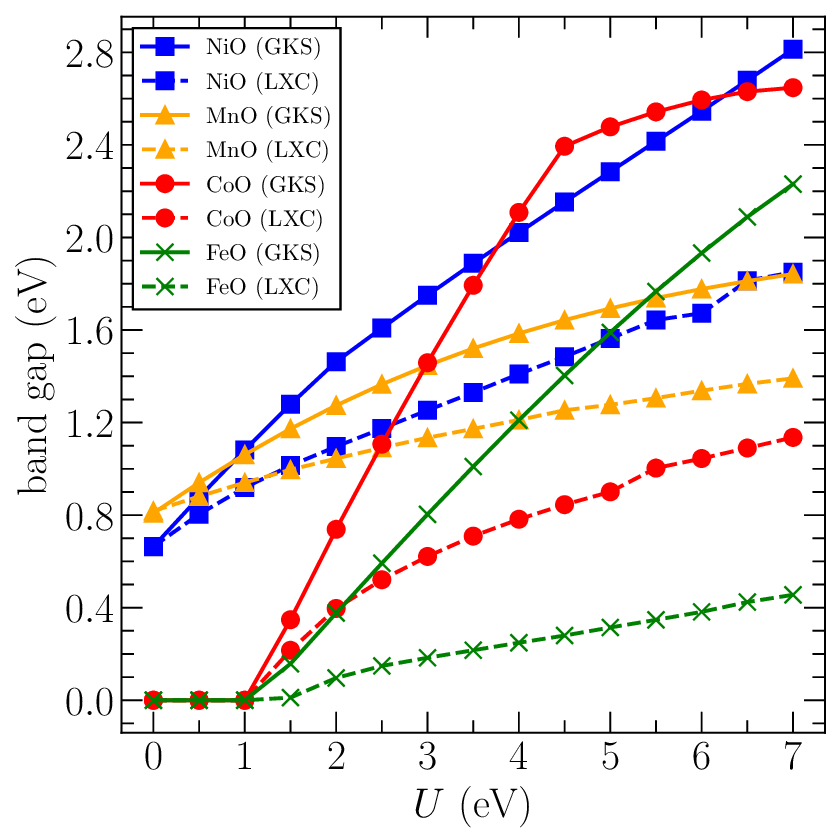}
	\caption{
		Computed band gap as a function of Hubbard-$U$ applied to $d$ orbitals for $\mathrm{X}O$ where $\mathrm{X}$ is Co,Fe,Ni,Mn (rocksalt structures).
		GKS refers to the LDA+$U$ gap calculated using GKS while LXC is obtained via inversion of the LDA+$U$ density.
		The experimental gaps are 2.5~eV, 2.4~eV, 4.0~eV and 3.9~eV for CoO\cite{CoO_band_gap_experimental}, FeO\cite{FeO_band_gap_experimental}, NiO\cite{NiO_band_gap_experimental} and MnO\cite{MnO_band_gap_experimental} respectively.
	}
	\label{fig:tmo_LDA_U_Gaps}
\end{figure}

We performed LDA+$U$ band structure calculations for (Co,Mn,Ni,Fe)O in the rocksalt structure (with the same lattice parameters as in the previous calculations)
with a Hubbard-$U$ applied to the $d$-orbitals of the respective metal cation across a range of $U$ from 0--7 eV.
We then inverted the resulting LDA+$U$ densities $\rho_\mathrm{t}^{\mathrm{LDA}+U}(\vec{r})$ to find the LXC-LDA+$U$ potential $v_\mathrm{LXC}^{\mathrm{LDA}+U}(\vec{r})$.
In Fig. \ref{fig:tmo_LDA_U_Gaps}, we show the calculated LDA+$U$ and LXC-LDA+$U$ band gap as a function of the Hubbard-$U$ parameter all TMOs considered in this work.
In MnO and NiO, we were unable to obtain the experimental band gap with LDA+$U$ (treated in GKS). %
As a recurring theme throughout this work, the LXC band gap is always lower than the LDA+$U$ band gap with the LXC-LDA+$U$ gap for $U\geq 2.5 \text{ eV}$ being about 75\%--80\% of the LDA+$U$ gap for MnO while it was 65\%--73\% of the LDA+$U$ gap in NiO.
We note that for MnO, the rate of change in the band gap as a function of $U$ starts to decrease at high $U$ as there is a limit to the localisation that can be achieved in the $d$-electrons.
We would expect similar behaviour in even higher $U$ in NiO as well.
For completeness, we point out that there is an `optimal' or appropriate value for Hubbard-$U$ for each TMO
although the determination of this value lies beyond the scope of this work and we instead draw attention to methods developed for this purpose elsewhere based on linear response(see Refs. \cite{Hubbard_U_linear_response_2005,Hubbard_U_linear_response_2006}).
More recently, these methods have been reformulated using the machinery of density functional perturbation theory (DFPT)\cite{Collinear_U_DFPT_Marzari_2018,Non_collinear_U_DFPT_Marzari_2023}.

Both FeO and CoO are falsely predicted to be metallic by the LDA. When LDA+$U$ is used, there exists a ``critical'' value of $U$, denoted $U_0$, at which both systems exhibit a metal-insulator transition.
The variation of the band gap as a function of Hubbard-$U$ is shown in Fig. \ref{fig:tmo_LDA_U_Gaps} for CoO and FeO.
The value of $U_0=1 \text{ eV}$  resulting in a non-zero gap is the same for both the LXC-LDA+$U$ and LDA+$U$  calculations in CoO.
However, in FeO, the value of $U_0$ for LXC-LDA+$U$ lagged behind the LDA+$U$ value with $U_0=1 \text{ eV}$ and $U_0=1.5 \text{ eV}$ for LDA+$U$ and LXC-LDA+$U$ respectively.
This behaviour, particularly in CoO, shows that $U_0$ causes a change in the density from the usual LDA density.
Consequently, the LXC potential obtained by the inversion of the LDA+$U$ target density necessarily differs from the non-local self-consistent LDA+$U$ potential.
We also note that while the gap opens up in the standard LDA+$U$ calculation at the same $U$, i.e. $U=1 \text{ eV}$,  for both FeO and CoO, the rate at which the gap increases is faster in CoO than FeO.
Moreover, it is possible to obtain the experimental band gap of 2.5 eV of CoO by using $U=4$ eV with LDA+$U$ but this is not possible with LXC-LDA+$U$; as a whole, we found that for LXC-LDA+$U$ band gap with $U\geq 1$ eV was around 40\% to 50\% of the LDA+$U$ band gap in CoO.
In contrast to the other TMOs, the LXC-LDA+$U$ band gap for FeO does not change significantly from $U=0 \text{ eV}$ and $7 \text{ eV}$, changing only by $0.46 \text{ eV}$ in FeO compared to $0.92 \text{ eV}$ in CoO.
In particular, we found that the LXC band gap for FeO was less than 25\% of the LDA+$U$ band gap.

\begin{figure}[t]
	\centering
	\includegraphics[width=0.85\linewidth]{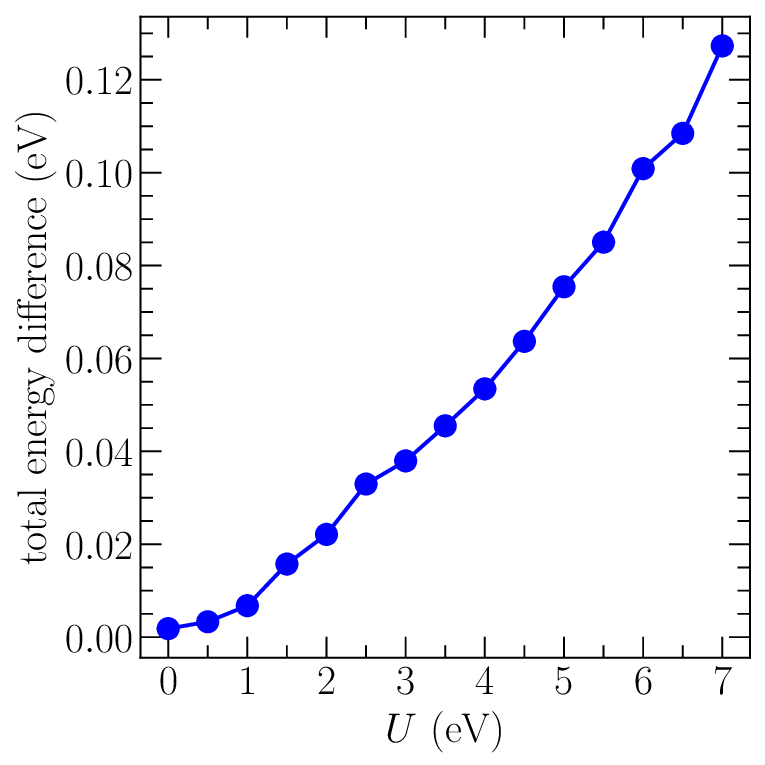}
	\caption{Variation of the total energy difference of FeO with Hubbard-$U$ for the LDA+$U$ energy functional when evaluated using the LXC orbitals and self-consistent orbitals.
	}
	\label{fig:FeO_eng_LDA+U}
\end{figure}

In a similar fashion to the HF and B3LYP functionals, total energy differences can be used to quantify the degree of non-locality of the LDA+$U$ potential.
We calculate the total energy difference according to Eq. \eqref{eq:inv_eng_diff} when the LDA+$U$ functional
is evaluated using the self-consistent orbitals and the orbitals that are eigenstates of the single-particle KS Hamiltonian with the LXC potential. In both instances, the same value of Hubbard-$U$ is utilised when evaluating the total energy functional.
\\Fig. \ref{fig:FeO_eng_LDA+U} shows the results of this procedure for FeO where the total energy difference, even at high $U\sim7 \text{ eV}$ is smaller than the difference between LFX and HF
suggesting LDA+$U$ has weaker non-locality.
This is not surprising as the non-local HF exchange operator $\op{V}{\mathrm{x}}$ acts on all occupied orbitals that comprise the HF determinant $\Phi_\mathrm{HF}$ while the Hubbard-$U$ is only applied to a subset of the orbitals, namely those with $d$-character, that comprise the KS determinant $\Phi_\mathrm{s}$.

\begin{figure*}[ht]
	\centering
	\includegraphics[width=0.90\linewidth]{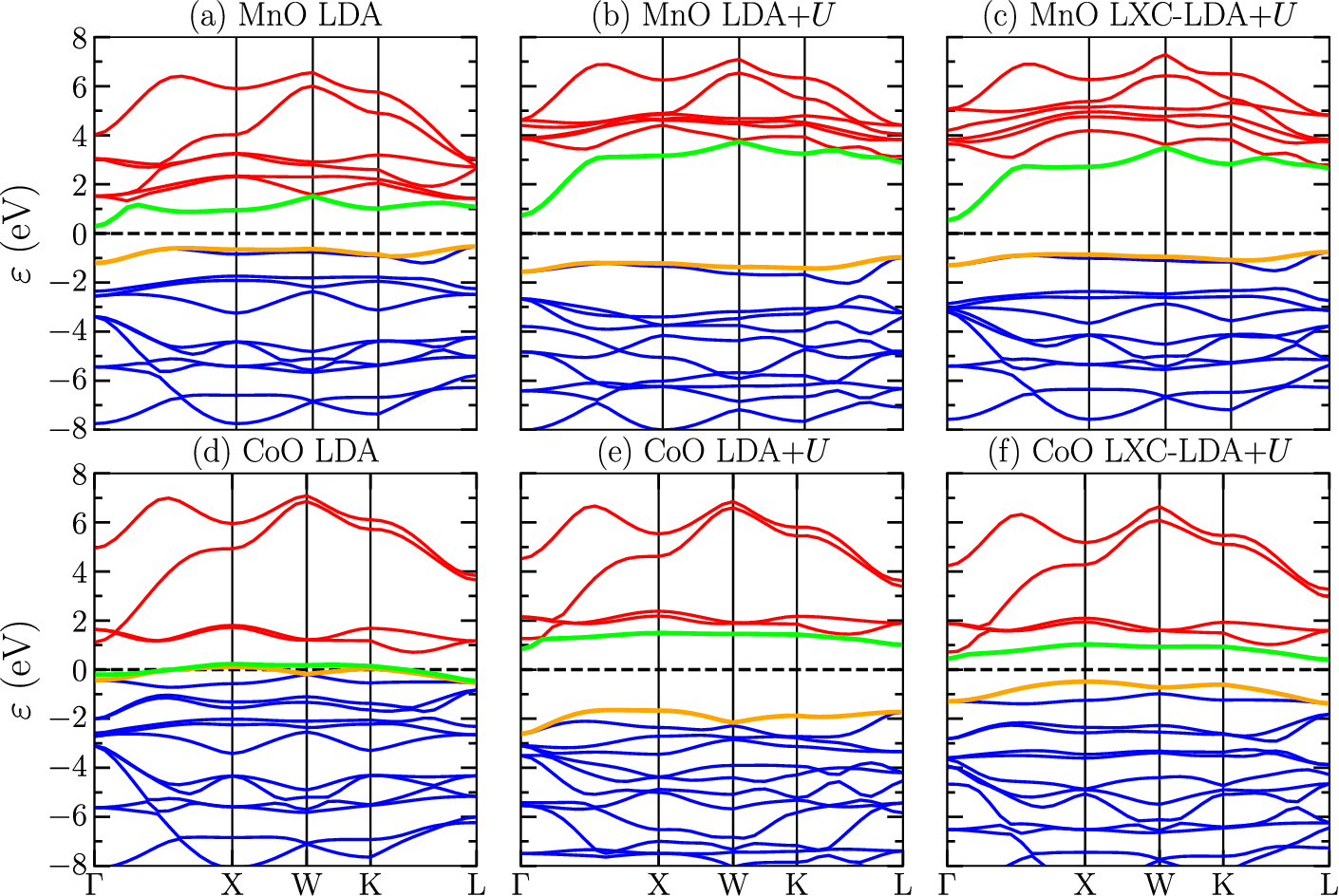}
	\caption{Computed band structure of MnO and CoO LDA, LDA+$U$ and LXC-LDA+$U$ with $U=5$ eV applied to $d$ orbitals.
		The Fermi energy has been set to 0 eV for all band structures.
		Colour scheme same as in Fig. \ref{fig:Si_HF_den}
	}
	\label{fig:MnO_CoO_LDA_U_Comparison}
\end{figure*}

Moreover, we now draw attention to the more noteworthy differences in the dispersion of the energy bands between LXC-LDA+$U$ and LDA+$U$ band structures.
In LDA+$U$, a sufficiently large $U$ leads to the excessive flattening of bands due to over-localisation of electrons to which the $U$ is applied (in this case the $d$ electrons).
This is shown in Fig. \ref{fig:MnO_CoO_LDA_U_Comparison} for MnO where the LDA, LDA+$U$ and LXC-LDA+$U$ band structures are plotted at $U=5 \text{ eV}$.
Intriguingly, the inversion restores some of the original energy dispersion of the LDA band structure for the occupied states although the LXC-LDA+$U$ gap is higher than the original LDA gap.
However, for unoccupied states, both the LXC and LDA+$U$ band structures have a similar conduction band width of $\sim 3.00 \text{ eV}$ while the LDA band structure has a conduction band width of $\sim 1.24 \textrm{ eV}$ in addition to all three having different band gaps as previously stated.
Examining the LDA+$U$ band structure of CoO shown in Fig. \ref{fig:MnO_CoO_LDA_U_Comparison} for $U=5$ eV, one can similarly find flat bands due to over-localisation.
However, the inversion does not nearly restore the energy dispersion of the occupied states in CoO compared to MnO.
In yet another contrast to MnO, the conduction band width of CoO was found to be similar across all three methods where the LDA band width was 0.68 eV, LDA+$U$ 0.66 eV, and LXC-LDA+$U$ 0.62 eV for $U=5$ eV.
On the other hand, the valence band width between the LXC-LDA+$U$ and LDA+$U$ band structures were similar, $0.90 \text{ eV}$ and $1.00 \text{ eV}$ respectively, but LDA had a valence band width of 0.65 eV.

\begin{figure}[th]
	\includegraphics[width=1.0\columnwidth]{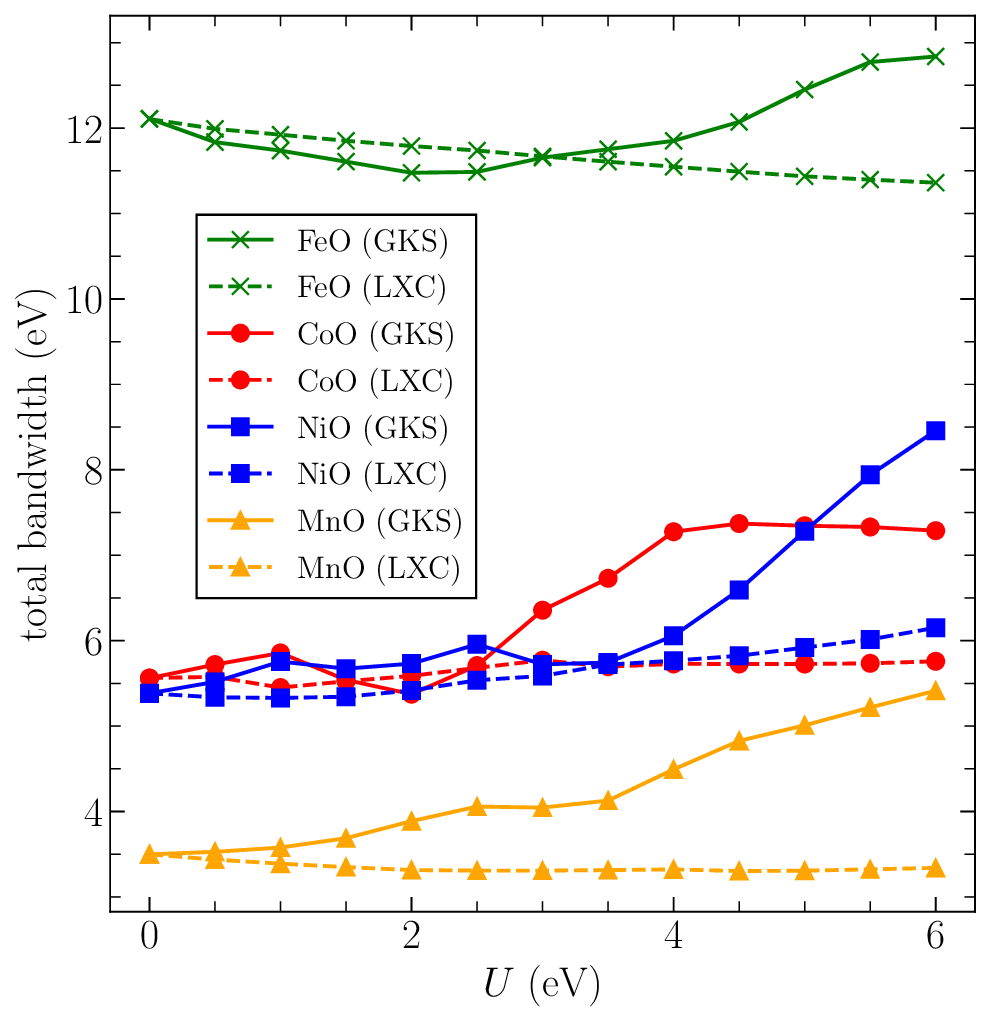}
	\caption{Total band width (see Eq. \eqref{eq:total_bandwidth} of main text) of bands with predominantly $d$ character. Compared to LDA+$U$, the LXC-LDA+$U$ is largely insensitive to $U$ and also does not significantly change from the LDA ($U=0$) value.}
	\label{fig:LDA+U_bandwidths}
\end{figure}

Investigating how the total occupied $d$ band width varies across these methods is non-trivial due to the difference in how the potentials act on the orbital. As previously stated, the LXC potential acts on all orbitals in an identical manner but the LDA+$U$ potential in general does not.
We calculated the projected density of states (PDOS) using \textsc{OptaDOS}\cite{optados,Yates_Optados_Adaptive_Smearing} using the population analysis methodology of Segall \textit{et al.}\cite{Segall_Population_Analysis} for both the LDA+$U$ and LXC-LDA+$U$
(PDOS calculations presented in supplemental material\cite{Supplementary_Material}) and found that the LXC potential largely preserves the character of each band, while LDA+$U$ does not, particularly where the valence band starts to acquire oxygen $p$ character in addition to that of the transition metal cation $d$ character due to hybridisation.
\textit{The number of bands that contribute to the total $d$-character is thus in general not the same between the LDA+$U$ and LXC-LDA+$U$ results.}
For these reasons, in Fig. \ref{fig:LDA+U_bandwidths}, we plot the total band width $\Delta \tilde{E}$ of $n_d$ bands below and including the valence band with (predominantly) $d$ character at the LDA level of theory, denoted $\{\tilde{\varepsilon}_{i\vec{k}}\}$,
\begin{equation}
	\tilde{E} = \sum_{i}^{n_d} \max_\vec{k} \{\tilde{\varepsilon}_{i\vec{k}}\}
	-
	\min_\vec{k}\{\tilde{\varepsilon}_{i\vec{k}}\}.
	\label{eq:total_bandwidth}
\end{equation}
In both the LDA and LXC-LDA+$U$ cases, $n_d$ corresponds to the number of $d$ electrons in the respective transition metal cation (although a notable exception is in FeO\cite{Supplementary_Material}).
We emphasise that unlike LDA+$U$, the total band width $\Delta \tilde{E}$ of the \textit{same} set of bands is insensitive to the value of Hubbard-$U$ which can be attributed to the local nature of the LXC potential which acts on all orbitals in the same way.
Moreover, we also point out that the $\Delta \tilde{E}$ does not change significantly from the LDA, i.e. $U=0$ value with changing $U$ in contrast to LDA+$U$, reflecting our previous observations for the band structures of CoO and MnO, cf. Fig. \ref{fig:MnO_CoO_LDA_U_Comparison}.

\subsection{Weak Non-Locality: Meta-GGA densities}\label{section:MGGAs}
The final class of densities we consider in this work are those generated via GKS calculations with meta-generalised gradient approximations (meta-GGAs), particularly using the rSCAN functional.
Like other orbital functionals which are implicit functionals of the density, meta-GGAs can also be treated via the OEP method similar to xOEP in which case the resulting KS potential is local $v_\mathrm{xc}^\mathrm{MGGA}(\vec{r})$.
Here, we consider a GKS treatment of MGGAs and investigate the degree of non-locality in the potential $\op{v}{\mathrm{xc}}{\mathrm{MGGA}}$.

In general, we found that the LXC-rSCAN band structures were similar to the GKS-rSCAN band structures (see Tables \ref{table:non_TMO_gaps} and \ref{table:TMO_gaps}). Indeed, we found that the MAD between the GKS-rSCAN and LXC-rSCAN band gaps was about 0.17 eV, suggesting that the non-locality of the GKS rSCAN potential $\op{v}{\mathrm{xc}}{\mathrm{rSCAN}}$ is weak.
As a recurring theme, Ge once again was noteworthy with LXC-rSCAN giving a metallic band structure while GKS-rSCAN gave it to be insulating with a band gap of 0.31 eV.

Interestingly, we also find that the LXC-rSCAN band gaps and band structures are similar to the LXC results for hybrid functionals in weakly correlated systems, see Table \ref{table:non_TMO_gaps}.
We found that the MAD between LXC-rSCAN and LXC-B3LYP was around 0.13 eV while the MARD between them was 7.18\%.
A similar result was found between LXC-PBE0 and LXC-HSE06, and LXC-rSCAN.
For comparison, the MAD between B3LYP and LXC-B3LYP was 1.15 eV with the MARD  being 33.7\% while the MAD between LXC-PBE0 and LXC-HSE06 with PBE was around 0.36 eV with a MARD of around 20\%.
This suggests that the densities of hybrid functionals are similar given that they yield similar LXC potentials and band structures.
The differences thus obtained in a GKS scheme are due to exchange, more precisely the degree of non-locality in the exchange potential $\op{v}{\mathrm{x}}{\mathrm{GKS}}$.
In the strongly correlated TMOs, the differences are larger and the LXC band gap for a hybrid functional is not close to the LXC-rSCAN band gap with the former typically being higher (apart from NiO).
Moreover, we also find a larger difference between the GKS-rSCAN and LXC-rSCAN band gaps and band structures in these systems.

\section{Conclusions}\label{section:CONCLUSIONS}
We have presented a thorough study across a range of solids of the KS potential with a local exchange-correlation (LXC) potential term obtained via the inversion of target densities from various DFAs employing a non-local potential in a GKS scheme.
Our calculations provide a means of quantifying the strength of the non-locality in various GKS schemes, which we achieved by comparing the computed KS band structures with GKS band structures using HF, hybrid functionals, LDA+$U$ functionals and meta-GGAs.

In general, we found that the exchange-only local Fock exchange (LFX) potential, obtained from the inversion of the HF density gives KS band gaps that
have the best agreement with experimental fundamental gaps out of any LXC potential (omitting the exchange and correlation (XC) derivative discontinuity).
Strictly, the LFX potential together with any other inverted LXC potential has zero XC discontinuity, since it is not the
functional derivative of an energy density functional. Nevertheless, because of the close similarity (at least in weakly correlated systems)
between the LFX and OEP potentials\cite{Hollins_LFX_2017,Hollins_Hylleraas_OEP,EXX_Semiconductors_Görling_1997,EXX_Semiconductors_Görling_1999,EXX_Delta_X_Grüning_Rubio_2006,Engel_2009_TMOs_EXX,EXX_RPA_Kresse_2014,Trushin_Görling_2019_EXX_perovskites},
we consider that LFX and LXC share the XC discontinuity of the OEP potential.
In this sense, the good agreement between the LFX bandgaps and experiment should be taken with caution as it is the result of a cancellation of two (small)
systematic errors.
One error is the over-localisation of the density from the omission of correlation in the self-consistent cycle, which tends to increase the value of the band gap.
The other error is the omission of $\Delta_{xc}^{OEP}$ which tends to underestimate the bandgap.
Specifically for FeO, we accounted for the overlocalisation error by running a HF + (PBE correlation) calculation and then inverting the density.
The result of the inversion gave the correct qualitative picture,
predicting insulating behaviour with a small gap, despite the uncompensated systematic underestimation of the bandgap from the remaining omission of $\Delta_{xc}$.

The LFX (KS) band gaps are also comparable in accuracy to those of hybrid DFAs when treated in GKS, with slightly reduced computational cost for the LFX band structure and associated spectroscopic calculations due to the use of a local potential.
We also note that to the best of our knowledge, the LFX potential and the xOEP are the only local potentials which qualitatively predict FeO to be insulating rather than metallic, highlighting that the failure of potentials from local and semi-local DFAs in strongly correlated systems is at least partially due to failure to correct for SIE.

In hybrid schemes, the reduced contribution from the Fock exchange term $\op{V}{\mathrm{x}}{\mathrm{HF}}$ results in GKS potential with weaker non-locality.
Consequently, although the LXC band gap is still lower than the corresponding GKS band gap for a given DFA, %
we find that the difference between the two is smaller.
In meta-GGAs, there is an even smaller difference between the LXC and GKS results highlighting that the non-locality for meta-GGAs %
is even weaker.

We also presented an alternative means of quantifying this non-locality by evaluating the total energy functional for a given DFA using the GKS and LXC orbitals and obtaining the energy difference between the two.
We found that the energy differences were typically an order of magnitude smaller for the B3LYP functional compared to the HF functional; naturally, the GKS orbitals yield a lower total energy minimum given that the minimisation is carried out without the constraint of a local potential.

Moreover, our results highlight that different DFAs can yield similar densities and thus similar LXC/KS potentials, although they have different GKS potentials.
For instance, in weakly correlated systems, we found that the LXC results for PBE0 and HSE06 were similar to those of PBE suggesting that the improved band gap prediction of these hybrid DFAs is due to the the non-locality in the GKS potential contributed by the non-local Fock exchange $\op{V}{\mathrm{x}}{\mathrm{HF}}$ term.
We observed similar behaviour between LXC-rSCAN and the LXC results for hybrid DFAs, notably LXC-B3LYP, suggesting that the improvement in the gap within GKS is a result of the stronger non-locality in B3LYP compared to rSCAN.

On the other hand, in strongly correlated systems, such as the anti-ferromagnetic TMOs considered in this work, LXC-PBE0 and LXC-HSE06 have a greater discrepancy with PBE. This is reflected in the larger total energy differences when the HF and B3LYP energy functionals are evaluated using GKS and LXC orbitals.
Taken as a whole, these results imply that GKS and KS will yield similar densities where exchange effects dominate over correlation.

The LXC-LDA+$U$ results for CoO and FeO in particular highlight the importance of correcting for self-interaction. At sufficiently large Hubbard-$U$, the LXC potential obtained from the inversion of the LDA+$U$ density does not qualitatively predict these systems to be falsely metallic, reflecting the role of the Hubbard-$U$ as a correction for self-interaction error\cite{Hubbard_U_linear_response_2005,Hubbard_U_linear_response_2006}.
The weaker non-locality from the Hubbard-$U$ can be seen with the smaller total energy difference in FeO for LDA+$U$ compared to HF across the range of $U$ we considered when evaluated using GKS and LXC orbitals.

Although we have specifically studied the KS potentials from various DFAs in this work, our algorithm is general and robust and can be readily applied to densities from other electronic structure methods, for example, quantum Monte Carlo (QMC), such that the local potential obtained via inversion of an accurate density can give insight into features of the exact KS potential\cite{den_inversion_Lucia_QMC_2023}.
Comparing the results of the inversion of QMC densities with the existing LXC results will enable the benchmarking of DFAs with regard to the densities they yield and, by extension, their LXC potentials in keeping with the original spirit of KS-DFT, namely to provide an accurate density through a mean-field local effective potential.

The data that supports the findings of this article is available through Durham collections\cite{data_repo}. The modifications to CASTEP have been merged into the main branch of the code and are available from CASTEP 23.1.1.
\begin{acknowledgments}
V Ravindran acknowledges the Durham Doctoral Scholarship programme for financial support. We acknowledge the use of the Durham Hamilton HPC service (Hamilton) and the UK national supercomputing facility (ARCHER2) funded by EPSRC grant EP/X035891/1.
\end{acknowledgments}

\bibliography{refs}

%apsrev4-2.bst 2019-01-14 (MD) hand-edited version of apsrev4-1.bst
%Control: key (0)
%Control: author (8) initials jnrlst
%Control: editor formatted (1) identically to author
%Control: production of article title (0) allowed
%Control: page (0) single
%Control: year (1) truncated
%Control: production of eprint (0) enabled
\begin{thebibliography}{151}%
\makeatletter
\providecommand \@ifxundefined [1]{%
 \@ifx{#1\undefined}
}%
\providecommand \@ifnum [1]{%
 \ifnum #1\expandafter \@firstoftwo
 \else \expandafter \@secondoftwo
 \fi
}%
\providecommand \@ifx [1]{%
 \ifx #1\expandafter \@firstoftwo
 \else \expandafter \@secondoftwo
 \fi
}%
\providecommand \natexlab [1]{#1}%
\providecommand \enquote  [1]{``#1''}%
\providecommand \bibnamefont  [1]{#1}%
\providecommand \bibfnamefont [1]{#1}%
\providecommand \citenamefont [1]{#1}%
\providecommand \href@noop [0]{\@secondoftwo}%
\providecommand \href [0]{\begingroup \@sanitize@url \@href}%
\providecommand \@href[1]{\@@startlink{#1}\@@href}%
\providecommand \@@href[1]{\endgroup#1\@@endlink}%
\providecommand \@sanitize@url [0]{\catcode `\\12\catcode `\$12\catcode
  `\&12\catcode `\#12\catcode `\^12\catcode `\_12\catcode `\%12\relax}%
\providecommand \@@startlink[1]{}%
\providecommand \@@endlink[0]{}%
\providecommand \url  [0]{\begingroup\@sanitize@url \@url }%
\providecommand \@url [1]{\endgroup\@href {#1}{\urlprefix }}%
\providecommand \urlprefix  [0]{URL }%
\providecommand \Eprint [0]{\href }%
\providecommand \doibase [0]{https://doi.org/}%
\providecommand \selectlanguage [0]{\@gobble}%
\providecommand \bibinfo  [0]{\@secondoftwo}%
\providecommand \bibfield  [0]{\@secondoftwo}%
\providecommand \translation [1]{[#1]}%
\providecommand \BibitemOpen [0]{}%
\providecommand \bibitemStop [0]{}%
\providecommand \bibitemNoStop [0]{.\EOS\space}%
\providecommand \EOS [0]{\spacefactor3000\relax}%
\providecommand \BibitemShut  [1]{\csname bibitem#1\endcsname}%
\let\auto@bib@innerbib\@empty
%</preamble>
\bibitem [{\citenamefont {Kohn}\ and\ \citenamefont
  {Sham}(1965)}]{Kohn-Sham_DFT}%
  \BibitemOpen
  \bibfield  {author} {\bibinfo {author} {\bibfnamefont {W.}~\bibnamefont
  {Kohn}}\ and\ \bibinfo {author} {\bibfnamefont {L.~J.}\ \bibnamefont
  {Sham}},\ }\bibfield  {title} {\bibinfo {title} {{Self-Consistent Equations
  Including Exchange and Correlation Effects}},\ }\href
  {https://doi.org/10.1103/PhysRev.140.A1133} {\bibfield  {journal} {\bibinfo
  {journal} {Phys. Rev.}\ }\textbf {\bibinfo {volume} {140}},\ \bibinfo {pages}
  {A1133} (\bibinfo {year} {1965})}\BibitemShut {NoStop}%
\bibitem [{\citenamefont {Sham}\ and\ \citenamefont
  {Schlüter}(1983)}]{delta_xc_Sham_Schuelter_1983}%
  \BibitemOpen
  \bibfield  {author} {\bibinfo {author} {\bibfnamefont {L.~J.}\ \bibnamefont
  {Sham}}\ and\ \bibinfo {author} {\bibfnamefont {M.}~\bibnamefont
  {Schlüter}},\ }\bibfield  {title} {\bibinfo {title} {{Density-Functional
  Theory of the Energy Gap}},\ }\href
  {https://doi.org/10.1103/PhysRevLett.51.1888} {\bibfield  {journal} {\bibinfo
   {journal} {Phys. Rev. Lett.}\ }\textbf {\bibinfo {volume} {51}},\ \bibinfo
  {pages} {1888} (\bibinfo {year} {1983})}\BibitemShut {NoStop}%
\bibitem [{\citenamefont {Godby}\ \emph {et~al.}(1986)\citenamefont {Godby},
  \citenamefont {Schlüter},\ and\ \citenamefont
  {Sham}}]{delta_xc_Sham_Schuelter_1986_Si}%
  \BibitemOpen
  \bibfield  {author} {\bibinfo {author} {\bibfnamefont {R.~W.}\ \bibnamefont
  {Godby}}, \bibinfo {author} {\bibfnamefont {M.}~\bibnamefont {Schlüter}},\
  and\ \bibinfo {author} {\bibfnamefont {L.~J.}\ \bibnamefont {Sham}},\
  }\bibfield  {title} {\bibinfo {title} {{Accurate Exchange-Correlation
  Potential for Silicon and Its Discontinuity on Addition of an Electron}},\
  }\href {https://doi.org/10.1103/PhysRevLett.56.2415} {\bibfield  {journal}
  {\bibinfo  {journal} {Phys. Rev. Lett.}\ }\textbf {\bibinfo {volume} {56}},\
  \bibinfo {pages} {2415} (\bibinfo {year} {1986})}\BibitemShut {NoStop}%
\bibitem [{\citenamefont {Godby}\ \emph
  {et~al.}(1987{\natexlab{a}})\citenamefont {Godby}, \citenamefont
  {Schlüter},\ and\ \citenamefont {Sham}}]{Sham_Schuelter_1987_arsenides}%
  \BibitemOpen
  \bibfield  {author} {\bibinfo {author} {\bibfnamefont {R.~W.}\ \bibnamefont
  {Godby}}, \bibinfo {author} {\bibfnamefont {M.}~\bibnamefont {Schlüter}},\
  and\ \bibinfo {author} {\bibfnamefont {L.~J.}\ \bibnamefont {Sham}},\
  }\bibfield  {title} {\bibinfo {title} {{Quasiparticle energies in GaAs and
  AlAs}},\ }\href {https://doi.org/10.1103/PhysRevB.35.4170} {\bibfield
  {journal} {\bibinfo  {journal} {Phys. Rev. B}\ }\textbf {\bibinfo {volume}
  {35}},\ \bibinfo {pages} {4170} (\bibinfo {year}
  {1987}{\natexlab{a}})}\BibitemShut {NoStop}%
\bibitem [{\citenamefont {Godby}\ \emph
  {et~al.}(1987{\natexlab{b}})\citenamefont {Godby}, \citenamefont
  {Schlüter},\ and\ \citenamefont {Sham}}]{Sham_Schuelter_1987_trends_xc}%
  \BibitemOpen
  \bibfield  {author} {\bibinfo {author} {\bibfnamefont {R.~W.}\ \bibnamefont
  {Godby}}, \bibinfo {author} {\bibfnamefont {M.}~\bibnamefont {Schlüter}},\
  and\ \bibinfo {author} {\bibfnamefont {L.~J.}\ \bibnamefont {Sham}},\
  }\bibfield  {title} {\bibinfo {title} {Trends in self-energy operators and
  their corresponding exchange-correlation potentials},\ }\href
  {https://doi.org/10.1103/PhysRevB.36.6497} {\bibfield  {journal} {\bibinfo
  {journal} {Phys. Rev. B}\ }\textbf {\bibinfo {volume} {36}},\ \bibinfo
  {pages} {6497} (\bibinfo {year} {1987}{\natexlab{b}})}\BibitemShut {NoStop}%
\bibitem [{\citenamefont {Godby}\ \emph {et~al.}(1988)\citenamefont {Godby},
  \citenamefont {M.},\ and\ \citenamefont
  {Sham}}]{delta_xc_Sham_Schuelter_1988}%
  \BibitemOpen
  \bibfield  {author} {\bibinfo {author} {\bibfnamefont {R.~W.}\ \bibnamefont
  {Godby}}, \bibinfo {author} {\bibfnamefont {S.}~\bibnamefont {M.}},\ and\
  \bibinfo {author} {\bibfnamefont {L.~J.}\ \bibnamefont {Sham}},\ }\bibfield
  {title} {\bibinfo {title} {{Self-energy operators and exchange-correlation
  potentials in semiconductors}},\ }\href
  {https://doi.org/10.1103/PhysRevB.37.10159} {\bibfield  {journal} {\bibinfo
  {journal} {Phys. Rev. B}\ }\textbf {\bibinfo {volume} {37}},\ \bibinfo
  {pages} {10159} (\bibinfo {year} {1988})}\BibitemShut {NoStop}%
\bibitem [{\citenamefont {Hohenberg}\ and\ \citenamefont
  {Kohn}(1964)}]{Hohenberg-Kohn_theorems}%
  \BibitemOpen
  \bibfield  {author} {\bibinfo {author} {\bibfnamefont {P.}~\bibnamefont
  {Hohenberg}}\ and\ \bibinfo {author} {\bibfnamefont {W.}~\bibnamefont
  {Kohn}},\ }\bibfield  {title} {\bibinfo {title} {{Inhomogeneous Electron
  Gas}},\ }\href {https://doi.org/10.1103/PhysRev.136.B864} {\bibfield
  {journal} {\bibinfo  {journal} {Phys. Rev.}\ }\textbf {\bibinfo {volume}
  {136}},\ \bibinfo {pages} {B864} (\bibinfo {year} {1964})}\BibitemShut
  {NoStop}%
\bibitem [{\citenamefont {Gidopoulos}(2011)}]{Nikitas_Interface_WFT_DFT}%
  \BibitemOpen
  \bibfield  {author} {\bibinfo {author} {\bibfnamefont {N.~I.}\ \bibnamefont
  {Gidopoulos}},\ }\bibfield  {title} {\bibinfo {title} {Progress at the
  interface of wave-function and density-functional theories},\ }\href
  {https://doi.org/10.1103/PhysRevA.83.040502} {\bibfield  {journal} {\bibinfo
  {journal} {Phys. Rev. A}\ }\textbf {\bibinfo {volume} {83}},\ \bibinfo
  {pages} {040502} (\bibinfo {year} {2011})}\BibitemShut {NoStop}%
\bibitem [{\citenamefont {Aouina}\ \emph {et~al.}(2023)\citenamefont {Aouina},
  \citenamefont {Gatti}, \citenamefont {Chen}, \citenamefont {Zhang},\ and\
  \citenamefont {Reining}}]{den_inversion_Lucia_QMC_2023}%
  \BibitemOpen
  \bibfield  {author} {\bibinfo {author} {\bibfnamefont {A.}~\bibnamefont
  {Aouina}}, \bibinfo {author} {\bibfnamefont {M.}~\bibnamefont {Gatti}},
  \bibinfo {author} {\bibfnamefont {S.}~\bibnamefont {Chen}}, \bibinfo {author}
  {\bibfnamefont {S.}~\bibnamefont {Zhang}},\ and\ \bibinfo {author}
  {\bibfnamefont {L.}~\bibnamefont {Reining}},\ }\bibfield  {title} {\bibinfo
  {title} {{Accurate Kohn-Sham auxiliary system from the ground-state density
  of solids}},\ }\href {https://doi.org/10.1103/PhysRevB.107.195123} {\bibfield
   {journal} {\bibinfo  {journal} {Phys. Rev. B}\ }\textbf {\bibinfo {volume}
  {107}},\ \bibinfo {pages} {195123} (\bibinfo {year} {2023})}\BibitemShut
  {NoStop}%
\bibitem [{\citenamefont {Perdew}\ and\ \citenamefont
  {Zunger}(1981)}]{LDA_CASTEP}%
  \BibitemOpen
  \bibfield  {author} {\bibinfo {author} {\bibfnamefont {J.~P.}\ \bibnamefont
  {Perdew}}\ and\ \bibinfo {author} {\bibfnamefont {A.}~\bibnamefont
  {Zunger}},\ }\bibfield  {title} {\bibinfo {title} {Self-interaction
  correction to density-functional approximations for many-electron systems},\
  }\href {https://doi.org/10.1103/PhysRevB.23.5048} {\bibfield  {journal}
  {\bibinfo  {journal} {Phys. Rev. B}\ }\textbf {\bibinfo {volume} {23}},\
  \bibinfo {pages} {5048} (\bibinfo {year} {1981})}\BibitemShut {NoStop}%
\bibitem [{\citenamefont {Svane}\ and\ \citenamefont
  {Gunnarsson}(1990)}]{Self_Interaction_Corrected_TMOs_1990_Svane}%
  \BibitemOpen
  \bibfield  {author} {\bibinfo {author} {\bibfnamefont {A.}~\bibnamefont
  {Svane}}\ and\ \bibinfo {author} {\bibfnamefont {O.}~\bibnamefont
  {Gunnarsson}},\ }\bibfield  {title} {\bibinfo {title} {Transition-metal
  oxides in the self-interaction--corrected density-functional formalism},\
  }\href {https://doi.org/10.1103/PhysRevLett.65.1148} {\bibfield  {journal}
  {\bibinfo  {journal} {Phys. Rev. Lett.}\ }\textbf {\bibinfo {volume} {65}},\
  \bibinfo {pages} {1148} (\bibinfo {year} {1990})}\BibitemShut {NoStop}%
\bibitem [{\citenamefont {Mori-S\'anchez}\ \emph {et~al.}(2008)\citenamefont
  {Mori-S\'anchez}, \citenamefont {Cohen},\ and\ \citenamefont
  {Yang}}]{Localisation_Delocalisation_Error_Yang_2008}%
  \BibitemOpen
  \bibfield  {author} {\bibinfo {author} {\bibfnamefont {P.}~\bibnamefont
  {Mori-S\'anchez}}, \bibinfo {author} {\bibfnamefont {A.~J.}\ \bibnamefont
  {Cohen}},\ and\ \bibinfo {author} {\bibfnamefont {W.}~\bibnamefont {Yang}},\
  }\bibfield  {title} {\bibinfo {title} {Localization and delocalization errors
  in density functional theory and implications for band-gap prediction},\
  }\href {https://doi.org/10.1103/PhysRevLett.100.146401} {\bibfield  {journal}
  {\bibinfo  {journal} {Phys. Rev. Lett.}\ }\textbf {\bibinfo {volume} {100}},\
  \bibinfo {pages} {146401} (\bibinfo {year} {2008})}\BibitemShut {NoStop}%
\bibitem [{\citenamefont {Bryenton}\ \emph {et~al.}(2023)\citenamefont
  {Bryenton}, \citenamefont {Adeleke}, \citenamefont {Dale},\ and\
  \citenamefont {Johnson}}]{delocalisation_error_perspective_2023}%
  \BibitemOpen
  \bibfield  {author} {\bibinfo {author} {\bibfnamefont {K.~R.}\ \bibnamefont
  {Bryenton}}, \bibinfo {author} {\bibfnamefont {A.~A.}\ \bibnamefont
  {Adeleke}}, \bibinfo {author} {\bibfnamefont {S.~G.}\ \bibnamefont {Dale}},\
  and\ \bibinfo {author} {\bibfnamefont {E.~R.}\ \bibnamefont {Johnson}},\
  }\bibfield  {title} {\bibinfo {title} {Delocalization error: The greatest
  outstanding challenge in density-functional theory},\ }\href
  {https://doi.org/https://doi.org/10.1002/wcms.1631} {\bibfield  {journal}
  {\bibinfo  {journal} {WIREs Computational Molecular Science}\ }\textbf
  {\bibinfo {volume} {13}},\ \bibinfo {pages} {e1631} (\bibinfo {year}
  {2023})}\BibitemShut {NoStop}%
\bibitem [{\citenamefont {Perdew}\ \emph {et~al.}(2017)\citenamefont {Perdew},
  \citenamefont {Yang}, \citenamefont {Burke}, \citenamefont {Yang},
  \citenamefont {Gross}, \citenamefont {Scheffler}, \citenamefont {Scuseria},
  \citenamefont {Henderson}, \citenamefont {Zhang}, \citenamefont {Ruzsinszky},
  \citenamefont {Peng}, \citenamefont {Sun}, \citenamefont {Trushin},\ and\
  \citenamefont {Görling}}]{band_gaps_in_GKS_2017}%
  \BibitemOpen
  \bibfield  {author} {\bibinfo {author} {\bibfnamefont {J.~P.}\ \bibnamefont
  {Perdew}}, \bibinfo {author} {\bibfnamefont {W.}~\bibnamefont {Yang}},
  \bibinfo {author} {\bibfnamefont {K.}~\bibnamefont {Burke}}, \bibinfo
  {author} {\bibfnamefont {Z.}~\bibnamefont {Yang}}, \bibinfo {author}
  {\bibfnamefont {E.~K.~U.}\ \bibnamefont {Gross}}, \bibinfo {author}
  {\bibfnamefont {M.}~\bibnamefont {Scheffler}}, \bibinfo {author}
  {\bibfnamefont {G.~E.}\ \bibnamefont {Scuseria}}, \bibinfo {author}
  {\bibfnamefont {T.~M.}\ \bibnamefont {Henderson}}, \bibinfo {author}
  {\bibfnamefont {I.~Y.}\ \bibnamefont {Zhang}}, \bibinfo {author}
  {\bibfnamefont {A.}~\bibnamefont {Ruzsinszky}}, \bibinfo {author}
  {\bibfnamefont {H.}~\bibnamefont {Peng}}, \bibinfo {author} {\bibfnamefont
  {J.}~\bibnamefont {Sun}}, \bibinfo {author} {\bibfnamefont {E.}~\bibnamefont
  {Trushin}},\ and\ \bibinfo {author} {\bibfnamefont {A.}~\bibnamefont
  {Görling}},\ }\bibfield  {title} {\bibinfo {title} {{Understanding band gaps
  of solids in generalized Kohn–Sham theory}},\ }\href
  {https://doi.org/10.1073/pnas.1621352114} {\bibfield  {journal} {\bibinfo
  {journal} {Proc. Natl. Acad. Sci. USA}\ }\textbf {\bibinfo {volume} {114}},\
  \bibinfo {pages} {2801} (\bibinfo {year} {2017})}\BibitemShut {NoStop}%
\bibitem [{\citenamefont {Wang}\ \emph {et~al.}(2021)\citenamefont {Wang} \emph
  {et~al.}}]{Uncertainty_in_DFT_energy_Corrections_2021}%
  \BibitemOpen
  \bibfield  {author} {\bibinfo {author} {\bibfnamefont {A.}~\bibnamefont
  {Wang}} \emph {et~al.},\ }\bibfield  {title} {\bibinfo {title} {A framework
  for quantifying uncertainty in {DFT} energy corrections},\ }\href@noop {}
  {\bibfield  {journal} {\bibinfo  {journal} {Sci. Rep.}\ }\textbf {\bibinfo
  {volume} {11}},\ \bibinfo {pages} {15496} (\bibinfo {year}
  {2021})}\BibitemShut {NoStop}%
\bibitem [{\citenamefont {Cohen}\ \emph {et~al.}(2012)\citenamefont {Cohen},
  \citenamefont {Mori-Sánchez},\ and\ \citenamefont
  {Yang}}]{DFT_Errors_2012_Wang}%
  \BibitemOpen
  \bibfield  {author} {\bibinfo {author} {\bibfnamefont {A.~J.}\ \bibnamefont
  {Cohen}}, \bibinfo {author} {\bibfnamefont {P.}~\bibnamefont
  {Mori-Sánchez}},\ and\ \bibinfo {author} {\bibfnamefont {W.}~\bibnamefont
  {Yang}},\ }\bibfield  {title} {\bibinfo {title} {Challenges for {Density
  Functional Theory}},\ }\href {https://doi.org/10.1021/cr200107z} {\bibfield
  {journal} {\bibinfo  {journal} {Chem. Rev.}\ }\textbf {\bibinfo {volume}
  {112}},\ \bibinfo {pages} {289–320} (\bibinfo {year} {2012})}\BibitemShut
  {NoStop}%
\bibitem [{\citenamefont {Burke}(2012)}]{Perspective_Burke_2012}%
  \BibitemOpen
  \bibfield  {author} {\bibinfo {author} {\bibfnamefont {K.}~\bibnamefont
  {Burke}},\ }\bibfield  {title} {\bibinfo {title} {{Perspective on density
  functional theory}},\ }\href {https://doi.org/10.1063/1.4704546} {\bibfield
  {journal} {\bibinfo  {journal} {The Journal of Chemical Physics}\ }\textbf
  {\bibinfo {volume} {136}},\ \bibinfo {pages} {150901} (\bibinfo {year}
  {2012})}\BibitemShut {NoStop}%
\bibitem [{\citenamefont {Vuckovic}\ and\ \citenamefont
  {Burke}(2020)}]{Geometry_Errors_Vuckovic2020}%
  \BibitemOpen
  \bibfield  {author} {\bibinfo {author} {\bibfnamefont {S.}~\bibnamefont
  {Vuckovic}}\ and\ \bibinfo {author} {\bibfnamefont {K.}~\bibnamefont
  {Burke}},\ }\bibfield  {title} {\bibinfo {title} {{Quantifying and
  Understanding Errors in Molecular Geometries}},\ }\href
  {https://doi.org/10.1021/acs.jpclett.0c03034} {\bibfield  {journal} {\bibinfo
   {journal} {J. Chem. Phys. Lett.}\ }\textbf {\bibinfo {volume} {11}},\
  \bibinfo {pages} {9957} (\bibinfo {year} {2020})}\BibitemShut {NoStop}%
\bibitem [{\citenamefont {Vuckovic}(2022)}]{Geometry_Errors_Vuckovic2022}%
  \BibitemOpen
  \bibfield  {author} {\bibinfo {author} {\bibfnamefont {S.}~\bibnamefont
  {Vuckovic}},\ }\bibfield  {title} {\bibinfo {title} {{Quantification of
  Geometric Errors Made Simple: Application to Main-Group Molecular
  Structures}},\ }\href {https://doi.org/10.1021/acs.jpca.1c10688} {\bibfield
  {journal} {\bibinfo  {journal} {J. Phys. Chem. A}\ }\textbf {\bibinfo
  {volume} {126}},\ \bibinfo {pages} {1300} (\bibinfo {year}
  {2022})}\BibitemShut {NoStop}%
\bibitem [{\citenamefont {De~Waele}\ \emph {et~al.}(2016)\citenamefont
  {De~Waele}, \citenamefont {Lejaeghere}, \citenamefont {Sluydts},\ and\
  \citenamefont {Cottenier}}]{Errors_for_Surfaces_2016}%
  \BibitemOpen
  \bibfield  {author} {\bibinfo {author} {\bibfnamefont {S.}~\bibnamefont
  {De~Waele}}, \bibinfo {author} {\bibfnamefont {K.}~\bibnamefont
  {Lejaeghere}}, \bibinfo {author} {\bibfnamefont {M.}~\bibnamefont
  {Sluydts}},\ and\ \bibinfo {author} {\bibfnamefont {S.}~\bibnamefont
  {Cottenier}},\ }\bibfield  {title} {\bibinfo {title} {Error estimates for
  density-functional theory predictions of surface energy and work function},\
  }\href {https://doi.org/10.1103/PhysRevB.94.235418} {\bibfield  {journal}
  {\bibinfo  {journal} {Phys. Rev. B}\ }\textbf {\bibinfo {volume} {94}},\
  \bibinfo {pages} {235418} (\bibinfo {year} {2016})}\BibitemShut {NoStop}%
\bibitem [{\citenamefont {Brittain}\ \emph {et~al.}(2009)\citenamefont
  {Brittain}, \citenamefont {Lin}, \citenamefont {Gilbert}, \citenamefont
  {Izgorodina}, \citenamefont {Gill},\ and\ \citenamefont
  {Coote}}]{systematic_exchange_errors_2009}%
  \BibitemOpen
  \bibfield  {author} {\bibinfo {author} {\bibfnamefont {D.~R.~B.}\
  \bibnamefont {Brittain}}, \bibinfo {author} {\bibfnamefont {C.~Y.}\
  \bibnamefont {Lin}}, \bibinfo {author} {\bibfnamefont {A.~T.~B.}\
  \bibnamefont {Gilbert}}, \bibinfo {author} {\bibfnamefont {E.~I.}\
  \bibnamefont {Izgorodina}}, \bibinfo {author} {\bibfnamefont {P.~M.~W.}\
  \bibnamefont {Gill}},\ and\ \bibinfo {author} {\bibfnamefont {M.~L.}\
  \bibnamefont {Coote}},\ }\bibfield  {title} {\bibinfo {title} {{The role of
  exchange in systematic DFT errors for some organic reactions}},\ }\href
  {https://doi.org/10.1039/B818412G} {\bibfield  {journal} {\bibinfo  {journal}
  {Phys. Chem. Chem. Phys.}\ }\textbf {\bibinfo {volume} {11}},\ \bibinfo
  {pages} {1138–1142} (\bibinfo {year} {2009})}\BibitemShut {NoStop}%
\bibitem [{\citenamefont {Gidopoulos}\ and\ \citenamefont
  {Lathiotakis}(2012)}]{Gidopoulos_Lathiotakis_2012_Screening_Charge}%
  \BibitemOpen
  \bibfield  {author} {\bibinfo {author} {\bibfnamefont {N.~I.}\ \bibnamefont
  {Gidopoulos}}\ and\ \bibinfo {author} {\bibfnamefont {N.~N.}\ \bibnamefont
  {Lathiotakis}},\ }\bibfield  {title} {\bibinfo {title} {Constraining density
  functional approximations to yield self-interaction free potentials},\ }\href
  {https://doi.org/10.1063/1.4728156} {\bibfield  {journal} {\bibinfo
  {journal} {J. Chem. Phys.}\ }\textbf {\bibinfo {volume} {136}},\ \bibinfo
  {pages} {224109} (\bibinfo {year} {2012})}\BibitemShut {NoStop}%
\bibitem [{\citenamefont {Gidopoulos}\ and\ \citenamefont
  {Lathiotakis}(2015)}]{Gidopoulos_Lathiotakis_SC_Charge_2015}%
  \BibitemOpen
  \bibfield  {author} {\bibinfo {author} {\bibfnamefont {N.}~\bibnamefont
  {Gidopoulos}}\ and\ \bibinfo {author} {\bibfnamefont {N.~N.}\ \bibnamefont
  {Lathiotakis}},\ }\bibfield  {title} {\bibinfo {title} {{Chapter Six -
  Constrained Local Potentials for Self-Interaction Correction in
  \textit{Advances in Atomic, Molecular, and Optical Physics}}}\ }(\bibinfo
  {publisher} {Academic Press},\ \bibinfo {year} {2015})\ p.\ \bibinfo {pages}
  {129–142}\BibitemShut {NoStop}%
\bibitem [{\citenamefont {Pitts}\ and\ \citenamefont
  {Lathiotakis}(2018)}]{Pitts_screening_charge_2018}%
  \BibitemOpen
  \bibfield  {author} {\bibinfo {author} {\bibfnamefont {T.}~\bibnamefont
  {Pitts}}\ and\ \bibinfo {author} {\bibfnamefont {N.~N.}\ \bibnamefont
  {Lathiotakis}},\ }\bibfield  {title} {\bibinfo {title} {{Performance of the
  constrained minimization of the total energy in density functional
  approximations: The electron repulsion density and potential}},\ }\href
  {https://doi.org/10.1140/epjb/e2018-90123-8} {\bibfield  {journal} {\bibinfo
  {journal} {Eur. Phys. J. B}\ }\textbf {\bibinfo {volume} {91}},\ \bibinfo
  {pages} {130} (\bibinfo {year} {2018})}\BibitemShut {NoStop}%
\bibitem [{\citenamefont {Callow}\ \emph
  {et~al.}(2020{\natexlab{a}})\citenamefont {Callow}, \citenamefont {Pearce},
  \citenamefont {Pitts}, \citenamefont {Lathiotakis}, \citenamefont {Hodgson},\
  and\ \citenamefont {Gidopoulos}}]{Callow_Gidopoulos_Faraday_Discussion}%
  \BibitemOpen
  \bibfield  {author} {\bibinfo {author} {\bibfnamefont {T.~J.}\ \bibnamefont
  {Callow}}, \bibinfo {author} {\bibfnamefont {B.~J.}\ \bibnamefont {Pearce}},
  \bibinfo {author} {\bibfnamefont {T.}~\bibnamefont {Pitts}}, \bibinfo
  {author} {\bibfnamefont {N.}~\bibnamefont {Lathiotakis}}, \bibinfo {author}
  {\bibfnamefont {M.~J.~P.}\ \bibnamefont {Hodgson}},\ and\ \bibinfo {author}
  {\bibfnamefont {N.~I.}\ \bibnamefont {Gidopoulos}},\ }\bibfield  {title}
  {\bibinfo {title} {Improving the exchange and correlation potential in
  density-functional approximations through constraints},\ }\href
  {https://doi.org/10.1039/D0FD00069H} {\bibfield  {journal} {\bibinfo
  {journal} {Faraday Discuss.}\ }\textbf {\bibinfo {volume} {224}},\ \bibinfo
  {pages} {126} (\bibinfo {year} {2020}{\natexlab{a}})}\BibitemShut {NoStop}%
\bibitem [{\citenamefont {Kim}\ \emph {et~al.}(2013)\citenamefont {Kim},
  \citenamefont {Sim},\ and\ \citenamefont
  {Burke}}]{density_driven_errors_Burke_2013}%
  \BibitemOpen
  \bibfield  {author} {\bibinfo {author} {\bibfnamefont {M.-C.}\ \bibnamefont
  {Kim}}, \bibinfo {author} {\bibfnamefont {E.}~\bibnamefont {Sim}},\ and\
  \bibinfo {author} {\bibfnamefont {K.}~\bibnamefont {Burke}},\ }\bibfield
  {title} {\bibinfo {title} {{Understanding and Reducing Errors in Density
  Functional Calculations}},\ }\href
  {https://doi.org/10.1103/PhysRevLett.111.073003} {\bibfield  {journal}
  {\bibinfo  {journal} {Phys. Rev. Lett.}\ }\textbf {\bibinfo {volume} {111}},\
  \bibinfo {pages} {073003} (\bibinfo {year} {2013})}\BibitemShut {NoStop}%
\bibitem [{\citenamefont {Kim}\ \emph {et~al.}(2014)\citenamefont {Kim},
  \citenamefont {Sim},\ and\ \citenamefont
  {Burke}}]{density_driven_errors_Burke_2014}%
  \BibitemOpen
  \bibfield  {author} {\bibinfo {author} {\bibfnamefont {M.-C.}\ \bibnamefont
  {Kim}}, \bibinfo {author} {\bibfnamefont {E.}~\bibnamefont {Sim}},\ and\
  \bibinfo {author} {\bibfnamefont {K.}~\bibnamefont {Burke}},\ }\bibfield
  {title} {\bibinfo {title} {{Ions in solution: Density corrected density
  functional theory (DC-DFT)}},\ }\href {https://doi.org/10.1063/1.4869189}
  {\bibfield  {journal} {\bibinfo  {journal} {J. Chem. Phys.}\ }\textbf
  {\bibinfo {volume} {140}},\ \bibinfo {pages} {18A528} (\bibinfo {year}
  {2014})}\BibitemShut {NoStop}%
\bibitem [{\citenamefont {Vuckovic}\ \emph {et~al.}(2019)\citenamefont
  {Vuckovic}, \citenamefont {Song}, \citenamefont {Kozlowski}, \citenamefont
  {Sim},\ and\ \citenamefont {Burke}}]{density_driven_errors_Burke_2019}%
  \BibitemOpen
  \bibfield  {author} {\bibinfo {author} {\bibfnamefont {S.}~\bibnamefont
  {Vuckovic}}, \bibinfo {author} {\bibfnamefont {S.}~\bibnamefont {Song}},
  \bibinfo {author} {\bibfnamefont {J.}~\bibnamefont {Kozlowski}}, \bibinfo
  {author} {\bibfnamefont {E.}~\bibnamefont {Sim}},\ and\ \bibinfo {author}
  {\bibfnamefont {K.}~\bibnamefont {Burke}},\ }\bibfield  {title} {\bibinfo
  {title} {{Density Functional Analysis: The Theory of Density-Corrected
  DFT}},\ }\href {https://doi.org/10.1021/acs.jctc.9b00826} {\bibfield
  {journal} {\bibinfo  {journal} {J. Chem. Theory Comput}\ }\textbf {\bibinfo
  {volume} {15}},\ \bibinfo {pages} {6636–6646} (\bibinfo {year}
  {2019})}\BibitemShut {NoStop}%
\bibitem [{\citenamefont {Sim}\ \emph {et~al.}(2022)\citenamefont {Sim},
  \citenamefont {Song}, \citenamefont {Vuckovic},\ and\ \citenamefont
  {Burke}}]{density_driven_errors_Burke_2022}%
  \BibitemOpen
  \bibfield  {author} {\bibinfo {author} {\bibfnamefont {E.}~\bibnamefont
  {Sim}}, \bibinfo {author} {\bibfnamefont {S.}~\bibnamefont {Song}}, \bibinfo
  {author} {\bibfnamefont {S.}~\bibnamefont {Vuckovic}},\ and\ \bibinfo
  {author} {\bibfnamefont {K.}~\bibnamefont {Burke}},\ }\bibfield  {title}
  {\bibinfo {title} {{Improving Results by Improving Densities:
  Density-Corrected Density Functional Theory}},\ }\href
  {https://doi.org/10.1021/jacs.1c11506} {\bibfield  {journal} {\bibinfo
  {journal} {J. Am. Chem. Soc}\ }\textbf {\bibinfo {volume} {144}},\ \bibinfo
  {pages} {6625–6639} (\bibinfo {year} {2022})}\BibitemShut {NoStop}%
\bibitem [{\citenamefont {Song}\ \emph {et~al.}(2022)\citenamefont {Song},
  \citenamefont {Vuckovic}, \citenamefont {Sim},\ and\ \citenamefont
  {Burke}}]{DC_DFT_explained_Burke_2022}%
  \BibitemOpen
  \bibfield  {author} {\bibinfo {author} {\bibfnamefont {S.}~\bibnamefont
  {Song}}, \bibinfo {author} {\bibfnamefont {S.}~\bibnamefont {Vuckovic}},
  \bibinfo {author} {\bibfnamefont {E.}~\bibnamefont {Sim}},\ and\ \bibinfo
  {author} {\bibfnamefont {K.}~\bibnamefont {Burke}},\ }\bibfield  {title}
  {\bibinfo {title} {{Density-Corrected DFT Explained: Questions and
  Answers}},\ }\href {https://doi.org/10.1021/acs.jctc.1c01045} {\bibfield
  {journal} {\bibinfo  {journal} {J. Chem. Theory Comput}\ }\textbf {\bibinfo
  {volume} {18}},\ \bibinfo {pages} {817} (\bibinfo {year} {2022})}\BibitemShut
  {NoStop}%
\bibitem [{\citenamefont {Nam}\ \emph {et~al.}(2020)\citenamefont {Nam},
  \citenamefont {Song}, \citenamefont {Sim},\ and\ \citenamefont
  {Burke}}]{density_driven_errors_inversion_Burke_2020}%
  \BibitemOpen
  \bibfield  {author} {\bibinfo {author} {\bibfnamefont {S.}~\bibnamefont
  {Nam}}, \bibinfo {author} {\bibfnamefont {S.}~\bibnamefont {Song}}, \bibinfo
  {author} {\bibfnamefont {E.}~\bibnamefont {Sim}},\ and\ \bibinfo {author}
  {\bibfnamefont {K.}~\bibnamefont {Burke}},\ }\bibfield  {title} {\bibinfo
  {title} {{Measuring Density-Driven Errors Using Kohn–Sham Inversion}},\
  }\href {https://doi.org/10.1021/acs.jctc.0c00391} {\bibfield  {journal}
  {\bibinfo  {journal} {J. Chem. Theory Comput}\ }\textbf {\bibinfo {volume}
  {16}},\ \bibinfo {pages} {5014} (\bibinfo {year} {2020})}\BibitemShut
  {NoStop}%
\bibitem [{\citenamefont {Dasgupta}\ \emph {et~al.}(2021)\citenamefont
  {Dasgupta}, \citenamefont {Lambros}, \citenamefont {Perdew},\ and\
  \citenamefont {Paesani}}]{DC_SCAN_Water_Clusters_Perdew_2021}%
  \BibitemOpen
  \bibfield  {author} {\bibinfo {author} {\bibfnamefont {S.}~\bibnamefont
  {Dasgupta}}, \bibinfo {author} {\bibfnamefont {E.}~\bibnamefont {Lambros}},
  \bibinfo {author} {\bibfnamefont {J.~P.}\ \bibnamefont {Perdew}},\ and\
  \bibinfo {author} {\bibfnamefont {F.}~\bibnamefont {Paesani}},\ }\bibfield
  {title} {\bibinfo {title} {{Elevating density functional theory to chemical
  accuracy for water simulations through a density-corrected many-body
  formalism}},\ }\href {https://doi.org/10.1038/s41467-021-26618-9} {\bibfield
  {journal} {\bibinfo  {journal} {Nat Commun}\ }\textbf {\bibinfo {volume}
  {12}},\ \bibinfo {pages} {6359} (\bibinfo {year} {2021})}\BibitemShut
  {NoStop}%
\bibitem [{\citenamefont {Dasgupta}\ \emph {et~al.}(2022)\citenamefont
  {Dasgupta}, \citenamefont {Shahi}, \citenamefont {Bhetwal}, \citenamefont
  {Perdew},\ and\ \citenamefont {Paesani}}]{DC_SCAN_Perdew_2022}%
  \BibitemOpen
  \bibfield  {author} {\bibinfo {author} {\bibfnamefont {S.}~\bibnamefont
  {Dasgupta}}, \bibinfo {author} {\bibfnamefont {C.}~\bibnamefont {Shahi}},
  \bibinfo {author} {\bibfnamefont {P.}~\bibnamefont {Bhetwal}}, \bibinfo
  {author} {\bibfnamefont {J.~P.}\ \bibnamefont {Perdew}},\ and\ \bibinfo
  {author} {\bibfnamefont {F.}~\bibnamefont {Paesani}},\ }\bibfield  {title}
  {\bibinfo {title} {{How Good Is the Density-Corrected SCAN Functional for
  Neutral and Ionic Aqueous Systems, and What Is So Right about the
  Hartree–Fock Density?}},\ }\href {https://doi.org/10.1021/acs.jctc.2c00313}
  {\bibfield  {journal} {\bibinfo  {journal} {J. Chem. Theory Comput}\ }\textbf
  {\bibinfo {volume} {18}},\ \bibinfo {pages} {4745–4761} (\bibinfo {year}
  {2022})}\BibitemShut {NoStop}%
\bibitem [{\citenamefont {Kaplan}\ \emph {et~al.}(2023)\citenamefont {Kaplan},
  \citenamefont {Shahi}, \citenamefont {Bhetwal}, \citenamefont {Sah},\ and\
  \citenamefont {Perdew}}]{DC_DFT_Reaction_Barriers_2023_Perdew}%
  \BibitemOpen
  \bibfield  {author} {\bibinfo {author} {\bibfnamefont {A.~D.}\ \bibnamefont
  {Kaplan}}, \bibinfo {author} {\bibfnamefont {C.}~\bibnamefont {Shahi}},
  \bibinfo {author} {\bibfnamefont {P.}~\bibnamefont {Bhetwal}}, \bibinfo
  {author} {\bibfnamefont {R.~K.}\ \bibnamefont {Sah}},\ and\ \bibinfo {author}
  {\bibfnamefont {J.~P.}\ \bibnamefont {Perdew}},\ }\bibfield  {title}
  {\bibinfo {title} {{Understanding Density-Driven Errors for Reaction Barrier
  Heights}},\ }\href {https://doi.org/10.1021/acs.jctc.2c00953} {\bibfield
  {journal} {\bibinfo  {journal} {J. Chem. Theory Comput}\ }\textbf {\bibinfo
  {volume} {19}},\ \bibinfo {pages} {532} (\bibinfo {year} {2023})}\BibitemShut
  {NoStop}%
\bibitem [{\citenamefont {Song}\ \emph {et~al.}(2023)\citenamefont {Song},
  \citenamefont {Vuckovic}, \citenamefont {Kim}, \citenamefont {Yu},
  \citenamefont {Sim},\ and\ \citenamefont
  {Burke}}]{DC_SCAN_Water_Clusters_Burke_2023}%
  \BibitemOpen
  \bibfield  {author} {\bibinfo {author} {\bibfnamefont {S.}~\bibnamefont
  {Song}}, \bibinfo {author} {\bibfnamefont {S.}~\bibnamefont {Vuckovic}},
  \bibinfo {author} {\bibfnamefont {Y.}~\bibnamefont {Kim}}, \bibinfo {author}
  {\bibfnamefont {H.}~\bibnamefont {Yu}}, \bibinfo {author} {\bibfnamefont
  {E.}~\bibnamefont {Sim}},\ and\ \bibinfo {author} {\bibfnamefont
  {K.}~\bibnamefont {Burke}},\ }\bibfield  {title} {\bibinfo {title} {Extending
  density functional theory with near chemical accuracy beyond pure water},\
  }\href {https://doi.org/10.1038/s41467-023-36094-y} {\bibfield  {journal}
  {\bibinfo  {journal} {Nat. Commun}\ }\textbf {\bibinfo {volume} {14}},\
  \bibinfo {pages} {799} (\bibinfo {year} {2023})}\BibitemShut {NoStop}%
\bibitem [{\citenamefont {Yu}\ \emph {et~al.}(2023)\citenamefont {Yu},
  \citenamefont {Song}, \citenamefont {Nam}, \citenamefont {Burke},\ and\
  \citenamefont {Sim}}]{DC_DFT_Burke_2023_spin_contamination}%
  \BibitemOpen
  \bibfield  {author} {\bibinfo {author} {\bibfnamefont {H.}~\bibnamefont
  {Yu}}, \bibinfo {author} {\bibfnamefont {S.}~\bibnamefont {Song}}, \bibinfo
  {author} {\bibfnamefont {S.}~\bibnamefont {Nam}}, \bibinfo {author}
  {\bibfnamefont {K.}~\bibnamefont {Burke}},\ and\ \bibinfo {author}
  {\bibfnamefont {E.}~\bibnamefont {Sim}},\ }\bibfield  {title} {\bibinfo
  {title} {{Density-Corrected Density Functional Theory for Open Shells: How to
  Deal with Spin Contamination}},\ }\href
  {https://doi.org/10.1021/acs.jpclett.3c02017} {\bibfield  {journal} {\bibinfo
   {journal} {J. Phys. Chem. Lett.}\ }\textbf {\bibinfo {volume} {14}},\
  \bibinfo {pages} {9230} (\bibinfo {year} {2023})}\BibitemShut {NoStop}%
\bibitem [{\citenamefont {Jensen}\ and\ \citenamefont
  {Wasserman}(2018)}]{den_inversion_numerical_issues_Jensen_Wasserman_2018}%
  \BibitemOpen
  \bibfield  {author} {\bibinfo {author} {\bibfnamefont {D.~S.}\ \bibnamefont
  {Jensen}}\ and\ \bibinfo {author} {\bibfnamefont {A.}~\bibnamefont
  {Wasserman}},\ }\bibfield  {title} {\bibinfo {title} {Numerical methods for
  the inverse problem of density functional theory},\ }\href
  {https://doi.org/https://doi.org/10.1002/qua.25425} {\bibfield  {journal}
  {\bibinfo  {journal} {Int. J. Quantum Chem}\ }\textbf {\bibinfo {volume}
  {118}},\ \bibinfo {pages} {e25425} (\bibinfo {year} {2018})}\BibitemShut
  {NoStop}%
\bibitem [{\citenamefont {Shi}\ and\ \citenamefont
  {Wasserman}(2021)}]{den_inversion_numerical_issues_Shi_Wasserman_2021}%
  \BibitemOpen
  \bibfield  {author} {\bibinfo {author} {\bibfnamefont {Y.}~\bibnamefont
  {Shi}}\ and\ \bibinfo {author} {\bibfnamefont {A.}~\bibnamefont
  {Wasserman}},\ }\bibfield  {title} {\bibinfo {title} {{Inverse Kohn–Sham
  Density Functional Theory: Progress and Challenges}},\ }\href
  {https://doi.org/10.1021/acs.jpclett.1c00752} {\bibfield  {journal} {\bibinfo
   {journal} {J. Phys. Chem. Lett}\ }\textbf {\bibinfo {volume} {12}},\
  \bibinfo {pages} {5308} (\bibinfo {year} {2021})}\BibitemShut {NoStop}%
\bibitem [{\citenamefont {Werden}\ and\ \citenamefont
  {Davidson}(1984)}]{den_inversion_Werden_Davidson_1984}%
  \BibitemOpen
  \bibfield  {author} {\bibinfo {author} {\bibfnamefont {S.~H.}\ \bibnamefont
  {Werden}}\ and\ \bibinfo {author} {\bibfnamefont {E.~R.}\ \bibnamefont
  {Davidson}},\ }\bibinfo {title} {{On the Calculation of Potentials from
  Densities}},\ in\ \href {https://doi.org/10.1007/978-1-4899-2142-0_3} {\emph
  {\bibinfo {booktitle} {{Local Density Approximations in Quantum Chemistry and
  Solid State Physics}}}},\ \bibinfo {editor} {edited by\ \bibinfo {editor}
  {\bibfnamefont {J.~P.}\ \bibnamefont {Dahl}}\ and\ \bibinfo {editor}
  {\bibfnamefont {J.}~\bibnamefont {Avery}}}\ (\bibinfo  {publisher} {Springer
  US},\ \bibinfo {address} {Boston, MA},\ \bibinfo {year} {1984})\ pp.\
  \bibinfo {pages} {33--42}\BibitemShut {NoStop}%
\bibitem [{\citenamefont {Chan}\ \emph {et~al.}(1997)\citenamefont {Chan},
  \citenamefont {Tozer},\ and\ \citenamefont
  {Handy}}]{den_inversion_Tozer_1995}%
  \BibitemOpen
  \bibfield  {author} {\bibinfo {author} {\bibfnamefont {G.~K.-L.}\
  \bibnamefont {Chan}}, \bibinfo {author} {\bibfnamefont {D.~J.}\ \bibnamefont
  {Tozer}},\ and\ \bibinfo {author} {\bibfnamefont {N.~C.}\ \bibnamefont
  {Handy}},\ }\bibfield  {title} {\bibinfo {title} {{Correlation potentials and
  functionals in Hartree-Fock-Kohn-Sham theory}},\ }\href
  {https://doi.org/10.1063/1.474506} {\bibfield  {journal} {\bibinfo  {journal}
  {J. Chem. Phys.}\ }\textbf {\bibinfo {volume} {107}},\ \bibinfo {pages}
  {1536} (\bibinfo {year} {1997})}\BibitemShut {NoStop}%
\bibitem [{\citenamefont {G\"orling}(1992)}]{den_inversion_Görling_1992}%
  \BibitemOpen
  \bibfield  {author} {\bibinfo {author} {\bibfnamefont {A.}~\bibnamefont
  {G\"orling}},\ }\bibfield  {title} {\bibinfo {title} {{Kohn-Sham potentials
  and wave functions from electron densities}},\ }\href
  {https://doi.org/10.1103/PhysRevA.46.3753} {\bibfield  {journal} {\bibinfo
  {journal} {Phys. Rev. A}\ }\textbf {\bibinfo {volume} {46}},\ \bibinfo
  {pages} {3753} (\bibinfo {year} {1992})}\BibitemShut {NoStop}%
\bibitem [{\citenamefont {van Leeuwen}\ and\ \citenamefont
  {Baerends}(1994)}]{den_inversion_Leeuwen_Baerends_1994}%
  \BibitemOpen
  \bibfield  {author} {\bibinfo {author} {\bibfnamefont {R.}~\bibnamefont {van
  Leeuwen}}\ and\ \bibinfo {author} {\bibfnamefont {E.~J.}\ \bibnamefont
  {Baerends}},\ }\bibfield  {title} {\bibinfo {title} {Exchange-correlation
  potential with correct asymptotic behavior},\ }\href
  {https://doi.org/10.1103/PhysRevA.49.2421} {\bibfield  {journal} {\bibinfo
  {journal} {Phys. Rev. A}\ }\textbf {\bibinfo {volume} {49}},\ \bibinfo
  {pages} {2421} (\bibinfo {year} {1994})}\BibitemShut {NoStop}%
\bibitem [{\citenamefont {Zhao}\ \emph {et~al.}(1994)\citenamefont {Zhao},
  \citenamefont {Morrison},\ and\ \citenamefont
  {Parr}}]{den_inversion_Zhao_Morrison_Parr_1994}%
  \BibitemOpen
  \bibfield  {author} {\bibinfo {author} {\bibfnamefont {Q.}~\bibnamefont
  {Zhao}}, \bibinfo {author} {\bibfnamefont {R.~C.}\ \bibnamefont {Morrison}},\
  and\ \bibinfo {author} {\bibfnamefont {R.~G.}\ \bibnamefont {Parr}},\
  }\bibfield  {title} {\bibinfo {title} {{From electron densities to Kohn-Sham
  kinetic energies, orbital energies, exchange-correlation potentials, and
  exchange-correlation energies}},\ }\href
  {https://doi.org/10.1103/PhysRevA.50.2138} {\bibfield  {journal} {\bibinfo
  {journal} {Phys. Rev. A}\ }\textbf {\bibinfo {volume} {50}},\ \bibinfo
  {pages} {2138} (\bibinfo {year} {1994})}\BibitemShut {NoStop}%
\bibitem [{\citenamefont {Savin}\ \emph {et~al.}(1998)\citenamefont {Savin},
  \citenamefont {Umrigar},\ and\ \citenamefont
  {Gonze}}]{den_inversion_Savin_Umrigar_Gonze_1998}%
  \BibitemOpen
  \bibfield  {author} {\bibinfo {author} {\bibfnamefont {A.}~\bibnamefont
  {Savin}}, \bibinfo {author} {\bibfnamefont {C.}~\bibnamefont {Umrigar}},\
  and\ \bibinfo {author} {\bibfnamefont {X.}~\bibnamefont {Gonze}},\ }\bibfield
   {title} {\bibinfo {title} {{Relationship of Kohn–Sham eigenvalues to
  excitation energies}},\ }\href
  {https://doi.org/10.1016/S0009-2614(98)00316-9} {\bibfield  {journal}
  {\bibinfo  {journal} {Chemical Physics Letters}\ }\textbf {\bibinfo {volume}
  {288}},\ \bibinfo {pages} {391–395} (\bibinfo {year} {1998})}\BibitemShut
  {NoStop}%
\bibitem [{\citenamefont {Wu}\ and\ \citenamefont
  {Yang}(2003)}]{den_inversion_Wu_Yang_2003}%
  \BibitemOpen
  \bibfield  {author} {\bibinfo {author} {\bibfnamefont {Q.}~\bibnamefont
  {Wu}}\ and\ \bibinfo {author} {\bibfnamefont {W.}~\bibnamefont {Yang}},\
  }\bibfield  {title} {\bibinfo {title} {{A direct optimization method for
  calculating density functionals and exchange–correlation potentials from
  electron densities}},\ }\href {https://doi.org/10.1063/1.1535422} {\bibfield
  {journal} {\bibinfo  {journal} {J. Chem. Phys.}\ }\textbf {\bibinfo {volume}
  {118}},\ \bibinfo {pages} {2498} (\bibinfo {year} {2003})}\BibitemShut
  {NoStop}%
\bibitem [{\citenamefont {Peirs}\ \emph {et~al.}(2003)\citenamefont {Peirs},
  \citenamefont {Van~Neck},\ and\ \citenamefont
  {Waroquier}}]{den_inversion_Peirs_Van_Neck_Waroquier_2003}%
  \BibitemOpen
  \bibfield  {author} {\bibinfo {author} {\bibfnamefont {K.}~\bibnamefont
  {Peirs}}, \bibinfo {author} {\bibfnamefont {D.}~\bibnamefont {Van~Neck}},\
  and\ \bibinfo {author} {\bibfnamefont {M.}~\bibnamefont {Waroquier}},\
  }\bibfield  {title} {\bibinfo {title} {Algorithm to derive exact
  exchange-correlation potentials from correlated densities in atoms},\ }\href
  {https://doi.org/10.1103/PhysRevA.67.012505} {\bibfield  {journal} {\bibinfo
  {journal} {Phys. Rev. A}\ }\textbf {\bibinfo {volume} {67}},\ \bibinfo
  {pages} {012505} (\bibinfo {year} {2003})}\BibitemShut {NoStop}%
\bibitem [{\citenamefont {Kadantsev}\ and\ \citenamefont
  {Stott}(2004)}]{den_inversion_Kadantsev_Stott_2004}%
  \BibitemOpen
  \bibfield  {author} {\bibinfo {author} {\bibfnamefont {E.~S.}\ \bibnamefont
  {Kadantsev}}\ and\ \bibinfo {author} {\bibfnamefont {M.~J.}\ \bibnamefont
  {Stott}},\ }\bibfield  {title} {\bibinfo {title} {{Variational method for
  inverting the Kohn-Sham procedure}},\ }\href
  {https://doi.org/10.1103/PhysRevA.69.012502} {\bibfield  {journal} {\bibinfo
  {journal} {Phys. Rev. A}\ }\textbf {\bibinfo {volume} {69}},\ \bibinfo
  {pages} {012502} (\bibinfo {year} {2004})}\BibitemShut {NoStop}%
\bibitem [{\citenamefont {Ryabinkin}\ and\ \citenamefont
  {Staroverov}(2012)}]{den_inversion_Ryabinkin_Staroverov_2012}%
  \BibitemOpen
  \bibfield  {author} {\bibinfo {author} {\bibfnamefont {I.~G.}\ \bibnamefont
  {Ryabinkin}}\ and\ \bibinfo {author} {\bibfnamefont {V.~N.}\ \bibnamefont
  {Staroverov}},\ }\bibfield  {title} {\bibinfo {title} {{Determination of
  Kohn–Sham effective potentials from electron densities using the
  differential virial theorem}},\ }\href {https://doi.org/10.1063/1.4763481}
  {\bibfield  {journal} {\bibinfo  {journal} {J. Chem. Phys.}\ }\textbf
  {\bibinfo {volume} {137}},\ \bibinfo {pages} {164113} (\bibinfo {year}
  {2012})}\BibitemShut {NoStop}%
\bibitem [{\citenamefont {Kumar}\ \emph {et~al.}(2019)\citenamefont {Kumar},
  \citenamefont {Singh},\ and\ \citenamefont
  {Harbola}}]{den_inversion_Kumar_2019}%
  \BibitemOpen
  \bibfield  {author} {\bibinfo {author} {\bibfnamefont {A.}~\bibnamefont
  {Kumar}}, \bibinfo {author} {\bibfnamefont {R.}~\bibnamefont {Singh}},\ and\
  \bibinfo {author} {\bibfnamefont {M.~K.}\ \bibnamefont {Harbola}},\
  }\bibfield  {title} {\bibinfo {title} {{Universal nature of different methods
  of obtaining the exact Kohn–Sham exchange-correlation potential for a given
  density}},\ }\href {https://doi.org/10.1088/1361-6455/ab04e8} {\bibfield
  {journal} {\bibinfo  {journal} {J. Phys. B: At. Mol. Opt. Phys.}\ }\textbf
  {\bibinfo {volume} {52}},\ \bibinfo {pages} {075007} (\bibinfo {year}
  {2019})}\BibitemShut {NoStop}%
\bibitem [{\citenamefont {Callow}\ \emph
  {et~al.}(2020{\natexlab{b}})\citenamefont {Callow}, \citenamefont
  {Lathiotakis},\ and\ \citenamefont
  {Gidopoulos}}]{den_inversion_Tim_Callow_2020}%
  \BibitemOpen
  \bibfield  {author} {\bibinfo {author} {\bibfnamefont {T.~J.}\ \bibnamefont
  {Callow}}, \bibinfo {author} {\bibfnamefont {N.~N.}\ \bibnamefont
  {Lathiotakis}},\ and\ \bibinfo {author} {\bibfnamefont {N.~I.}\ \bibnamefont
  {Gidopoulos}},\ }\bibfield  {title} {\bibinfo {title} {{Density-inversion
  method for the Kohn–Sham potential: Role of the screening density}},\
  }\href {https://doi.org/10.1063/5.0005781} {\bibfield  {journal} {\bibinfo
  {journal} {J. Chem. Phys.}\ }\textbf {\bibinfo {volume} {152}},\ \bibinfo
  {pages} {164114} (\bibinfo {year} {2020}{\natexlab{b}})}\BibitemShut
  {NoStop}%
\bibitem [{\citenamefont {Bousiadi}\ \emph {et~al.}(2022)\citenamefont
  {Bousiadi}, \citenamefont {Gidopoulos},\ and\ \citenamefont
  {Lathiotakis}}]{den_inversion_Sofia_Extension_to_Tim_Callow_2022}%
  \BibitemOpen
  \bibfield  {author} {\bibinfo {author} {\bibfnamefont {S.}~\bibnamefont
  {Bousiadi}}, \bibinfo {author} {\bibfnamefont {N.~I.}\ \bibnamefont
  {Gidopoulos}},\ and\ \bibinfo {author} {\bibfnamefont {N.~N.}\ \bibnamefont
  {Lathiotakis}},\ }\bibfield  {title} {\bibinfo {title} {{Density inversion
  method for local basis sets without potential auxiliary functions: inverting
  densities from RDMFT}},\ }\href {https://doi.org/10.1039/D2CP01866G}
  {\bibfield  {journal} {\bibinfo  {journal} {Phys. Chem. Chem. Phys.}\
  }\textbf {\bibinfo {volume} {24}},\ \bibinfo {pages} {19279} (\bibinfo {year}
  {2022})}\BibitemShut {NoStop}%
\bibitem [{\citenamefont {G\"orling}\ and\ \citenamefont
  {Ernzerhof}(1995)}]{HF_inv_Görling_1995}%
  \BibitemOpen
  \bibfield  {author} {\bibinfo {author} {\bibfnamefont {A.}~\bibnamefont
  {G\"orling}}\ and\ \bibinfo {author} {\bibfnamefont {M.}~\bibnamefont
  {Ernzerhof}},\ }\bibfield  {title} {\bibinfo {title} {{Energy differences
  between Kohn-Sham and Hartree-Fock wave functions yielding the same electron
  density}},\ }\href {https://doi.org/10.1103/PhysRevA.51.4501} {\bibfield
  {journal} {\bibinfo  {journal} {Phys. Rev. A}\ }\textbf {\bibinfo {volume}
  {51}},\ \bibinfo {pages} {4501} (\bibinfo {year} {1995})}\BibitemShut
  {NoStop}%
\bibitem [{\citenamefont {Ryabinkin}\ \emph {et~al.}(2013)\citenamefont
  {Ryabinkin}, \citenamefont {Kananenka},\ and\ \citenamefont
  {Staroverov}}]{HF_inv_Staroverov_2013}%
  \BibitemOpen
  \bibfield  {author} {\bibinfo {author} {\bibfnamefont {I.~G.}\ \bibnamefont
  {Ryabinkin}}, \bibinfo {author} {\bibfnamefont {A.~A.}\ \bibnamefont
  {Kananenka}},\ and\ \bibinfo {author} {\bibfnamefont {V.~N.}\ \bibnamefont
  {Staroverov}},\ }\bibfield  {title} {\bibinfo {title} {{Accurate and
  Efficient Approximation to the Optimized Effective Potential for Exchange}},\
  }\href {https://doi.org/10.1103/PhysRevLett.111.013001} {\bibfield  {journal}
  {\bibinfo  {journal} {J. Chem. Phys.}\ }\textbf {\bibinfo {volume} {111}},\
  \bibinfo {pages} {013001} (\bibinfo {year} {2013})}\BibitemShut {NoStop}%
\bibitem [{\citenamefont {Kohut}\ \emph {et~al.}(2014)\citenamefont {Kohut},
  \citenamefont {Ryabinkin},\ and\ \citenamefont
  {Staroverov}}]{HF_inv_Staroverov_2014}%
  \BibitemOpen
  \bibfield  {author} {\bibinfo {author} {\bibfnamefont {S.~V.}\ \bibnamefont
  {Kohut}}, \bibinfo {author} {\bibfnamefont {I.~G.}\ \bibnamefont
  {Ryabinkin}},\ and\ \bibinfo {author} {\bibfnamefont {V.~N.}\ \bibnamefont
  {Staroverov}},\ }\bibfield  {title} {\bibinfo {title} {{Hierarchy of model
  Kohn–Sham potentials for orbital-dependent functionals: A practical
  alternative to the optimized effective potential method}},\ }\href
  {https://doi.org/10.1063/1.4871500} {\bibfield  {journal} {\bibinfo
  {journal} {J. Chem. Phys.}\ }\textbf {\bibinfo {volume} {140}},\ \bibinfo
  {pages} {18A535} (\bibinfo {year} {2014})}\BibitemShut {NoStop}%
\bibitem [{\citenamefont {Holas}\ \emph {et~al.}(1993)\citenamefont {Holas},
  \citenamefont {March}, \citenamefont {Takahashi},\ and\ \citenamefont
  {Zhang}}]{HF_inv_as_density_func_problem_Holas_1993}%
  \BibitemOpen
  \bibfield  {author} {\bibinfo {author} {\bibfnamefont {A.}~\bibnamefont
  {Holas}}, \bibinfo {author} {\bibfnamefont {N.~H.}\ \bibnamefont {March}},
  \bibinfo {author} {\bibfnamefont {Y.}~\bibnamefont {Takahashi}},\ and\
  \bibinfo {author} {\bibfnamefont {C.}~\bibnamefont {Zhang}},\ }\bibfield
  {title} {\bibinfo {title} {{Hartree-Fock method posed as a density-functional
  theory: Application to the Be atom}},\ }\href
  {https://doi.org/10.1103/PhysRevA.48.2708} {\bibfield  {journal} {\bibinfo
  {journal} {Phys. Rev. A}\ }\textbf {\bibinfo {volume} {48}},\ \bibinfo
  {pages} {2708} (\bibinfo {year} {1993})}\BibitemShut {NoStop}%
\bibitem [{\citenamefont
  {Nagy}(1997)}]{HF_inv_as_density_func_problem_Nagy_1997}%
  \BibitemOpen
  \bibfield  {author} {\bibinfo {author} {\bibfnamefont {A.}~\bibnamefont
  {Nagy}},\ }\bibfield  {title} {\bibinfo {title} {{Alternative derivation of
  the Krieger-Li-Iafrate approximation to the optimized-effective-potential
  method}},\ }\href {https://doi.org/10.1103/PhysRevA.55.3465} {\bibfield
  {journal} {\bibinfo  {journal} {Phys. Rev. A}\ }\textbf {\bibinfo {volume}
  {55}},\ \bibinfo {pages} {3465} (\bibinfo {year} {1997})}\BibitemShut
  {NoStop}%
\bibitem [{\citenamefont {{Jiqiang Chen}}\ and\ \citenamefont
  {Stott}(1994)}]{Stott_HF_inv_1994_small_atoms}%
  \BibitemOpen
  \bibfield  {author} {\bibinfo {author} {\bibfnamefont {R.~O.~E.}\
  \bibnamefont {{Jiqiang Chen}}}\ and\ \bibinfo {author} {\bibfnamefont
  {M.~J.}\ \bibnamefont {Stott}},\ }\bibfield  {title} {\bibinfo {title}
  {Exchange-correlation potential for small atoms},\ }\href
  {https://doi.org/10.1080/01418639408240169} {\bibfield  {journal} {\bibinfo
  {journal} {Philos. mag., B}\ }\textbf {\bibinfo {volume} {69}},\ \bibinfo
  {pages} {1001–1009} (\bibinfo {year} {1994})}\BibitemShut {NoStop}%
\bibitem [{\citenamefont {Hollins}\ \emph {et~al.}(2017)\citenamefont
  {Hollins}, \citenamefont {Clark}, \citenamefont {Refson},\ and\ \citenamefont
  {Gidopoulos}}]{Hollins_LFX_2017}%
  \BibitemOpen
  \bibfield  {author} {\bibinfo {author} {\bibfnamefont {T.~W.}\ \bibnamefont
  {Hollins}}, \bibinfo {author} {\bibfnamefont {S.~J.}\ \bibnamefont {Clark}},
  \bibinfo {author} {\bibfnamefont {K.}~\bibnamefont {Refson}},\ and\ \bibinfo
  {author} {\bibfnamefont {N.~I.}\ \bibnamefont {Gidopoulos}},\ }\bibfield
  {title} {\bibinfo {title} {{A local Fock-exchange potential in Kohn–Sham
  equations}},\ }\href {https://doi.org/10.1088/1361-648X/29/4/04LT01}
  {\bibfield  {journal} {\bibinfo  {journal} {J. Phys.: Condens. Matter}\
  }\textbf {\bibinfo {volume} {29}},\ \bibinfo {pages} {04LT01} (\bibinfo
  {year} {2017})}\BibitemShut {NoStop}%
\bibitem [{\citenamefont {Fletcher}\ and\ \citenamefont
  {Reeves}(1964)}]{Fletcher_Reeves_CG}%
  \BibitemOpen
  \bibfield  {author} {\bibinfo {author} {\bibfnamefont {R.}~\bibnamefont
  {Fletcher}}\ and\ \bibinfo {author} {\bibfnamefont {C.~M.}\ \bibnamefont
  {Reeves}},\ }\bibfield  {title} {\bibinfo {title} {{Function minimization by
  conjugate gradients}},\ }\href {https://doi.org/10.1093/comjnl/7.2.149}
  {\bibfield  {journal} {\bibinfo  {journal} {Comput. J}\ }\textbf {\bibinfo
  {volume} {7}},\ \bibinfo {pages} {149} (\bibinfo {year} {1964})}\BibitemShut
  {NoStop}%
\bibitem [{\citenamefont {Kohn}(1999)}]{Kohn_Nobel_Lecture_1999}%
  \BibitemOpen
  \bibfield  {author} {\bibinfo {author} {\bibfnamefont {W.}~\bibnamefont
  {Kohn}},\ }\bibfield  {title} {\bibinfo {title} {Nobel lecture: Electronic
  structure of matter---wave functions and density functionals},\ }\href
  {https://doi.org/10.1103/RevModPhys.71.1253} {\bibfield  {journal} {\bibinfo
  {journal} {Rev. Mod. Phys.}\ }\textbf {\bibinfo {volume} {71}},\ \bibinfo
  {pages} {1253} (\bibinfo {year} {1999})}\BibitemShut {NoStop}%
\bibitem [{\citenamefont {Kümmel}\ and\ \citenamefont
  {Kronik}(2008)}]{Kuemmel_Orbital_Dependent_Review_2008}%
  \BibitemOpen
  \bibfield  {author} {\bibinfo {author} {\bibfnamefont {S.}~\bibnamefont
  {Kümmel}}\ and\ \bibinfo {author} {\bibfnamefont {L.}~\bibnamefont
  {Kronik}},\ }\bibfield  {title} {\bibinfo {title} {{Orbital-dependent density
  functionals: Theory and applications}},\ }\href
  {https://doi.org/10.1103/RevModPhys.80.3} {\bibfield  {journal} {\bibinfo
  {journal} {Rev. Mod. Phys.}\ }\textbf {\bibinfo {volume} {80}},\ \bibinfo
  {pages} {3} (\bibinfo {year} {2008})}\BibitemShut {NoStop}%
\bibitem [{\citenamefont {Yang}\ \emph {et~al.}(2016)\citenamefont {Yang},
  \citenamefont {Peng}, \citenamefont {Sun},\ and\ \citenamefont
  {Perdew}}]{GKS_OEP_SCAN_Perdew_2016}%
  \BibitemOpen
  \bibfield  {author} {\bibinfo {author} {\bibfnamefont {Z.-H.}\ \bibnamefont
  {Yang}}, \bibinfo {author} {\bibfnamefont {H.}~\bibnamefont {Peng}}, \bibinfo
  {author} {\bibfnamefont {J.}~\bibnamefont {Sun}},\ and\ \bibinfo {author}
  {\bibfnamefont {J.~P.}\ \bibnamefont {Perdew}},\ }\bibfield  {title}
  {\bibinfo {title} {More realistic band gaps from meta-generalized gradient
  approximations: Only in a generalized {Kohn-Sham} scheme},\ }\href
  {https://doi.org/10.1103/PhysRevB.93.205205} {\bibfield  {journal} {\bibinfo
  {journal} {Phys. Rev. B}\ }\textbf {\bibinfo {volume} {93}},\ \bibinfo
  {pages} {205205} (\bibinfo {year} {2016})}\BibitemShut {NoStop}%
\bibitem [{\citenamefont {Clark}\ \emph {et~al.}(2017)\citenamefont {Clark},
  \citenamefont {Hollins}, \citenamefont {Refson},\ and\ \citenamefont
  {Gidopoulos}}]{LFX_Metals_2017}%
  \BibitemOpen
  \bibfield  {author} {\bibinfo {author} {\bibfnamefont {S.~J.}\ \bibnamefont
  {Clark}}, \bibinfo {author} {\bibfnamefont {T.~W.}\ \bibnamefont {Hollins}},
  \bibinfo {author} {\bibfnamefont {K.}~\bibnamefont {Refson}},\ and\ \bibinfo
  {author} {\bibfnamefont {N.~I.}\ \bibnamefont {Gidopoulos}},\ }\bibfield
  {title} {\bibinfo {title} {Self-interaction free local exchange potentials
  applied to metallic systems},\ }\href
  {https://doi.org/10.1088/1361-648X/aa7ba6} {\bibfield  {journal} {\bibinfo
  {journal} {J. Phys.: Condens. Matter}\ }\textbf {\bibinfo {volume} {29}},\
  \bibinfo {pages} {374002} (\bibinfo {year} {2017})}\BibitemShut {NoStop}%
\bibitem [{\citenamefont {St\"adele}\ \emph {et~al.}(1997)\citenamefont
  {St\"adele}, \citenamefont {Majewski}, \citenamefont {Vogl},\ and\
  \citenamefont {G\"orling}}]{EXX_Semiconductors_Görling_1997}%
  \BibitemOpen
  \bibfield  {author} {\bibinfo {author} {\bibfnamefont {M.}~\bibnamefont
  {St\"adele}}, \bibinfo {author} {\bibfnamefont {J.~A.}\ \bibnamefont
  {Majewski}}, \bibinfo {author} {\bibfnamefont {P.}~\bibnamefont {Vogl}},\
  and\ \bibinfo {author} {\bibfnamefont {A.}~\bibnamefont {G\"orling}},\
  }\bibfield  {title} {\bibinfo {title} {{Exact Kohn-Sham Exchange Potential in
  Semiconductors}},\ }\href {https://doi.org/10.1103/PhysRevLett.79.2089}
  {\bibfield  {journal} {\bibinfo  {journal} {Phys. Rev. Lett.}\ }\textbf
  {\bibinfo {volume} {79}},\ \bibinfo {pages} {2089} (\bibinfo {year}
  {1997})}\BibitemShut {NoStop}%
\bibitem [{\citenamefont {St\"adele}\ \emph {et~al.}(1999)\citenamefont
  {St\"adele}, \citenamefont {Moukara}, \citenamefont {Majewski}, \citenamefont
  {Vogl},\ and\ \citenamefont {G\"orling}}]{EXX_Semiconductors_Görling_1999}%
  \BibitemOpen
  \bibfield  {author} {\bibinfo {author} {\bibfnamefont {M.}~\bibnamefont
  {St\"adele}}, \bibinfo {author} {\bibfnamefont {M.}~\bibnamefont {Moukara}},
  \bibinfo {author} {\bibfnamefont {J.~A.}\ \bibnamefont {Majewski}}, \bibinfo
  {author} {\bibfnamefont {P.}~\bibnamefont {Vogl}},\ and\ \bibinfo {author}
  {\bibfnamefont {A.}~\bibnamefont {G\"orling}},\ }\bibfield  {title} {\bibinfo
  {title} {Exact exchange {Kohn-Sham} formalism applied to semiconductors},\
  }\href {https://doi.org/10.1103/PhysRevB.59.10031} {\bibfield  {journal}
  {\bibinfo  {journal} {Phys. Rev. B}\ }\textbf {\bibinfo {volume} {59}},\
  \bibinfo {pages} {10031} (\bibinfo {year} {1999})}\BibitemShut {NoStop}%
\bibitem [{\citenamefont {Grüning}\ \emph {et~al.}(2006)\citenamefont
  {Grüning}, \citenamefont {Marini},\ and\ \citenamefont
  {Rubio}}]{EXX_Delta_X_Grüning_Rubio_2006}%
  \BibitemOpen
  \bibfield  {author} {\bibinfo {author} {\bibfnamefont {M.}~\bibnamefont
  {Grüning}}, \bibinfo {author} {\bibfnamefont {A.}~\bibnamefont {Marini}},\
  and\ \bibinfo {author} {\bibfnamefont {A.}~\bibnamefont {Rubio}},\ }\bibfield
   {title} {\bibinfo {title} {{Density functionals from many-body perturbation
  theory: The band gap for semiconductors and insulators}},\ }\href
  {https://doi.org/10.1063/1.2189226} {\bibfield  {journal} {\bibinfo
  {journal} {J. Chem. Phys.}\ }\textbf {\bibinfo {volume} {124}},\ \bibinfo
  {pages} {154108} (\bibinfo {year} {2006})}\BibitemShut {NoStop}%
\bibitem [{\citenamefont {Klimeš}\ and\ \citenamefont
  {Kresse}(2014)}]{EXX_RPA_Kresse_2014}%
  \BibitemOpen
  \bibfield  {author} {\bibinfo {author} {\bibfnamefont {J.}~\bibnamefont
  {Klimeš}}\ and\ \bibinfo {author} {\bibfnamefont {G.}~\bibnamefont
  {Kresse}},\ }\bibfield  {title} {\bibinfo {title} {{Kohn-Sham band gaps and
  potentials of solids from the optimised effective potential method within the
  random phase approximation}},\ }\href {https://doi.org/10.1063/1.4863502}
  {\bibfield  {journal} {\bibinfo  {journal} {J. Chem. Phys.}\ }\textbf
  {\bibinfo {volume} {140}},\ \bibinfo {pages} {054516} (\bibinfo {year}
  {2014})}\BibitemShut {NoStop}%
\bibitem [{\citenamefont {Hollins}\ \emph {et~al.}(2012)\citenamefont
  {Hollins}, \citenamefont {Clark}, \citenamefont {Refson},\ and\ \citenamefont
  {Gidopoulos}}]{Hollins_Hylleraas_OEP}%
  \BibitemOpen
  \bibfield  {author} {\bibinfo {author} {\bibfnamefont {T.~W.}\ \bibnamefont
  {Hollins}}, \bibinfo {author} {\bibfnamefont {S.~J.}\ \bibnamefont {Clark}},
  \bibinfo {author} {\bibfnamefont {K.}~\bibnamefont {Refson}},\ and\ \bibinfo
  {author} {\bibfnamefont {N.~I.}\ \bibnamefont {Gidopoulos}},\ }\bibfield
  {title} {\bibinfo {title} {{Optimized effective potential using the Hylleraas
  variational method}},\ }\href {https://doi.org/10.1103/PhysRevB.85.235126}
  {\bibfield  {journal} {\bibinfo  {journal} {Phys. Rev. B}\ }\textbf {\bibinfo
  {volume} {85}},\ \bibinfo {pages} {235126} (\bibinfo {year}
  {2012})}\BibitemShut {NoStop}%
\bibitem [{\citenamefont {Trushin}\ \emph {et~al.}(2019)\citenamefont
  {Trushin}, \citenamefont {Fromm},\ and\ \citenamefont
  {G\"orling}}]{Trushin_Görling_2019_EXX_perovskites}%
  \BibitemOpen
  \bibfield  {author} {\bibinfo {author} {\bibfnamefont {E.}~\bibnamefont
  {Trushin}}, \bibinfo {author} {\bibfnamefont {L.}~\bibnamefont {Fromm}},\
  and\ \bibinfo {author} {\bibfnamefont {A.}~\bibnamefont {G\"orling}},\
  }\bibfield  {title} {\bibinfo {title} {{Assessment of the exact-exchange-only
  Kohn-Sham method for the calculation of band structures for transition metal
  oxide and metal halide perovskites}},\ }\href
  {https://doi.org/10.1103/PhysRevB.100.075205} {\bibfield  {journal} {\bibinfo
   {journal} {Phys. Rev. B}\ }\textbf {\bibinfo {volume} {100}},\ \bibinfo
  {pages} {075205} (\bibinfo {year} {2019})}\BibitemShut {NoStop}%
\bibitem [{\citenamefont {Kraisler}\ and\ \citenamefont
  {Kronik}(2013)}]{Kraisler_Kronik_2013_Delta_XC}%
  \BibitemOpen
  \bibfield  {author} {\bibinfo {author} {\bibfnamefont {E.}~\bibnamefont
  {Kraisler}}\ and\ \bibinfo {author} {\bibfnamefont {L.}~\bibnamefont
  {Kronik}},\ }\bibfield  {title} {\bibinfo {title} {{Piecewise Linearity of
  Approximate Density Functionals Revisited: Implications for Frontier Orbital
  Energies}},\ }\href {https://doi.org/10.1103/PhysRevLett.110.126403}
  {\bibfield  {journal} {\bibinfo  {journal} {Phys. Rev. Lett.}\ }\textbf
  {\bibinfo {volume} {110}},\ \bibinfo {pages} {126403} (\bibinfo {year}
  {2013})}\BibitemShut {NoStop}%
\bibitem [{\citenamefont {Kraisler}\ and\ \citenamefont
  {Kronik}(2014)}]{Kraisler_Kronik_2014_Delta_XC}%
  \BibitemOpen
  \bibfield  {author} {\bibinfo {author} {\bibfnamefont {E.}~\bibnamefont
  {Kraisler}}\ and\ \bibinfo {author} {\bibfnamefont {L.}~\bibnamefont
  {Kronik}},\ }\bibfield  {title} {\bibinfo {title} {{Fundamental gaps with
  approximate density functionals: The derivative discontinuity revealed from
  ensemble considerations}},\ }\href {https://doi.org/10.1063/1.4871462}
  {\bibfield  {journal} {\bibinfo  {journal} {J. Chem. Phys.}\ }\textbf
  {\bibinfo {volume} {140}},\ \bibinfo {pages} {18A540} (\bibinfo {year}
  {2014})}\BibitemShut {NoStop}%
\bibitem [{\citenamefont {Hollins}(2014)}]{HollinsPhDThesis}%
  \BibitemOpen
  \bibfield  {author} {\bibinfo {author} {\bibfnamefont {T.~W.}\ \bibnamefont
  {Hollins}},\ }\emph {\bibinfo {title} {Local Exchange Potentials In Density
  Functional Theory}},\ \href@noop {} {\bibinfo {type} {{PhD} thesis}},\
  \bibinfo  {school} {University of Durham} (\bibinfo {year}
  {2014})\BibitemShut {NoStop}%
\bibitem [{\citenamefont {Becke}(1993)}]{B3LYP_1}%
  \BibitemOpen
  \bibfield  {author} {\bibinfo {author} {\bibfnamefont {A.~D.}\ \bibnamefont
  {Becke}},\ }\bibfield  {title} {\bibinfo {title} {{Density‐functional
  thermochemistry. III. The role of exact exchange}},\ }\href@noop {}
  {\bibfield  {journal} {\bibinfo  {journal} {J. Chem. Phys.}\ }\textbf
  {\bibinfo {volume} {98}},\ \bibinfo {pages} {5648} (\bibinfo {year}
  {1993})}\BibitemShut {NoStop}%
\bibitem [{\citenamefont {Perdew}\ \emph
  {et~al.}(1996{\natexlab{a}})\citenamefont {Perdew}, \citenamefont
  {Ernzerhof},\ and\ \citenamefont {Burke}}]{PBE0}%
  \BibitemOpen
  \bibfield  {author} {\bibinfo {author} {\bibfnamefont {J.~P.}\ \bibnamefont
  {Perdew}}, \bibinfo {author} {\bibfnamefont {M.}~\bibnamefont {Ernzerhof}},\
  and\ \bibinfo {author} {\bibfnamefont {K.}~\bibnamefont {Burke}},\ }\bibfield
   {title} {\bibinfo {title} {{Rationale for mixing exact exchange with density
  functional approximations}},\ }\href {https://doi.org/10.1063/1.472933}
  {\bibfield  {journal} {\bibinfo  {journal} {J. Chem. Phys.}\ }\textbf
  {\bibinfo {volume} {105}},\ \bibinfo {pages} {9982} (\bibinfo {year}
  {1996}{\natexlab{a}})}\BibitemShut {NoStop}%
\bibitem [{\citenamefont {Heyd}\ \emph {et~al.}(2003)\citenamefont {Heyd},
  \citenamefont {Scuseria},\ and\ \citenamefont {Ernzerhof}}]{HSE03}%
  \BibitemOpen
  \bibfield  {author} {\bibinfo {author} {\bibfnamefont {J.}~\bibnamefont
  {Heyd}}, \bibinfo {author} {\bibfnamefont {G.~E.}\ \bibnamefont {Scuseria}},\
  and\ \bibinfo {author} {\bibfnamefont {M.}~\bibnamefont {Ernzerhof}},\
  }\bibfield  {title} {\bibinfo {title} {{Hybrid functionals based on a
  screened Coulomb potential}},\ }\href {https://doi.org/10.1063/1.1564060}
  {\bibfield  {journal} {\bibinfo  {journal} {J. Chem. Phys.}\ }\textbf
  {\bibinfo {volume} {118}},\ \bibinfo {pages} {8207} (\bibinfo {year}
  {2003})}\BibitemShut {NoStop}%
\bibitem [{\citenamefont {Cococcioni}\ and\ \citenamefont
  {de~Gironcoli}(2005)}]{Hubbard_U_linear_response_2005}%
  \BibitemOpen
  \bibfield  {author} {\bibinfo {author} {\bibfnamefont {M.}~\bibnamefont
  {Cococcioni}}\ and\ \bibinfo {author} {\bibfnamefont {S.}~\bibnamefont
  {de~Gironcoli}},\ }\bibfield  {title} {\bibinfo {title} {{Linear response
  approach to the calculation of the effective interaction parameters in the
  $\mathrm{LDA}+{U}$ method}},\ }\href
  {https://doi.org/10.1103/PhysRevB.71.035105} {\bibfield  {journal} {\bibinfo
  {journal} {Phys. Rev. B}\ }\textbf {\bibinfo {volume} {71}},\ \bibinfo
  {pages} {035105} (\bibinfo {year} {2005})}\BibitemShut {NoStop}%
\bibitem [{\citenamefont {Kulik}\ \emph {et~al.}(2006)\citenamefont {Kulik},
  \citenamefont {Cococcioni}, \citenamefont {Scherlis},\ and\ \citenamefont
  {Marzari}}]{Hubbard_U_linear_response_2006}%
  \BibitemOpen
  \bibfield  {author} {\bibinfo {author} {\bibfnamefont {H.~J.}\ \bibnamefont
  {Kulik}}, \bibinfo {author} {\bibfnamefont {M.}~\bibnamefont {Cococcioni}},
  \bibinfo {author} {\bibfnamefont {D.~A.}\ \bibnamefont {Scherlis}},\ and\
  \bibinfo {author} {\bibfnamefont {N.}~\bibnamefont {Marzari}},\ }\bibfield
  {title} {\bibinfo {title} {{Density Functional Theory in Transition-Metal
  Chemistry: A Self-Consistent Hubbard $U$ Approach}},\ }\href
  {https://doi.org/10.1103/PhysRevLett.97.103001} {\bibfield  {journal}
  {\bibinfo  {journal} {Phys. Rev. Lett.}\ }\textbf {\bibinfo {volume} {97}},\
  \bibinfo {pages} {103001} (\bibinfo {year} {2006})}\BibitemShut {NoStop}%
\bibitem [{\citenamefont {Suhai}(1983)}]{MP2_Solids_1983_Suhai}%
  \BibitemOpen
  \bibfield  {author} {\bibinfo {author} {\bibfnamefont {S.}~\bibnamefont
  {Suhai}},\ }\bibfield  {title} {\bibinfo {title} {Quasiparticle energy-band
  structures in semiconducting polymers: {Correlation} effects on the band gap
  in polyacetylene},\ }\href {https://doi.org/10.1103/PhysRevB.27.3506}
  {\bibfield  {journal} {\bibinfo  {journal} {Phys. Rev. B}\ }\textbf {\bibinfo
  {volume} {27}},\ \bibinfo {pages} {3506} (\bibinfo {year}
  {1983})}\BibitemShut {NoStop}%
\bibitem [{\citenamefont {Grüneis}\ \emph {et~al.}(2010)\citenamefont
  {Grüneis}, \citenamefont {Marsman},\ and\ \citenamefont
  {Kresse}}]{MP2_Kresse_2010}%
  \BibitemOpen
  \bibfield  {author} {\bibinfo {author} {\bibfnamefont {A.}~\bibnamefont
  {Grüneis}}, \bibinfo {author} {\bibfnamefont {M.}~\bibnamefont {Marsman}},\
  and\ \bibinfo {author} {\bibfnamefont {G.}~\bibnamefont {Kresse}},\
  }\bibfield  {title} {\bibinfo {title} {{Second-order Møller–Plesset
  perturbation theory applied to extended systems. II. Structural and energetic
  properties}},\ }\href {https://doi.org/10.1063/1.3466765} {\bibfield
  {journal} {\bibinfo  {journal} {J. Chem. Phys.}\ }\textbf {\bibinfo {volume}
  {133}},\ \bibinfo {pages} {074107} (\bibinfo {year} {2010})}\BibitemShut
  {NoStop}%
\bibitem [{\citenamefont {Lange}\ and\ \citenamefont
  {Berkelbach}(2021)}]{MP2_Berkelbach_2021}%
  \BibitemOpen
  \bibfield  {author} {\bibinfo {author} {\bibfnamefont {M.~F.}\ \bibnamefont
  {Lange}}\ and\ \bibinfo {author} {\bibfnamefont {T.~C.}\ \bibnamefont
  {Berkelbach}},\ }\bibfield  {title} {\bibinfo {title} {{Improving MP2
  bandgaps with low-scaling approximations to EOM-CCSD}},\ }\href
  {https://doi.org/10.1063/5.0061242} {\bibfield  {journal} {\bibinfo
  {journal} {J. Chem. Phys.}\ }\textbf {\bibinfo {volume} {155}},\ \bibinfo
  {pages} {081101} (\bibinfo {year} {2021})}\BibitemShut {NoStop}%
\bibitem [{\citenamefont {Clark}\ \emph {et~al.}(2005)\citenamefont {Clark},
  \citenamefont {Segall}, \citenamefont {Pickard}, \citenamefont {Hasnip},
  \citenamefont {Probert}, \citenamefont {Refson},\ and\ \citenamefont
  {Payne}}]{CASTEP}%
  \BibitemOpen
  \bibfield  {author} {\bibinfo {author} {\bibfnamefont {S.~J.}\ \bibnamefont
  {Clark}}, \bibinfo {author} {\bibfnamefont {M.~D.}\ \bibnamefont {Segall}},
  \bibinfo {author} {\bibfnamefont {C.~J.}\ \bibnamefont {Pickard}}, \bibinfo
  {author} {\bibfnamefont {P.~J.}\ \bibnamefont {Hasnip}}, \bibinfo {author}
  {\bibfnamefont {M.~I.~J.}\ \bibnamefont {Probert}}, \bibinfo {author}
  {\bibfnamefont {K.}~\bibnamefont {Refson}},\ and\ \bibinfo {author}
  {\bibfnamefont {M.~C.}\ \bibnamefont {Payne}},\ }\bibfield  {title} {\bibinfo
  {title} {First principles methods using {CASTEP}},\ }\href
  {https://doi.org/10.1524/zkri.220.5.567.65075} {\bibfield  {journal}
  {\bibinfo  {journal} {Z. Kristall.}\ }\textbf {\bibinfo {volume} {220}},\
  \bibinfo {pages} {567–570} (\bibinfo {year} {2005})}\BibitemShut {NoStop}%
\bibitem [{\citenamefont {Car}\ and\ \citenamefont
  {Parrinello}(1985)}]{Car_Parrinello_1985}%
  \BibitemOpen
  \bibfield  {author} {\bibinfo {author} {\bibfnamefont {R.}~\bibnamefont
  {Car}}\ and\ \bibinfo {author} {\bibfnamefont {M.}~\bibnamefont
  {Parrinello}},\ }\bibfield  {title} {\bibinfo {title} {{Unified Approach for
  Molecular Dynamics and Density-Functional Theory}},\ }\href
  {https://doi.org/10.1103/PhysRevLett.55.2471} {\bibfield  {journal} {\bibinfo
   {journal} {Phys. Rev. Lett.}\ }\textbf {\bibinfo {volume} {55}},\ \bibinfo
  {pages} {2471} (\bibinfo {year} {1985})}\BibitemShut {NoStop}%
\bibitem [{\citenamefont {Monkhorst}\ and\ \citenamefont
  {Pack}(1976)}]{Mohkhorst_Pack_Grid}%
  \BibitemOpen
  \bibfield  {author} {\bibinfo {author} {\bibfnamefont {H.~J.}\ \bibnamefont
  {Monkhorst}}\ and\ \bibinfo {author} {\bibfnamefont {J.~D.}\ \bibnamefont
  {Pack}},\ }\bibfield  {title} {\bibinfo {title} {Special points for
  {Brillouin}-zone integrations},\ }\href
  {https://doi.org/10.1103/PhysRevB.13.5188} {\bibfield  {journal} {\bibinfo
  {journal} {Phys. Rev. B}\ }\textbf {\bibinfo {volume} {13}},\ \bibinfo
  {pages} {5188} (\bibinfo {year} {1976})}\BibitemShut {NoStop}%
\bibitem [{\citenamefont {Lejaeghere}\ \emph {et~al.}(2016)\citenamefont
  {Lejaeghere} \emph {et~al.}}]{CASTEP_NCP19_Reference}%
  \BibitemOpen
  \bibfield  {author} {\bibinfo {author} {\bibfnamefont {K.}~\bibnamefont
  {Lejaeghere}} \emph {et~al.},\ }\bibfield  {title} {\bibinfo {title}
  {Reproducibility in density functional theory calculations of solids},\
  }\href {https://doi.org/10.1126/science.aad3000} {\bibfield  {journal}
  {\bibinfo  {journal} {Science}\ }\textbf {\bibinfo {volume} {351}},\ \bibinfo
  {pages} {aad3000} (\bibinfo {year} {2016})}\BibitemShut {NoStop}%
\bibitem [{\citenamefont {Lee}\ \emph {et~al.}(1988)\citenamefont {Lee},
  \citenamefont {Yang},\ and\ \citenamefont {Parr}}]{B3LYP_2}%
  \BibitemOpen
  \bibfield  {author} {\bibinfo {author} {\bibfnamefont {C.}~\bibnamefont
  {Lee}}, \bibinfo {author} {\bibfnamefont {W.}~\bibnamefont {Yang}},\ and\
  \bibinfo {author} {\bibfnamefont {R.~G.}\ \bibnamefont {Parr}},\ }\bibfield
  {title} {\bibinfo {title} {Development of the {Colle-Salvetti}
  correlation-energy formula into a functional of the electron density},\
  }\href {https://doi.org/10.1103/PhysRevB.37.785} {\bibfield  {journal}
  {\bibinfo  {journal} {Phys. Rev. B}\ }\textbf {\bibinfo {volume} {37}},\
  \bibinfo {pages} {785} (\bibinfo {year} {1988})}\BibitemShut {NoStop}%
\bibitem [{\citenamefont {Stephens}\ \emph {et~al.}(1994)\citenamefont
  {Stephens}, \citenamefont {Devlin}, \citenamefont {Chabalowski},\ and\
  \citenamefont {Frisch}}]{B3LYP_3}%
  \BibitemOpen
  \bibfield  {author} {\bibinfo {author} {\bibfnamefont {P.~J.}\ \bibnamefont
  {Stephens}}, \bibinfo {author} {\bibfnamefont {F.~J.}\ \bibnamefont
  {Devlin}}, \bibinfo {author} {\bibfnamefont {C.~F.}\ \bibnamefont
  {Chabalowski}},\ and\ \bibinfo {author} {\bibfnamefont {M.~J.}\ \bibnamefont
  {Frisch}},\ }\bibfield  {title} {\bibinfo {title} {Ab initio calculation of
  vibrational absorption and circular dichroism spectra using density
  functional force fields},\ }\href {https://doi.org/10.1021/j100096a001}
  {\bibfield  {journal} {\bibinfo  {journal} {J. Phys. Chem.}\ }\textbf
  {\bibinfo {volume} {98}},\ \bibinfo {pages} {11623–11627} (\bibinfo {year}
  {1994})}\BibitemShut {NoStop}%
\bibitem [{\citenamefont {Heyd}\ \emph {et~al.}(2006)\citenamefont {Heyd},
  \citenamefont {Scuseria},\ and\ \citenamefont {Ernzerhof}}]{HSE06}%
  \BibitemOpen
  \bibfield  {author} {\bibinfo {author} {\bibfnamefont {J.}~\bibnamefont
  {Heyd}}, \bibinfo {author} {\bibfnamefont {G.~E.}\ \bibnamefont {Scuseria}},\
  and\ \bibinfo {author} {\bibfnamefont {M.}~\bibnamefont {Ernzerhof}},\
  }\bibfield  {title} {\bibinfo {title} {{Erratum: `Hybrid functionals based on
  a screened Coulomb potential' [J. Chem. Phys. 118, 8207 (2003)]}},\ }\href
  {https://doi.org/10.1063/1.2204597} {\bibfield  {journal} {\bibinfo
  {journal} {J. Chem. Phys.}\ }\textbf {\bibinfo {volume} {124}} (\bibinfo
  {year} {2006})}\BibitemShut {NoStop}%
\bibitem [{\citenamefont {Perdew}\ \emph
  {et~al.}(1996{\natexlab{b}})\citenamefont {Perdew}, \citenamefont {Burke},\
  and\ \citenamefont {Ernzerhof}}]{PBE}%
  \BibitemOpen
  \bibfield  {author} {\bibinfo {author} {\bibfnamefont {J.~P.}\ \bibnamefont
  {Perdew}}, \bibinfo {author} {\bibfnamefont {K.}~\bibnamefont {Burke}},\ and\
  \bibinfo {author} {\bibfnamefont {M.}~\bibnamefont {Ernzerhof}},\ }\bibfield
  {title} {\bibinfo {title} {{Generalized Gradient Approximation Made
  Simple}},\ }\href {https://doi.org/10.1103/PhysRevLett.77.3865} {\bibfield
  {journal} {\bibinfo  {journal} {Phys. Rev. Lett.}\ }\textbf {\bibinfo
  {volume} {77}},\ \bibinfo {pages} {3865} (\bibinfo {year}
  {1996}{\natexlab{b}})}\BibitemShut {NoStop}%
\bibitem [{\citenamefont {Madelung}(1996)}]{Madelung_Basic_Data}%
  \BibitemOpen
  \bibfield  {author} {\bibinfo {author} {\bibfnamefont {O.}~\bibnamefont
  {Madelung}},\ }\href@noop {} {\emph {\bibinfo {title} {Semiconductors --
  Basic Data}}},\ \bibinfo {edition} {2nd}\ ed.\ (\bibinfo  {publisher}
  {Springer Berlin / Heidelberg},\ \bibinfo {address} {Berlin, Heidelberg},\
  \bibinfo {year} {1996})\BibitemShut {NoStop}%
\bibitem [{\citenamefont {Perdew}\ and\ \citenamefont
  {Schmidt}(2001)}]{Jacobs_ladder}%
  \BibitemOpen
  \bibfield  {author} {\bibinfo {author} {\bibfnamefont {J.~P.}\ \bibnamefont
  {Perdew}}\ and\ \bibinfo {author} {\bibfnamefont {K.}~\bibnamefont
  {Schmidt}},\ }\bibfield  {title} {\bibinfo {title} {{Jacob’s ladder of
  density functional approximations for the exchange-correlation energy}},\
  }\href {https://doi.org/10.1063/1.1390175} {\bibfield  {journal} {\bibinfo
  {journal} {AIP Conf Proc}\ }\textbf {\bibinfo {volume} {577}},\ \bibinfo
  {pages} {1} (\bibinfo {year} {2001})}\BibitemShut {NoStop}%
\bibitem [{Note1()}]{Note1}%
  \BibitemOpen
  \bibinfo {note} {For LDA target densities, the LXC potential was initialised
  to the PBE potential calculated from the target density. If the LXC potential
  was instead initialised to the LDA potential calculated from the target
  density like elsewhere in this work, the inversion is already converged
  without any further iteration, and more importantly remains
  converged.}\BibitemShut {Stop}%
\bibitem [{Sup()}]{Supplementary_Material}%
  \BibitemOpen
  \href@noop {} {}\bibinfo {note} {See supplemental material at DOI:XXXXXXX for
  the inversion of the LDA density and projected density of states that show
  how we selected bands with predominantly $d$-character in calculations for
  transition metal oxides.}\BibitemShut {Stop}%
\bibitem [{\citenamefont {Cardona}\ and\ \citenamefont
  {Thewalt}(2005)}]{Cardona_Thewalt_ZPR_values}%
  \BibitemOpen
  \bibfield  {author} {\bibinfo {author} {\bibfnamefont {M.}~\bibnamefont
  {Cardona}}\ and\ \bibinfo {author} {\bibfnamefont {M.~L.~W.}\ \bibnamefont
  {Thewalt}},\ }\bibfield  {title} {\bibinfo {title} {Isotope effects on the
  optical spectra of semiconductors},\ }\href
  {https://doi.org/10.1103/RevModPhys.77.1173} {\bibfield  {journal} {\bibinfo
  {journal} {Rev. Mod. Phys.}\ }\textbf {\bibinfo {volume} {77}},\ \bibinfo
  {pages} {1173} (\bibinfo {year} {2005})}\BibitemShut {NoStop}%
\bibitem [{\citenamefont {Kang}\ \emph {et~al.}(2019)\citenamefont {Kang},
  \citenamefont {Li}, \citenamefont {Wu}, \citenamefont {Nguyen},\ and\
  \citenamefont {Hu}}]{BAs_band_gap}%
  \BibitemOpen
  \bibfield  {author} {\bibinfo {author} {\bibfnamefont {J.~S.}\ \bibnamefont
  {Kang}}, \bibinfo {author} {\bibfnamefont {M.}~\bibnamefont {Li}}, \bibinfo
  {author} {\bibfnamefont {H.}~\bibnamefont {Wu}}, \bibinfo {author}
  {\bibfnamefont {H.}~\bibnamefont {Nguyen}},\ and\ \bibinfo {author}
  {\bibfnamefont {Y.}~\bibnamefont {Hu}},\ }\bibfield  {title} {\bibinfo
  {title} {{Basic physical properties of cubic boron arsenide}},\ }\href
  {https://doi.org/10.1063/1.5116025} {\bibfield  {journal} {\bibinfo
  {journal} {Appl. Phys. Lett.}\ }\textbf {\bibinfo {volume} {115}} (\bibinfo
  {year} {2019})}\BibitemShut {NoStop}%
\bibitem [{\citenamefont {Li}\ \emph {et~al.}(2004)\citenamefont {Li},
  \citenamefont {Kennedy}, \citenamefont {Kubota}, \citenamefont {Kato},\ and\
  \citenamefont {Garrett}}]{BaHfO3_Lat_Param}%
  \BibitemOpen
  \bibfield  {author} {\bibinfo {author} {\bibfnamefont {L.}~\bibnamefont
  {Li}}, \bibinfo {author} {\bibfnamefont {B.~J.}\ \bibnamefont {Kennedy}},
  \bibinfo {author} {\bibfnamefont {Y.}~\bibnamefont {Kubota}}, \bibinfo
  {author} {\bibfnamefont {K.}~\bibnamefont {Kato}},\ and\ \bibinfo {author}
  {\bibfnamefont {R.~F.}\ \bibnamefont {Garrett}},\ }\bibfield  {title}
  {\bibinfo {title} {Structures and phase transitions in
  {$\textrm{Sr}_{1-x}\textrm{Ba}_x\textrm{HfO}_3$} perovskites},\ }\href
  {https://doi.org/10.1039/B308258J} {\bibfield  {journal} {\bibinfo  {journal}
  {J. Mater. Chem.}\ }\textbf {\bibinfo {volume} {14}},\ \bibinfo {pages} {263}
  (\bibinfo {year} {2004})}\BibitemShut {NoStop}%
\bibitem [{\citenamefont {Kim}\ \emph {et~al.}(2017)\citenamefont {Kim},
  \citenamefont {Park}, \citenamefont {Ha}, \citenamefont {Kim}, \citenamefont
  {Kim}, \citenamefont {Shin}, \citenamefont {Kim}, \citenamefont {Yu},
  \citenamefont {Kim},\ and\ \citenamefont {Char}}]{BaHfO3_band_gap}%
  \BibitemOpen
  \bibfield  {author} {\bibinfo {author} {\bibfnamefont {Y.~M.}\ \bibnamefont
  {Kim}}, \bibinfo {author} {\bibfnamefont {C.}~\bibnamefont {Park}}, \bibinfo
  {author} {\bibfnamefont {T.}~\bibnamefont {Ha}}, \bibinfo {author}
  {\bibfnamefont {U.}~\bibnamefont {Kim}}, \bibinfo {author} {\bibfnamefont
  {N.}~\bibnamefont {Kim}}, \bibinfo {author} {\bibfnamefont {J.}~\bibnamefont
  {Shin}}, \bibinfo {author} {\bibfnamefont {Y.}~\bibnamefont {Kim}}, \bibinfo
  {author} {\bibfnamefont {J.}~\bibnamefont {Yu}}, \bibinfo {author}
  {\bibfnamefont {J.~H.}\ \bibnamefont {Kim}},\ and\ \bibinfo {author}
  {\bibfnamefont {K.}~\bibnamefont {Char}},\ }\bibfield  {title} {\bibinfo
  {title} {{High-k perovskite gate oxide BaHfO\textsubscript{3}}},\ }\href
  {https://doi.org/10.1063/1.4974864} {\bibfield  {journal} {\bibinfo
  {journal} {APL Mater}\ }\textbf {\bibinfo {volume} {5}} (\bibinfo {year}
  {2017})},\ \bibinfo {note} {016104}\BibitemShut {NoStop}%
\bibitem [{\citenamefont {Hellwege}\ and\ \citenamefont
  {Hellwege}(1969)}]{BaTiO3_Lat_Param_1}%
  \BibitemOpen
  \bibfield  {author} {\bibinfo {author} {\bibfnamefont {K.~H.}\ \bibnamefont
  {Hellwege}}\ and\ \bibinfo {author} {\bibfnamefont {A.~M.}\ \bibnamefont
  {Hellwege}},\ }\bibfield  {title} {\bibinfo {title} {Ferroelectrics and
  related substances},\ }in\ \href@noop {} {\emph {\bibinfo {booktitle}
  {Landolt-Börnstein}}},\ Vol.~\bibinfo {volume} {3}\ (\bibinfo  {publisher}
  {Springer-Verlag},\ \bibinfo {year} {1969})\BibitemShut {NoStop}%
\bibitem [{\citenamefont {Wang}\ \emph {et~al.}(2010)\citenamefont {Wang},
  \citenamefont {Meng}, \citenamefont {Ma}, \citenamefont {Xu},\ and\
  \citenamefont {Chen}}]{BaTiO3_Lat_Param_2}%
  \BibitemOpen
  \bibfield  {author} {\bibinfo {author} {\bibfnamefont {J.~J.}\ \bibnamefont
  {Wang}}, \bibinfo {author} {\bibfnamefont {F.~Y.}\ \bibnamefont {Meng}},
  \bibinfo {author} {\bibfnamefont {X.~Q.}\ \bibnamefont {Ma}}, \bibinfo
  {author} {\bibfnamefont {M.~X.}\ \bibnamefont {Xu}},\ and\ \bibinfo {author}
  {\bibfnamefont {L.~Q.}\ \bibnamefont {Chen}},\ }\bibfield  {title} {\bibinfo
  {title} {{Lattice, elastic, polarization, and electrostrictive properties of
  BaTiO\textsubscript{3} from first-principles}},\ }\href
  {https://doi.org/10.1063/1.3462441} {\bibfield  {journal} {\bibinfo
  {journal} {J. Appl. Phys}\ }\textbf {\bibinfo {volume} {108}} (\bibinfo
  {year} {2010})},\ \bibinfo {note} {034107}\BibitemShut {NoStop}%
\bibitem [{\citenamefont {Wemple}(1970)}]{BaTiO3_band_gap}%
  \BibitemOpen
  \bibfield  {author} {\bibinfo {author} {\bibfnamefont {S.~H.}\ \bibnamefont
  {Wemple}},\ }\bibfield  {title} {\bibinfo {title} {{Polarization Fluctuations
  and the Optical-Absorption Edge in BaTi${\mathrm{O}}_{3}$}},\ }\href
  {https://doi.org/10.1103/PhysRevB.2.2679} {\bibfield  {journal} {\bibinfo
  {journal} {Phys. Rev. B}\ }\textbf {\bibinfo {volume} {2}},\ \bibinfo {pages}
  {2679} (\bibinfo {year} {1970})}\BibitemShut {NoStop}%
\bibitem [{\citenamefont {Levin}\ \emph {et~al.}(2003)\citenamefont {Levin},
  \citenamefont {Amos}, \citenamefont {Bell}, \citenamefont {Farber},
  \citenamefont {Vanderah}, \citenamefont {Roth},\ and\ \citenamefont
  {Toby}}]{BaZrO3_Lat_Param_1}%
  \BibitemOpen
  \bibfield  {author} {\bibinfo {author} {\bibfnamefont {I.}~\bibnamefont
  {Levin}}, \bibinfo {author} {\bibfnamefont {T.~G.}\ \bibnamefont {Amos}},
  \bibinfo {author} {\bibfnamefont {S.~M.}\ \bibnamefont {Bell}}, \bibinfo
  {author} {\bibfnamefont {L.}~\bibnamefont {Farber}}, \bibinfo {author}
  {\bibfnamefont {T.~A.}\ \bibnamefont {Vanderah}}, \bibinfo {author}
  {\bibfnamefont {R.~S.}\ \bibnamefont {Roth}},\ and\ \bibinfo {author}
  {\bibfnamefont {B.~H.}\ \bibnamefont {Toby}},\ }\bibfield  {title} {\bibinfo
  {title} {Phase equilibria, crystal structures, and dielectric anomaly in the
  {BaZrO\textsubscript{3}–CaZrO\textsubscript{3}} system},\ }\href
  {https://doi.org/https://doi.org/10.1016/S0022-4596(03)00220-2} {\bibfield
  {journal} {\bibinfo  {journal} {J. Chem. Phys.}\ }\textbf {\bibinfo {volume}
  {175}},\ \bibinfo {pages} {170} (\bibinfo {year} {2003})}\BibitemShut
  {NoStop}%
\bibitem [{\citenamefont {Evarestov}(2011)}]{BaZrO3_Lat_Param_2}%
  \BibitemOpen
  \bibfield  {author} {\bibinfo {author} {\bibfnamefont {R.~A.}\ \bibnamefont
  {Evarestov}},\ }\bibfield  {title} {\bibinfo {title} {{Hybrid density
  functional theory LCAO calculations on phonons in Ba(Ti,Zr,Hf)${}_{3}$}},\
  }\href {https://doi.org/10.1103/PhysRevB.83.014105} {\bibfield  {journal}
  {\bibinfo  {journal} {Phys. Rev. B}\ }\textbf {\bibinfo {volume} {83}},\
  \bibinfo {pages} {014105} (\bibinfo {year} {2011})}\BibitemShut {NoStop}%
\bibitem [{\citenamefont {Robertson}(2000)}]{BaZrO3_band_gap}%
  \BibitemOpen
  \bibfield  {author} {\bibinfo {author} {\bibfnamefont {J.}~\bibnamefont
  {Robertson}},\ }\bibfield  {title} {\bibinfo {title} {{Band offsets of
  wide-band-gap oxides and implications for future electronic devices}},\
  }\href {https://doi.org/10.1116/1.591472} {\bibfield  {journal} {\bibinfo
  {journal} {J. Vac. Sci. Technol. B}\ }\textbf {\bibinfo {volume} {18}},\
  \bibinfo {pages} {1785} (\bibinfo {year} {2000})}\BibitemShut {NoStop}%
\bibitem [{\citenamefont {Heaton}\ and\ \citenamefont
  {Lin}(1982)}]{KMgF3_Lat_Param_1_and_gap}%
  \BibitemOpen
  \bibfield  {author} {\bibinfo {author} {\bibfnamefont {R.~A.}\ \bibnamefont
  {Heaton}}\ and\ \bibinfo {author} {\bibfnamefont {C.~C.}\ \bibnamefont
  {Lin}},\ }\bibfield  {title} {\bibinfo {title} {Electronic energy-band
  structure of the {KMg${\mathrm{F}}_{3}$} crystal},\ }\href
  {https://doi.org/10.1103/PhysRevB.25.3538} {\bibfield  {journal} {\bibinfo
  {journal} {Phys. Rev. B}\ }\textbf {\bibinfo {volume} {25}},\ \bibinfo
  {pages} {3538} (\bibinfo {year} {1982})}\BibitemShut {NoStop}%
\bibitem [{\citenamefont {Lua\~na}\ \emph {et~al.}(1997)\citenamefont
  {Lua\~na}, \citenamefont {Costales},\ and\ \citenamefont
  {Mart\'{\i}n~Pend\'as}}]{KMgF3_Lat_Param_2}%
  \BibitemOpen
  \bibfield  {author} {\bibinfo {author} {\bibfnamefont {V.}~\bibnamefont
  {Lua\~na}}, \bibinfo {author} {\bibfnamefont {A.}~\bibnamefont {Costales}},\
  and\ \bibinfo {author} {\bibfnamefont {A.}~\bibnamefont
  {Mart\'{\i}n~Pend\'as}},\ }\bibfield  {title} {\bibinfo {title} {{Ions in
  crystals: The topology of the electron density in ionic materials.II. The
  cubic alkali halide perovskites}},\ }\href
  {https://doi.org/10.1103/PhysRevB.55.4285} {\bibfield  {journal} {\bibinfo
  {journal} {Phys. Rev. B}\ }\textbf {\bibinfo {volume} {55}},\ \bibinfo
  {pages} {4285} (\bibinfo {year} {1997})}\BibitemShut {NoStop}%
\bibitem [{\citenamefont {Abramov}\ \emph {et~al.}(1995)\citenamefont
  {Abramov}, \citenamefont {Tsirelson}, \citenamefont {Zavodnik}, \citenamefont
  {Ivanov},\ and\ \citenamefont {D.}}]{SrTiO3_Lat_Param}%
  \BibitemOpen
  \bibfield  {author} {\bibinfo {author} {\bibfnamefont {Y.~A.}\ \bibnamefont
  {Abramov}}, \bibinfo {author} {\bibfnamefont {V.~G.}\ \bibnamefont
  {Tsirelson}}, \bibinfo {author} {\bibfnamefont {V.~E.}\ \bibnamefont
  {Zavodnik}}, \bibinfo {author} {\bibfnamefont {S.~A.}\ \bibnamefont
  {Ivanov}},\ and\ \bibinfo {author} {\bibfnamefont {B.~I.}\ \bibnamefont
  {D.}},\ }\bibfield  {title} {\bibinfo {title} {{The chemical bond and atomic
  displacements in {SrTiO${\sb 3}$} from X-ray diffraction analysis}},\ }\href
  {https://doi.org/10.1107/S0108768195003752} {\bibfield  {journal} {\bibinfo
  {journal} {Acta Crystallographica Section B}\ }\textbf {\bibinfo {volume}
  {51}},\ \bibinfo {pages} {942} (\bibinfo {year} {1995})}\BibitemShut
  {NoStop}%
\bibitem [{\citenamefont {van Benthem}\ \emph {et~al.}(2001)\citenamefont {van
  Benthem}, \citenamefont {Elsässer},\ and\ \citenamefont
  {French}}]{SrTiO3_band_gap}%
  \BibitemOpen
  \bibfield  {author} {\bibinfo {author} {\bibfnamefont {K.}~\bibnamefont {van
  Benthem}}, \bibinfo {author} {\bibfnamefont {C.}~\bibnamefont {Elsässer}},\
  and\ \bibinfo {author} {\bibfnamefont {R.~H.}\ \bibnamefont {French}},\
  }\bibfield  {title} {\bibinfo {title} {{Bulk electronic structure of
  {SrTiO\textsubscript{3}}: {Experiment} and theory}},\ }\href
  {https://doi.org/10.1063/1.1415766} {\bibfield  {journal} {\bibinfo
  {journal} {J. Appl. Phys}\ }\textbf {\bibinfo {volume} {90}},\ \bibinfo
  {pages} {6156} (\bibinfo {year} {2001})}\BibitemShut {NoStop}%
\bibitem [{\citenamefont {Bhandari}\ \emph {et~al.}(2018)\citenamefont
  {Bhandari}, \citenamefont {van Schilfgaarde}, \citenamefont {Kotani},\ and\
  \citenamefont {Lambrecht}}]{ZPR_SrTiO3_Bhandari_Schlifgaarde}%
  \BibitemOpen
  \bibfield  {author} {\bibinfo {author} {\bibfnamefont {C.}~\bibnamefont
  {Bhandari}}, \bibinfo {author} {\bibfnamefont {M.}~\bibnamefont {van
  Schilfgaarde}}, \bibinfo {author} {\bibfnamefont {T.}~\bibnamefont
  {Kotani}},\ and\ \bibinfo {author} {\bibfnamefont {W.~R.~L.}\ \bibnamefont
  {Lambrecht}},\ }\bibfield  {title} {\bibinfo {title} {{All-electron
  quasiparticle self-consistent $\mathit{GW}$ band structures for
  ${\mathrm{SrTiO}}_{3}$ including lattice polarization corrections in
  different phases}},\ }\href
  {https://doi.org/10.1103/PhysRevMaterials.2.013807} {\bibfield  {journal}
  {\bibinfo  {journal} {Phys. Rev. Mater.}\ }\textbf {\bibinfo {volume} {2}},\
  \bibinfo {pages} {013807} (\bibinfo {year} {2018})}\BibitemShut {NoStop}%
\bibitem [{\citenamefont {Engel}\ \emph {et~al.}(2022)\citenamefont {Engel},
  \citenamefont {Miranda}, \citenamefont {Chaput}, \citenamefont {Togo},
  \citenamefont {Verdi}, \citenamefont {Marsman},\ and\ \citenamefont
  {Kresse}}]{Kresse_ZPR_values}%
  \BibitemOpen
  \bibfield  {author} {\bibinfo {author} {\bibfnamefont {M.}~\bibnamefont
  {Engel}}, \bibinfo {author} {\bibfnamefont {H.}~\bibnamefont {Miranda}},
  \bibinfo {author} {\bibfnamefont {L.}~\bibnamefont {Chaput}}, \bibinfo
  {author} {\bibfnamefont {A.}~\bibnamefont {Togo}}, \bibinfo {author}
  {\bibfnamefont {C.}~\bibnamefont {Verdi}}, \bibinfo {author} {\bibfnamefont
  {M.}~\bibnamefont {Marsman}},\ and\ \bibinfo {author} {\bibfnamefont
  {G.}~\bibnamefont {Kresse}},\ }\bibfield  {title} {\bibinfo {title}
  {Zero-point renormalization of the band gap of semiconductors and insulators
  using the projector augmented wave method},\ }\href
  {https://doi.org/10.1103/PhysRevB.106.094316} {\bibfield  {journal} {\bibinfo
   {journal} {Phys. Rev. B}\ }\textbf {\bibinfo {volume} {106}},\ \bibinfo
  {pages} {094316} (\bibinfo {year} {2022})}\BibitemShut {NoStop}%
\bibitem [{\citenamefont {Ashcroft}\ and\ \citenamefont
  {Mermin}(1976)}]{Ashcroft_Mermin}%
  \BibitemOpen
  \bibfield  {author} {\bibinfo {author} {\bibfnamefont {N.~W.}\ \bibnamefont
  {Ashcroft}}\ and\ \bibinfo {author} {\bibfnamefont {N.~D.}\ \bibnamefont
  {Mermin}},\ }\href@noop {} {\emph {\bibinfo {title} {Solid State Physics}}}\
  (\bibinfo  {publisher} {Saunders College Publishing},\ \bibinfo {year}
  {1976})\BibitemShut {NoStop}%
\bibitem [{\citenamefont {Blair}\ \emph {et~al.}(2015)\citenamefont {Blair},
  \citenamefont {Kroukis},\ and\ \citenamefont
  {Gidopoulos}}]{Nikitas_hyper_HF}%
  \BibitemOpen
  \bibfield  {author} {\bibinfo {author} {\bibfnamefont {A.~I.}\ \bibnamefont
  {Blair}}, \bibinfo {author} {\bibfnamefont {A.}~\bibnamefont {Kroukis}},\
  and\ \bibinfo {author} {\bibfnamefont {N.~I.}\ \bibnamefont {Gidopoulos}},\
  }\bibfield  {title} {\bibinfo {title} {{A correction for the Hartree-Fock
  density of states for jellium without screening}},\ }\href
  {https://doi.org/10.1063/1.4909519} {\bibfield  {journal} {\bibinfo
  {journal} {J. Chem. Phys.}\ }\textbf {\bibinfo {volume} {142}} (\bibinfo
  {year} {2015})}\BibitemShut {NoStop}%
\bibitem [{\citenamefont {Bachelet}\ and\ \citenamefont
  {Christensen}(1985)}]{Bachelet_1985_Ge_metallic}%
  \BibitemOpen
  \bibfield  {author} {\bibinfo {author} {\bibfnamefont {G.~B.}\ \bibnamefont
  {Bachelet}}\ and\ \bibinfo {author} {\bibfnamefont {N.~E.}\ \bibnamefont
  {Christensen}},\ }\bibfield  {title} {\bibinfo {title} {Relativistic and
  core-relaxation effects on the energy bands of gallium arsenide and
  germanium},\ }\href {https://doi.org/10.1103/PhysRevB.31.879} {\bibfield
  {journal} {\bibinfo  {journal} {Phys. Rev. B}\ }\textbf {\bibinfo {volume}
  {31}},\ \bibinfo {pages} {879} (\bibinfo {year} {1985})}\BibitemShut
  {NoStop}%
\bibitem [{\citenamefont {Ekuma}\ \emph {et~al.}(2013)\citenamefont {Ekuma},
  \citenamefont {Jarrell}, \citenamefont {Moreno},\ and\ \citenamefont
  {Bagayoko}}]{Ekuma_2013_Ge_metallic}%
  \BibitemOpen
  \bibfield  {author} {\bibinfo {author} {\bibfnamefont {C.}~\bibnamefont
  {Ekuma}}, \bibinfo {author} {\bibfnamefont {M.}~\bibnamefont {Jarrell}},
  \bibinfo {author} {\bibfnamefont {J.}~\bibnamefont {Moreno}},\ and\ \bibinfo
  {author} {\bibfnamefont {D.}~\bibnamefont {Bagayoko}},\ }\bibfield  {title}
  {\bibinfo {title} {Re-examining the electronic structure of germanium: {A}
  first-principle study},\ }\href
  {https://doi.org/https://doi.org/10.1016/j.physleta.2013.05.043} {\bibfield
  {journal} {\bibinfo  {journal} {Physics Letters A}\ }\textbf {\bibinfo
  {volume} {377}},\ \bibinfo {pages} {2172} (\bibinfo {year}
  {2013})}\BibitemShut {NoStop}%
\bibitem [{\citenamefont {Hedin}(1965)}]{GW_original_paper_Hedin_1965}%
  \BibitemOpen
  \bibfield  {author} {\bibinfo {author} {\bibfnamefont {L.}~\bibnamefont
  {Hedin}},\ }\bibfield  {title} {\bibinfo {title} {{New Method for Calculating
  the One-Particle Green's Function with Application to the Electron-Gas
  Problem}},\ }\href {https://doi.org/10.1103/PhysRev.139.A796} {\bibfield
  {journal} {\bibinfo  {journal} {Phys. Rev.}\ }\textbf {\bibinfo {volume}
  {139}},\ \bibinfo {pages} {A796} (\bibinfo {year} {1965})}\BibitemShut
  {NoStop}%
\bibitem [{\citenamefont {Rohlfing}\ \emph {et~al.}(1993)\citenamefont
  {Rohlfing}, \citenamefont {Kr\"uger},\ and\ \citenamefont
  {Pollmann}}]{GW_Ge_band_gap_correct_1_Rohlfing_1993}%
  \BibitemOpen
  \bibfield  {author} {\bibinfo {author} {\bibfnamefont {M.}~\bibnamefont
  {Rohlfing}}, \bibinfo {author} {\bibfnamefont {P.}~\bibnamefont {Kr\"uger}},\
  and\ \bibinfo {author} {\bibfnamefont {J.}~\bibnamefont {Pollmann}},\
  }\bibfield  {title} {\bibinfo {title} {Quasiparticle band-structure
  calculations for {C, Si, Ge, GaAs, and SiC using Gaussian-orbital} basis
  sets},\ }\href {https://doi.org/10.1103/PhysRevB.48.17791} {\bibfield
  {journal} {\bibinfo  {journal} {Phys. Rev. B}\ }\textbf {\bibinfo {volume}
  {48}},\ \bibinfo {pages} {17791} (\bibinfo {year} {1993})}\BibitemShut
  {NoStop}%
\bibitem [{\citenamefont
  {Fleszar}(2001)}]{GW_Ge_band_gap_correct_2_Fleszar_2001}%
  \BibitemOpen
  \bibfield  {author} {\bibinfo {author} {\bibfnamefont {A.}~\bibnamefont
  {Fleszar}},\ }\bibfield  {title} {\bibinfo {title} {{LDA, $GW$, and
  exact-exchange Kohn-Sham scheme calculations of the electronic structure of
  $\mathrm{sp}$ semiconductors}},\ }\href
  {https://doi.org/10.1103/PhysRevB.64.245204} {\bibfield  {journal} {\bibinfo
  {journal} {Phys. Rev. B}\ }\textbf {\bibinfo {volume} {64}},\ \bibinfo
  {pages} {245204} (\bibinfo {year} {2001})}\BibitemShut {NoStop}%
\bibitem [{\citenamefont {Betzinger}\ \emph {et~al.}(2012)\citenamefont
  {Betzinger}, \citenamefont {Friedrich}, \citenamefont {G\"orling},\ and\
  \citenamefont {Bl\"ugel}}]{Betzinger_Görling_2012_EXX_perovskites}%
  \BibitemOpen
  \bibfield  {author} {\bibinfo {author} {\bibfnamefont {M.}~\bibnamefont
  {Betzinger}}, \bibinfo {author} {\bibfnamefont {C.}~\bibnamefont
  {Friedrich}}, \bibinfo {author} {\bibfnamefont {A.}~\bibnamefont
  {G\"orling}},\ and\ \bibinfo {author} {\bibfnamefont {S.}~\bibnamefont
  {Bl\"ugel}},\ }\bibfield  {title} {\bibinfo {title} {{Precise response
  functions in all-electron methods: Application to the
  optimized-effective-potential approach}},\ }\href
  {https://doi.org/10.1103/PhysRevB.85.245124} {\bibfield  {journal} {\bibinfo
  {journal} {Phys. Rev. B}\ }\textbf {\bibinfo {volume} {85}},\ \bibinfo
  {pages} {245124} (\bibinfo {year} {2012})}\BibitemShut {NoStop}%
\bibitem [{\citenamefont {Karsai}\ \emph {et~al.}(2018)\citenamefont {Karsai},
  \citenamefont {Engel}, \citenamefont {Flage-Larsen},\ and\ \citenamefont
  {Kresse}}]{ZPR_Kresse_Semiconductors}%
  \BibitemOpen
  \bibfield  {author} {\bibinfo {author} {\bibfnamefont {F.}~\bibnamefont
  {Karsai}}, \bibinfo {author} {\bibfnamefont {M.}~\bibnamefont {Engel}},
  \bibinfo {author} {\bibfnamefont {E.}~\bibnamefont {Flage-Larsen}},\ and\
  \bibinfo {author} {\bibfnamefont {G.}~\bibnamefont {Kresse}},\ }\bibfield
  {title} {\bibinfo {title} {{Electron–phonon coupling in semiconductors
  within the $GW$ approximation}},\ }\href
  {https://doi.org/10.1088/1367-2630/aaf53f} {\bibfield  {journal} {\bibinfo
  {journal} {New J. Phys.}\ }\textbf {\bibinfo {volume} {20}},\ \bibinfo
  {pages} {123008} (\bibinfo {year} {2018})}\BibitemShut {NoStop}%
\bibitem [{\citenamefont {Becke}(1988)}]{B88}%
  \BibitemOpen
  \bibfield  {author} {\bibinfo {author} {\bibfnamefont {A.~D.}\ \bibnamefont
  {Becke}},\ }\bibfield  {title} {\bibinfo {title} {Density-functional
  exchange-energy approximation with correct asymptotic behavior},\ }\href
  {https://doi.org/10.1103/PhysRevA.38.3098} {\bibfield  {journal} {\bibinfo
  {journal} {Phys. Rev. A}\ }\textbf {\bibinfo {volume} {38}},\ \bibinfo
  {pages} {3098} (\bibinfo {year} {1988})}\BibitemShut {NoStop}%
\bibitem [{\citenamefont {Levy}\ and\ \citenamefont
  {Perdew}(1985)}]{exchange_virial_Levy_Perdew_1985}%
  \BibitemOpen
  \bibfield  {author} {\bibinfo {author} {\bibfnamefont {M.}~\bibnamefont
  {Levy}}\ and\ \bibinfo {author} {\bibfnamefont {J.~P.}\ \bibnamefont
  {Perdew}},\ }\bibfield  {title} {\bibinfo {title} {{Hellmann-Feynman, virial,
  and scaling requisites for the exact universal density functionals. Shape of
  the correlation potential and diamagnetic susceptibility for atoms}},\ }\href
  {https://doi.org/10.1103/PhysRevA.32.2010} {\bibfield  {journal} {\bibinfo
  {journal} {Phys. Rev. A}\ }\textbf {\bibinfo {volume} {32}},\ \bibinfo
  {pages} {2010} (\bibinfo {year} {1985})}\BibitemShut {NoStop}%
\bibitem [{\citenamefont {Sasaki}\ \emph {et~al.}(1979)\citenamefont {Sasaki},
  \citenamefont {Fujino},\ and\ \citenamefont {Tákeuchi}}]{TMO_lat_params}%
  \BibitemOpen
  \bibfield  {author} {\bibinfo {author} {\bibfnamefont {S.}~\bibnamefont
  {Sasaki}}, \bibinfo {author} {\bibfnamefont {K.}~\bibnamefont {Fujino}},\
  and\ \bibinfo {author} {\bibfnamefont {Y.}~\bibnamefont {Tákeuchi}},\
  }\bibfield  {title} {\bibinfo {title} {{X-Ray Determination of
  Electron-Density Distributions in Oxides, MgO, MnO, CoO, and NiO, and Atomic
  Scattering Factors of their Constituent Atoms}},\ }\href
  {https://doi.org/10.2183/pjab.55.43} {\bibfield  {journal} {\bibinfo
  {journal} {Proc. Jpn. Acad., Ser. B}\ }\textbf {\bibinfo {volume} {55}},\
  \bibinfo {pages} {43} (\bibinfo {year} {1979})}\BibitemShut {NoStop}%
\bibitem [{\citenamefont {van Elp}\ \emph {et~al.}(1991)\citenamefont {van Elp}
  \emph {et~al.}}]{CoO_band_gap_experimental}%
  \BibitemOpen
  \bibfield  {author} {\bibinfo {author} {\bibfnamefont {J.}~\bibnamefont {van
  Elp}} \emph {et~al.},\ }\bibfield  {title} {\bibinfo {title} {{Electronic
  structure of CoO, Li-doped CoO, and ${\mathrm{LiCoO}}_{2}$}},\ }\href
  {https://doi.org/10.1103/PhysRevB.44.6090} {\bibfield  {journal} {\bibinfo
  {journal} {Phys. Rev. B}\ }\textbf {\bibinfo {volume} {44}},\ \bibinfo
  {pages} {6090} (\bibinfo {year} {1991})}\BibitemShut {NoStop}%
\bibitem [{\citenamefont {Parmigiani}\ and\ \citenamefont
  {Sangaletti}(1999)}]{FeO_band_gap_experimental}%
  \BibitemOpen
  \bibfield  {author} {\bibinfo {author} {\bibfnamefont {F.}~\bibnamefont
  {Parmigiani}}\ and\ \bibinfo {author} {\bibfnamefont {L.}~\bibnamefont
  {Sangaletti}},\ }\bibfield  {title} {\bibinfo {title} {{Fine structures in
  the X-ray photoemission spectra of MnO, FeO, CoO, and NiO single crystals}},\
  }\href {https://doi.org/10.1016/S0368-2048(98)00294-1} {\bibfield  {journal}
  {\bibinfo  {journal} {J Electron Spectros Relat Phenomena.}\ }\textbf
  {\bibinfo {volume} {98-99}},\ \bibinfo {pages} {287–302} (\bibinfo {year}
  {1999})}\BibitemShut {NoStop}%
\bibitem [{\citenamefont {Tran}\ \emph {et~al.}(2006)\citenamefont {Tran},
  \citenamefont {Blaha}, \citenamefont {Schwarz},\ and\ \citenamefont
  {Nov\'ak}}]{TMOs_2006_Hybrid_Calculations}%
  \BibitemOpen
  \bibfield  {author} {\bibinfo {author} {\bibfnamefont {F.}~\bibnamefont
  {Tran}}, \bibinfo {author} {\bibfnamefont {P.}~\bibnamefont {Blaha}},
  \bibinfo {author} {\bibfnamefont {K.}~\bibnamefont {Schwarz}},\ and\ \bibinfo
  {author} {\bibfnamefont {P.}~\bibnamefont {Nov\'ak}},\ }\bibfield  {title}
  {\bibinfo {title} {Hybrid exchange-correlation energy functionals for
  strongly correlated electrons: {Applications} to transition-metal
  monoxides},\ }\href {https://doi.org/10.1103/PhysRevB.74.155108} {\bibfield
  {journal} {\bibinfo  {journal} {Phys. Rev. B}\ }\textbf {\bibinfo {volume}
  {74}},\ \bibinfo {pages} {155108} (\bibinfo {year} {2006})}\BibitemShut
  {NoStop}%
\bibitem [{\citenamefont {Zimmermann}\ \emph {et~al.}(1999)\citenamefont
  {Zimmermann} \emph {et~al.}}]{MnO_band_gap_experimental}%
  \BibitemOpen
  \bibfield  {author} {\bibinfo {author} {\bibfnamefont {R.}~\bibnamefont
  {Zimmermann}} \emph {et~al.},\ }\bibfield  {title} {\bibinfo {title}
  {Electronic structure of 3$d$-transition-metal oxides: on-site {Coulomb}
  repulsion versus covalency},\ }\href
  {https://doi.org/10.1088/0953-8984/11/7/002} {\bibfield  {journal} {\bibinfo
  {journal} {J. Phys. Condens. Matter}\ }\textbf {\bibinfo {volume} {11}},\
  \bibinfo {pages} {1657} (\bibinfo {year} {1999})}\BibitemShut {NoStop}%
\bibitem [{\citenamefont {Sawatzky}\ and\ \citenamefont
  {Allen}(1984)}]{NiO_band_gap_experimental}%
  \BibitemOpen
  \bibfield  {author} {\bibinfo {author} {\bibfnamefont {G.~A.}\ \bibnamefont
  {Sawatzky}}\ and\ \bibinfo {author} {\bibfnamefont {J.~W.}\ \bibnamefont
  {Allen}},\ }\bibfield  {title} {\bibinfo {title} {{Magnitude and Origin of
  the Band Gap in NiO}},\ }\href {https://doi.org/10.1103/PhysRevLett.53.2339}
  {\bibfield  {journal} {\bibinfo  {journal} {Phys. Rev. Lett.}\ }\textbf
  {\bibinfo {volume} {53}},\ \bibinfo {pages} {2339} (\bibinfo {year}
  {1984})}\BibitemShut {NoStop}%
\bibitem [{\citenamefont {Himmetoglu}\ \emph {et~al.}(2014)\citenamefont
  {Himmetoglu}, \citenamefont {Floris}, \citenamefont {de~Gironcoli},\ and\
  \citenamefont {Cococcioni}}]{DFT_U_Str_Correlation_Review_2014}%
  \BibitemOpen
  \bibfield  {author} {\bibinfo {author} {\bibfnamefont {B.}~\bibnamefont
  {Himmetoglu}}, \bibinfo {author} {\bibfnamefont {A.}~\bibnamefont {Floris}},
  \bibinfo {author} {\bibfnamefont {S.}~\bibnamefont {de~Gironcoli}},\ and\
  \bibinfo {author} {\bibfnamefont {M.}~\bibnamefont {Cococcioni}},\ }\bibfield
   {title} {\bibinfo {title} {Hubbard-corrected {DFT} energy functionals: The
  {LDA+$U$} description of correlated systems},\ }\href
  {https://doi.org/https://doi.org/10.1002/qua.24521} {\bibfield  {journal}
  {\bibinfo  {journal} {Int. J. Quantum Chem.}\ }\textbf {\bibinfo {volume}
  {114}},\ \bibinfo {pages} {14} (\bibinfo {year} {2014})}\BibitemShut
  {NoStop}%
\bibitem [{\citenamefont {Gillen}\ and\ \citenamefont
  {Robertson}(2013)}]{Gillen_Robertson_TMOs_Screened_Exchange_2013}%
  \BibitemOpen
  \bibfield  {author} {\bibinfo {author} {\bibfnamefont {R.}~\bibnamefont
  {Gillen}}\ and\ \bibinfo {author} {\bibfnamefont {J.}~\bibnamefont
  {Robertson}},\ }\bibfield  {title} {\bibinfo {title} {Accurate screened
  exchange band structures for the transition metal monoxides {MnO, FeO, CoO
  and NiO}},\ }\href {https://doi.org/10.1088/0953-8984/25/16/165502}
  {\bibfield  {journal} {\bibinfo  {journal} {J. Phys.: Condens. Matter}\
  }\textbf {\bibinfo {volume} {25}},\ \bibinfo {pages} {165502} (\bibinfo
  {year} {2013})}\BibitemShut {NoStop}%
\bibitem [{\citenamefont {Engel}\ and\ \citenamefont
  {Schmid}(2009)}]{Engel_2009_TMOs_EXX}%
  \BibitemOpen
  \bibfield  {author} {\bibinfo {author} {\bibfnamefont {E.}~\bibnamefont
  {Engel}}\ and\ \bibinfo {author} {\bibfnamefont {R.~N.}\ \bibnamefont
  {Schmid}},\ }\bibfield  {title} {\bibinfo {title} {Insulating ground states
  of transition-metal monoxides from exact exchange},\ }\href
  {https://doi.org/10.1103/PhysRevLett.103.036404} {\bibfield  {journal}
  {\bibinfo  {journal} {Phys. Rev. Lett.}\ }\textbf {\bibinfo {volume} {103}},\
  \bibinfo {pages} {036404} (\bibinfo {year} {2009})}\BibitemShut {NoStop}%
\bibitem [{\citenamefont {Sakuma}\ and\ \citenamefont
  {Aryasetiawan}(2013)}]{TMOs_2013_Dynamical_Screen_Coulomb_Int_cRPA}%
  \BibitemOpen
  \bibfield  {author} {\bibinfo {author} {\bibfnamefont {R.}~\bibnamefont
  {Sakuma}}\ and\ \bibinfo {author} {\bibfnamefont {F.}~\bibnamefont
  {Aryasetiawan}},\ }\bibfield  {title} {\bibinfo {title} {{First-principles
  calculations of dynamical screened interactions for the transition metal
  oxides $M$O ($M$=Mn, Fe, Co, Ni)}},\ }\href
  {https://doi.org/10.1103/PhysRevB.87.165118} {\bibfield  {journal} {\bibinfo
  {journal} {Phys. Rev. B}\ }\textbf {\bibinfo {volume} {87}},\ \bibinfo
  {pages} {165118} (\bibinfo {year} {2013})}\BibitemShut {NoStop}%
\bibitem [{\citenamefont {Kotliar}\ \emph {et~al.}(2006)\citenamefont
  {Kotliar}, \citenamefont {Savrasov}, \citenamefont {Haule}, \citenamefont
  {Oudovenko}, \citenamefont {Parcollet},\ and\ \citenamefont
  {Marianetti}}]{DMFT_Review_Kotliar_2006}%
  \BibitemOpen
  \bibfield  {author} {\bibinfo {author} {\bibfnamefont {G.}~\bibnamefont
  {Kotliar}}, \bibinfo {author} {\bibfnamefont {S.~Y.}\ \bibnamefont
  {Savrasov}}, \bibinfo {author} {\bibfnamefont {K.}~\bibnamefont {Haule}},
  \bibinfo {author} {\bibfnamefont {V.~S.}\ \bibnamefont {Oudovenko}}, \bibinfo
  {author} {\bibfnamefont {O.}~\bibnamefont {Parcollet}},\ and\ \bibinfo
  {author} {\bibfnamefont {C.~A.}\ \bibnamefont {Marianetti}},\ }\bibfield
  {title} {\bibinfo {title} {Electronic structure calculations with dynamical
  mean-field theory},\ }\href {https://doi.org/10.1103/RevModPhys.78.865}
  {\bibfield  {journal} {\bibinfo  {journal} {Rev. Mod. Phys.}\ }\textbf
  {\bibinfo {volume} {78}},\ \bibinfo {pages} {865} (\bibinfo {year}
  {2006})}\BibitemShut {NoStop}%
\bibitem [{\citenamefont {Kune{\v{s}}}\ \emph {et~al.}(2008)\citenamefont
  {Kune{\v{s}}}, \citenamefont {Lukoyanov}, \citenamefont {Anisimov},
  \citenamefont {Scalettar},\ and\ \citenamefont
  {Pickett}}]{MnO_LDA_DMFT_calculation_Kuneš_2008}%
  \BibitemOpen
  \bibfield  {author} {\bibinfo {author} {\bibfnamefont {J.}~\bibnamefont
  {Kune{\v{s}}}}, \bibinfo {author} {\bibfnamefont {A.~V.}\ \bibnamefont
  {Lukoyanov}}, \bibinfo {author} {\bibfnamefont {V.~I.}\ \bibnamefont
  {Anisimov}}, \bibinfo {author} {\bibfnamefont {R.~T.}\ \bibnamefont
  {Scalettar}},\ and\ \bibinfo {author} {\bibfnamefont {W.~E.}\ \bibnamefont
  {Pickett}},\ }\bibfield  {title} {\bibinfo {title} {{Collapse of magnetic
  moment drives the Mott transition in MnO}},\ }\href
  {https://doi.org/10.1038/nmat2115} {\bibfield  {journal} {\bibinfo  {journal}
  {Nature Materials}\ }\textbf {\bibinfo {volume} {7}},\ \bibinfo {pages} {198}
  (\bibinfo {year} {2008})}\BibitemShut {NoStop}%
\bibitem [{\citenamefont {Mandal}\ \emph
  {et~al.}(2019{\natexlab{a}})\citenamefont {Mandal}, \citenamefont {Haule},
  \citenamefont {Rabe},\ and\ \citenamefont
  {Vanderbilt}}]{Mandal_Vanderbilt_TMOs_eDMFT_2019}%
  \BibitemOpen
  \bibfield  {author} {\bibinfo {author} {\bibfnamefont {S.}~\bibnamefont
  {Mandal}}, \bibinfo {author} {\bibfnamefont {K.}~\bibnamefont {Haule}},
  \bibinfo {author} {\bibfnamefont {K.~M.}\ \bibnamefont {Rabe}},\ and\
  \bibinfo {author} {\bibfnamefont {D.}~\bibnamefont {Vanderbilt}},\ }\bibfield
   {title} {\bibinfo {title} {Systematic beyond-{DFT} study of binary
  transition metal oxides},\ }\href {https://doi.org/10.1038/s41524-019-0251-7}
  {\bibfield  {journal} {\bibinfo  {journal} {Npj Comput. Mater.}\ }\textbf
  {\bibinfo {volume} {5}},\ \bibinfo {pages} {115} (\bibinfo {year}
  {2019}{\natexlab{a}})}\BibitemShut {NoStop}%
\bibitem [{\citenamefont {Mandal}\ \emph
  {et~al.}(2019{\natexlab{b}})\citenamefont {Mandal}, \citenamefont {Haule},
  \citenamefont {Rabe},\ and\ \citenamefont
  {Vanderbilt}}]{Mandal_Vanderbilt_eDMFT_paramagnetic_TMOs_2019}%
  \BibitemOpen
  \bibfield  {author} {\bibinfo {author} {\bibfnamefont {S.}~\bibnamefont
  {Mandal}}, \bibinfo {author} {\bibfnamefont {K.}~\bibnamefont {Haule}},
  \bibinfo {author} {\bibfnamefont {K.~M.}\ \bibnamefont {Rabe}},\ and\
  \bibinfo {author} {\bibfnamefont {D.}~\bibnamefont {Vanderbilt}},\ }\bibfield
   {title} {\bibinfo {title} {Influence of magnetic ordering on the spectral
  properties of binary transition metal oxides},\ }\href
  {https://doi.org/10.1103/PhysRevB.100.245109} {\bibfield  {journal} {\bibinfo
   {journal} {Phys. Rev. B}\ }\textbf {\bibinfo {volume} {100}},\ \bibinfo
  {pages} {245109} (\bibinfo {year} {2019}{\natexlab{b}})}\BibitemShut
  {NoStop}%
\bibitem [{\citenamefont {Bartók}\ and\ \citenamefont
  {Yates}(2019)}]{RSCAN_CASTEP}%
  \BibitemOpen
  \bibfield  {author} {\bibinfo {author} {\bibfnamefont {A.~P.}\ \bibnamefont
  {Bartók}}\ and\ \bibinfo {author} {\bibfnamefont {J.~R.}\ \bibnamefont
  {Yates}},\ }\bibfield  {title} {\bibinfo {title} {{Regularized SCAN
  functional}},\ }\bibfield  {journal} {\bibinfo  {journal} {J. Chem. Phys.}\
  }\textbf {\bibinfo {volume} {150}},\ \href
  {https://doi.org/10.1063/1.5094646} {10.1063/1.5094646} (\bibinfo {year}
  {2019}),\ \bibinfo {note} {161101}\BibitemShut {NoStop}%
\bibitem [{\citenamefont {Sai~Gautam}\ and\ \citenamefont
  {Carter}(2018)}]{TMOs_SCAN_Performance_1_Carter_2018}%
  \BibitemOpen
  \bibfield  {author} {\bibinfo {author} {\bibfnamefont {G.}~\bibnamefont
  {Sai~Gautam}}\ and\ \bibinfo {author} {\bibfnamefont {E.~A.}\ \bibnamefont
  {Carter}},\ }\bibfield  {title} {\bibinfo {title} {{Evaluating transition
  metal oxides within DFT-SCAN and $\text{SCAN}+U$ frameworks for solar
  thermochemical applications}},\ }\href
  {https://doi.org/10.1103/PhysRevMaterials.2.095401} {\bibfield  {journal}
  {\bibinfo  {journal} {Phys. Rev. Mater.}\ }\textbf {\bibinfo {volume} {2}},\
  \bibinfo {pages} {095401} (\bibinfo {year} {2018})}\BibitemShut {NoStop}%
\bibitem [{\citenamefont {Long}\ \emph {et~al.}(2020)\citenamefont {Long},
  \citenamefont {Sai~Gautam},\ and\ \citenamefont
  {Carter}}]{TMOs_SCAN_Performance_2_Carter_2020}%
  \BibitemOpen
  \bibfield  {author} {\bibinfo {author} {\bibfnamefont {O.~Y.}\ \bibnamefont
  {Long}}, \bibinfo {author} {\bibfnamefont {G.}~\bibnamefont {Sai~Gautam}},\
  and\ \bibinfo {author} {\bibfnamefont {E.~A.}\ \bibnamefont {Carter}},\
  }\bibfield  {title} {\bibinfo {title} {{Evaluating optimal $U$ for $3d$
  transition-metal oxides within the SCAN+$U$ framework}},\ }\href
  {https://doi.org/10.1103/PhysRevMaterials.4.045401} {\bibfield  {journal}
  {\bibinfo  {journal} {Phys. Rev. Mater.}\ }\textbf {\bibinfo {volume} {4}},\
  \bibinfo {pages} {045401} (\bibinfo {year} {2020})}\BibitemShut {NoStop}%
\bibitem [{\citenamefont {Sun}\ \emph {et~al.}(2015)\citenamefont {Sun},
  \citenamefont {Ruzsinszky},\ and\ \citenamefont {Perdew}}]{SCAN}%
  \BibitemOpen
  \bibfield  {author} {\bibinfo {author} {\bibfnamefont {J.}~\bibnamefont
  {Sun}}, \bibinfo {author} {\bibfnamefont {A.}~\bibnamefont {Ruzsinszky}},\
  and\ \bibinfo {author} {\bibfnamefont {J.~P.}\ \bibnamefont {Perdew}},\
  }\bibfield  {title} {\bibinfo {title} {{Strongly Constrained and
  Appropriately Normed Semilocal Density Functional}},\ }\href
  {https://doi.org/10.1103/PhysRevLett.115.036402} {\bibfield  {journal}
  {\bibinfo  {journal} {Phys. Rev. Lett.}\ }\textbf {\bibinfo {volume} {115}},\
  \bibinfo {pages} {036402} (\bibinfo {year} {2015})}\BibitemShut {NoStop}%
\bibitem [{\citenamefont {Hubbard}\ and\ \citenamefont
  {Flowers}(1963)}]{Hubbard_Model_OG}%
  \BibitemOpen
  \bibfield  {author} {\bibinfo {author} {\bibfnamefont {J.}~\bibnamefont
  {Hubbard}}\ and\ \bibinfo {author} {\bibfnamefont {B.~H.}\ \bibnamefont
  {Flowers}},\ }\bibfield  {title} {\bibinfo {title} {Electron correlations in
  narrow energy bands},\ }\href {https://doi.org/10.1098/rspa.1963.0204}
  {\bibfield  {journal} {\bibinfo  {journal} {Proc. R. Soc. Lond. A}\ }\textbf
  {\bibinfo {volume} {276}},\ \bibinfo {pages} {238} (\bibinfo {year}
  {1963})}\BibitemShut {NoStop}%
\bibitem [{\citenamefont
  {Tasaki}(1998{\natexlab{a}})}]{Tasaki_1998_Review_Hubbard_intro}%
  \BibitemOpen
  \bibfield  {author} {\bibinfo {author} {\bibfnamefont {H.}~\bibnamefont
  {Tasaki}},\ }\bibfield  {title} {\bibinfo {title} {{From Nagaoka's
  Ferromagnetism to Flat-Band Ferromagnetism and Beyond: An Introduction to
  Ferromagnetism in the Hubbard Model}},\ }\href
  {https://doi.org/10.1143/PTP.99.489} {\bibfield  {journal} {\bibinfo
  {journal} {Prog. Theor. Phys.}\ }\textbf {\bibinfo {volume} {99}},\ \bibinfo
  {pages} {489} (\bibinfo {year} {1998}{\natexlab{a}})}\BibitemShut {NoStop}%
\bibitem [{\citenamefont
  {Tasaki}(1998{\natexlab{b}})}]{Tasaki_1998_Review_Hubbard_rigorous}%
  \BibitemOpen
  \bibfield  {author} {\bibinfo {author} {\bibfnamefont {H.}~\bibnamefont
  {Tasaki}},\ }\bibfield  {title} {\bibinfo {title} {The {Hubbard} model - an
  introduction and selected rigorous results},\ }\href
  {https://doi.org/10.1088/0953-8984/10/20/004} {\bibfield  {journal} {\bibinfo
   {journal} {J. Phys. Condens. Matter}\ }\textbf {\bibinfo {volume} {10}},\
  \bibinfo {pages} {4353} (\bibinfo {year} {1998}{\natexlab{b}})}\BibitemShut
  {NoStop}%
\bibitem [{\citenamefont {Dabo}\ \emph {et~al.}(2010)\citenamefont {Dabo},
  \citenamefont {Ferretti}, \citenamefont {Poilvert}, \citenamefont {Li},
  \citenamefont {Marzari},\ and\ \citenamefont
  {Cococcioni}}]{Koopmans_compliant_1_2010}%
  \BibitemOpen
  \bibfield  {author} {\bibinfo {author} {\bibfnamefont {I.}~\bibnamefont
  {Dabo}}, \bibinfo {author} {\bibfnamefont {A.}~\bibnamefont {Ferretti}},
  \bibinfo {author} {\bibfnamefont {N.}~\bibnamefont {Poilvert}}, \bibinfo
  {author} {\bibfnamefont {Y.}~\bibnamefont {Li}}, \bibinfo {author}
  {\bibfnamefont {N.}~\bibnamefont {Marzari}},\ and\ \bibinfo {author}
  {\bibfnamefont {M.}~\bibnamefont {Cococcioni}},\ }\bibfield  {title}
  {\bibinfo {title} {Koopmans' condition for density-functional theory},\
  }\href {https://doi.org/10.1103/PhysRevB.82.115121} {\bibfield  {journal}
  {\bibinfo  {journal} {Phys. Rev. B}\ }\textbf {\bibinfo {volume} {82}},\
  \bibinfo {pages} {115121} (\bibinfo {year} {2010})}\BibitemShut {NoStop}%
\bibitem [{\citenamefont {Nguyen}\ \emph {et~al.}(2018)\citenamefont {Nguyen},
  \citenamefont {Colonna}, \citenamefont {Ferretti},\ and\ \citenamefont
  {Marzari}}]{Koopmans_compliant_2_2018}%
  \BibitemOpen
  \bibfield  {author} {\bibinfo {author} {\bibfnamefont {N.~L.}\ \bibnamefont
  {Nguyen}}, \bibinfo {author} {\bibfnamefont {N.}~\bibnamefont {Colonna}},
  \bibinfo {author} {\bibfnamefont {A.}~\bibnamefont {Ferretti}},\ and\
  \bibinfo {author} {\bibfnamefont {N.}~\bibnamefont {Marzari}},\ }\bibfield
  {title} {\bibinfo {title} {Koopmans-compliant spectral functionals for
  extended systems},\ }\href {https://doi.org/10.1103/PhysRevX.8.021051}
  {\bibfield  {journal} {\bibinfo  {journal} {Phys. Rev. X}\ }\textbf {\bibinfo
  {volume} {8}},\ \bibinfo {pages} {021051} (\bibinfo {year}
  {2018})}\BibitemShut {NoStop}%
\bibitem [{\citenamefont {De~Gennaro}\ \emph {et~al.}(2022)\citenamefont
  {De~Gennaro}, \citenamefont {Colonna}, \citenamefont {Linscott},\ and\
  \citenamefont {Marzari}}]{Koopmans_compliant_3_2022}%
  \BibitemOpen
  \bibfield  {author} {\bibinfo {author} {\bibfnamefont {R.}~\bibnamefont
  {De~Gennaro}}, \bibinfo {author} {\bibfnamefont {N.}~\bibnamefont {Colonna}},
  \bibinfo {author} {\bibfnamefont {E.}~\bibnamefont {Linscott}},\ and\
  \bibinfo {author} {\bibfnamefont {N.}~\bibnamefont {Marzari}},\ }\bibfield
  {title} {\bibinfo {title} {{Bloch's theorem in orbital-density-dependent
  functionals: Band structures from Koopmans spectral functionals}},\ }\href
  {https://doi.org/10.1103/PhysRevB.106.035106} {\bibfield  {journal} {\bibinfo
   {journal} {Phys. Rev. B}\ }\textbf {\bibinfo {volume} {106}},\ \bibinfo
  {pages} {035106} (\bibinfo {year} {2022})}\BibitemShut {NoStop}%
\bibitem [{\citenamefont {Colonna}\ \emph {et~al.}(2019)\citenamefont
  {Colonna}, \citenamefont {Nguyen}, \citenamefont {Ferretti},\ and\
  \citenamefont {Marzari}}]{Koopmans_Compliant_Review_2019}%
  \BibitemOpen
  \bibfield  {author} {\bibinfo {author} {\bibfnamefont {N.}~\bibnamefont
  {Colonna}}, \bibinfo {author} {\bibfnamefont {N.~L.}\ \bibnamefont {Nguyen}},
  \bibinfo {author} {\bibfnamefont {A.}~\bibnamefont {Ferretti}},\ and\
  \bibinfo {author} {\bibfnamefont {N.}~\bibnamefont {Marzari}},\ }\bibfield
  {title} {\bibinfo {title} {{Koopmans-Compliant Functionals and Potentials and
  Their Application to the GW100 Test Set}},\ }\href
  {https://doi.org/10.1021/acs.jctc.8b00976} {\bibfield  {journal} {\bibinfo
  {journal} {J. Chem. Theory Comput}\ }\textbf {\bibinfo {volume} {15}},\
  \bibinfo {pages} {1905} (\bibinfo {year} {2019})},\ \bibinfo {note} {pMID:
  30640457}\BibitemShut {NoStop}%
\bibitem [{\citenamefont {Perdew}\ \emph {et~al.}(1982)\citenamefont {Perdew},
  \citenamefont {Parr}, \citenamefont {Levy},\ and\ \citenamefont
  {Balduz}}]{delta_xc_Perdew_Parr_Levy_Balduz_1982}%
  \BibitemOpen
  \bibfield  {author} {\bibinfo {author} {\bibfnamefont {J.~P.}\ \bibnamefont
  {Perdew}}, \bibinfo {author} {\bibfnamefont {R.~G.}\ \bibnamefont {Parr}},
  \bibinfo {author} {\bibfnamefont {M.}~\bibnamefont {Levy}},\ and\ \bibinfo
  {author} {\bibfnamefont {J.~L.}\ \bibnamefont {Balduz}},\ }\bibfield  {title}
  {\bibinfo {title} {{Density-Functional Theory for Fractional Particle Number:
  Derivative Discontinuities of the Energy}},\ }\href
  {https://doi.org/10.1103/PhysRevLett.49.1691} {\bibfield  {journal} {\bibinfo
   {journal} {Phys. Rev. Lett.}\ }\textbf {\bibinfo {volume} {49}},\ \bibinfo
  {pages} {1691} (\bibinfo {year} {1982})}\BibitemShut {NoStop}%
\bibitem [{\citenamefont {Timrov}\ \emph {et~al.}(2018)\citenamefont {Timrov},
  \citenamefont {Marzari},\ and\ \citenamefont
  {Cococcioni}}]{Collinear_U_DFPT_Marzari_2018}%
  \BibitemOpen
  \bibfield  {author} {\bibinfo {author} {\bibfnamefont {I.}~\bibnamefont
  {Timrov}}, \bibinfo {author} {\bibfnamefont {N.}~\bibnamefont {Marzari}},\
  and\ \bibinfo {author} {\bibfnamefont {M.}~\bibnamefont {Cococcioni}},\
  }\bibfield  {title} {\bibinfo {title} {Hubbard parameters from
  density-functional perturbation theory},\ }\href
  {https://doi.org/10.1103/PhysRevB.98.085127} {\bibfield  {journal} {\bibinfo
  {journal} {Phys. Rev. B}\ }\textbf {\bibinfo {volume} {98}},\ \bibinfo
  {pages} {085127} (\bibinfo {year} {2018})}\BibitemShut {NoStop}%
\bibitem [{\citenamefont {Binci}\ and\ \citenamefont
  {Marzari}(2023)}]{Non_collinear_U_DFPT_Marzari_2023}%
  \BibitemOpen
  \bibfield  {author} {\bibinfo {author} {\bibfnamefont {L.}~\bibnamefont
  {Binci}}\ and\ \bibinfo {author} {\bibfnamefont {N.}~\bibnamefont
  {Marzari}},\ }\bibfield  {title} {\bibinfo {title} {{Noncollinear
  $\mathrm{DFT}+U$ and Hubbard parameters with fully relativistic ultrasoft
  pseudopotentials}},\ }\href {https://doi.org/10.1103/PhysRevB.108.115157}
  {\bibfield  {journal} {\bibinfo  {journal} {Phys. Rev. B}\ }\textbf {\bibinfo
  {volume} {108}},\ \bibinfo {pages} {115157} (\bibinfo {year}
  {2023})}\BibitemShut {NoStop}%
\bibitem [{\citenamefont {Morris}\ \emph {et~al.}(2014)\citenamefont {Morris},
  \citenamefont {Nicholls}, \citenamefont {Pickard},\ and\ \citenamefont
  {Yates}}]{optados}%
  \BibitemOpen
  \bibfield  {author} {\bibinfo {author} {\bibfnamefont {A.~J.}\ \bibnamefont
  {Morris}}, \bibinfo {author} {\bibfnamefont {R.~J.}\ \bibnamefont
  {Nicholls}}, \bibinfo {author} {\bibfnamefont {C.~J.}\ \bibnamefont
  {Pickard}},\ and\ \bibinfo {author} {\bibfnamefont {J.~R.}\ \bibnamefont
  {Yates}},\ }\bibfield  {title} {\bibinfo {title} {{OptaDOS: A tool for
  obtaining density of states, core-level and optical spectra from electronic
  structure codes}},\ }\href
  {https://doi.org/https://doi.org/10.1016/j.cpc.2014.02.013} {\bibfield
  {journal} {\bibinfo  {journal} {Comp. Phys. Comm.}\ }\textbf {\bibinfo
  {volume} {185}},\ \bibinfo {pages} {1477} (\bibinfo {year}
  {2014})}\BibitemShut {NoStop}%
\bibitem [{\citenamefont {Yates}\ \emph {et~al.}(2007)\citenamefont {Yates},
  \citenamefont {Wang}, \citenamefont {Vanderbilt},\ and\ \citenamefont
  {Souza}}]{Yates_Optados_Adaptive_Smearing}%
  \BibitemOpen
  \bibfield  {author} {\bibinfo {author} {\bibfnamefont {J.~R.}\ \bibnamefont
  {Yates}}, \bibinfo {author} {\bibfnamefont {X.}~\bibnamefont {Wang}},
  \bibinfo {author} {\bibfnamefont {D.}~\bibnamefont {Vanderbilt}},\ and\
  \bibinfo {author} {\bibfnamefont {I.}~\bibnamefont {Souza}},\ }\bibfield
  {title} {\bibinfo {title} {{Spectral and Fermi surface properties from
  Wannier interpolation}},\ }\href {https://doi.org/10.1103/PhysRevB.75.195121}
  {\bibfield  {journal} {\bibinfo  {journal} {Phys. Rev. B}\ }\textbf {\bibinfo
  {volume} {75}},\ \bibinfo {pages} {195121} (\bibinfo {year}
  {2007})}\BibitemShut {NoStop}%
\bibitem [{\citenamefont {Segall}\ \emph {et~al.}(1996)\citenamefont {Segall},
  \citenamefont {Shah}, \citenamefont {Pickard},\ and\ \citenamefont
  {Payne}}]{Segall_Population_Analysis}%
  \BibitemOpen
  \bibfield  {author} {\bibinfo {author} {\bibfnamefont {M.~D.}\ \bibnamefont
  {Segall}}, \bibinfo {author} {\bibfnamefont {R.}~\bibnamefont {Shah}},
  \bibinfo {author} {\bibfnamefont {C.~J.}\ \bibnamefont {Pickard}},\ and\
  \bibinfo {author} {\bibfnamefont {M.~C.}\ \bibnamefont {Payne}},\ }\bibfield
  {title} {\bibinfo {title} {Population analysis of plane-wave electronic
  structure calculations of bulk materials},\ }\href
  {https://doi.org/10.1103/PhysRevB.54.16317} {\bibfield  {journal} {\bibinfo
  {journal} {Phys. Rev. B}\ }\textbf {\bibinfo {volume} {54}},\ \bibinfo
  {pages} {16317} (\bibinfo {year} {1996})}\BibitemShut {NoStop}%
\bibitem [{\citenamefont {Ravindran}\ \emph {et~al.}(2024)\citenamefont
  {Ravindran}, \citenamefont {Gidopoulos},\ and\ \citenamefont
  {Clark}}]{data_repo}%
  \BibitemOpen
  \bibfield  {author} {\bibinfo {author} {\bibfnamefont {V.}~\bibnamefont
  {Ravindran}}, \bibinfo {author} {\bibfnamefont {N.~I.}\ \bibnamefont
  {Gidopoulos}},\ and\ \bibinfo {author} {\bibfnamefont {S.~J.}\ \bibnamefont
  {Clark}},\ }\href {https://doi.org/10.15128/r18w32r562d} {\bibinfo {title}
  {{Local Exchange-Correlation Potentials by Density Inversion In Solids
  [dataset]}}},\ \bibinfo {howpublished}
  {\url{http://doi.org/10.15128/r18w32r562d}} (\bibinfo {year}
  {2024})\BibitemShut {NoStop}%
\end{thebibliography}%


%apsrev4-2.bst 2019-01-14 (MD) hand-edited version of apsrev4-1.bst
%Control: key (0)
%Control: author (8) initials jnrlst
%Control: editor formatted (1) identically to author
%Control: production of article title (0) allowed
%Control: page (0) single
%Control: year (1) truncated
%Control: production of eprint (0) enabled
\begin{thebibliography}{9}%
\makeatletter
\providecommand \@ifxundefined [1]{%
 \@ifx{#1\undefined}
}%
\providecommand \@ifnum [1]{%
 \ifnum #1\expandafter \@firstoftwo
 \else \expandafter \@secondoftwo
 \fi
}%
\providecommand \@ifx [1]{%
 \ifx #1\expandafter \@firstoftwo
 \else \expandafter \@secondoftwo
 \fi
}%
\providecommand \natexlab [1]{#1}%
\providecommand \enquote  [1]{``#1''}%
\providecommand \bibnamefont  [1]{#1}%
\providecommand \bibfnamefont [1]{#1}%
\providecommand \citenamefont [1]{#1}%
\providecommand \href@noop [0]{\@secondoftwo}%
\providecommand \href [0]{\begingroup \@sanitize@url \@href}%
\providecommand \@href[1]{\@@startlink{#1}\@@href}%
\providecommand \@@href[1]{\endgroup#1\@@endlink}%
\providecommand \@sanitize@url [0]{\catcode `\\12\catcode `\$12\catcode
  `\&12\catcode `\#12\catcode `\^12\catcode `\_12\catcode `\%12\relax}%
\providecommand \@@startlink[1]{}%
\providecommand \@@endlink[0]{}%
\providecommand \url  [0]{\begingroup\@sanitize@url \@url }%
\providecommand \@url [1]{\endgroup\@href {#1}{\urlprefix }}%
\providecommand \urlprefix  [0]{URL }%
\providecommand \Eprint [0]{\href }%
\providecommand \doibase [0]{https://doi.org/}%
\providecommand \selectlanguage [0]{\@gobble}%
\providecommand \bibinfo  [0]{\@secondoftwo}%
\providecommand \bibfield  [0]{\@secondoftwo}%
\providecommand \translation [1]{[#1]}%
\providecommand \BibitemOpen [0]{}%
\providecommand \bibitemStop [0]{}%
\providecommand \bibitemNoStop [0]{.\EOS\space}%
\providecommand \EOS [0]{\spacefactor3000\relax}%
\providecommand \BibitemShut  [1]{\csname bibitem#1\endcsname}%
\let\auto@bib@innerbib\@empty
%</preamble>
\bibitem [{\citenamefont {Hollins}\ \emph {et~al.}(2017)\citenamefont
  {Hollins}, \citenamefont {Clark}, \citenamefont {Refson},\ and\ \citenamefont
  {Gidopoulos}}]{Hollins_LFX_2017}%
  \BibitemOpen
  \bibfield  {author} {\bibinfo {author} {\bibfnamefont {T.~W.}\ \bibnamefont
  {Hollins}}, \bibinfo {author} {\bibfnamefont {S.~J.}\ \bibnamefont {Clark}},
  \bibinfo {author} {\bibfnamefont {K.}~\bibnamefont {Refson}},\ and\ \bibinfo
  {author} {\bibfnamefont {N.~I.}\ \bibnamefont {Gidopoulos}},\ }\bibfield
  {title} {\bibinfo {title} {{A local Fock-exchange potential in Kohn–Sham
  equations}},\ }\href {https://doi.org/10.1088/1361-648X/29/4/04LT01}
  {\bibfield  {journal} {\bibinfo  {journal} {J. Phys.: Condens. Matter}\
  }\textbf {\bibinfo {volume} {29}},\ \bibinfo {pages} {04LT01} (\bibinfo
  {year} {2017})}\BibitemShut {NoStop}%
\bibitem [{\citenamefont {Kohn}\ and\ \citenamefont
  {Sham}(1965)}]{Kohn-Sham_DFT}%
  \BibitemOpen
  \bibfield  {author} {\bibinfo {author} {\bibfnamefont {W.}~\bibnamefont
  {Kohn}}\ and\ \bibinfo {author} {\bibfnamefont {L.~J.}\ \bibnamefont
  {Sham}},\ }\bibfield  {title} {\bibinfo {title} {{Self-Consistent Equations
  Including Exchange and Correlation Effects}},\ }\href
  {https://doi.org/10.1103/PhysRev.140.A1133} {\bibfield  {journal} {\bibinfo
  {journal} {Phys. Rev.}\ }\textbf {\bibinfo {volume} {140}},\ \bibinfo {pages}
  {A1133} (\bibinfo {year} {1965})}\BibitemShut {NoStop}%
\bibitem [{\citenamefont {Perdew}\ and\ \citenamefont
  {Zunger}(1981)}]{LDA_CASTEP}%
  \BibitemOpen
  \bibfield  {author} {\bibinfo {author} {\bibfnamefont {J.~P.}\ \bibnamefont
  {Perdew}}\ and\ \bibinfo {author} {\bibfnamefont {A.}~\bibnamefont
  {Zunger}},\ }\bibfield  {title} {\bibinfo {title} {Self-interaction
  correction to density-functional approximations for many-electron systems},\
  }\href {https://doi.org/10.1103/PhysRevB.23.5048} {\bibfield  {journal}
  {\bibinfo  {journal} {Phys. Rev. B}\ }\textbf {\bibinfo {volume} {23}},\
  \bibinfo {pages} {5048} (\bibinfo {year} {1981})}\BibitemShut {NoStop}%
\bibitem [{\citenamefont {Hubbard}\ and\ \citenamefont
  {Flowers}(1963)}]{Hubbard_Model_OG}%
  \BibitemOpen
  \bibfield  {author} {\bibinfo {author} {\bibfnamefont {J.}~\bibnamefont
  {Hubbard}}\ and\ \bibinfo {author} {\bibfnamefont {B.~H.}\ \bibnamefont
  {Flowers}},\ }\bibfield  {title} {\bibinfo {title} {Electron correlations in
  narrow energy bands},\ }\href {https://doi.org/10.1098/rspa.1963.0204}
  {\bibfield  {journal} {\bibinfo  {journal} {Proc. R. Soc. Lond. A}\ }\textbf
  {\bibinfo {volume} {276}},\ \bibinfo {pages} {238} (\bibinfo {year}
  {1963})}\BibitemShut {NoStop}%
\bibitem [{\citenamefont
  {Tasaki}(1998{\natexlab{a}})}]{Tasaki_1998_Review_Hubbard_intro}%
  \BibitemOpen
  \bibfield  {author} {\bibinfo {author} {\bibfnamefont {H.}~\bibnamefont
  {Tasaki}},\ }\bibfield  {title} {\bibinfo {title} {{From Nagaoka's
  Ferromagnetism to Flat-Band Ferromagnetism and Beyond: An Introduction to
  Ferromagnetism in the Hubbard Model}},\ }\href
  {https://doi.org/10.1143/PTP.99.489} {\bibfield  {journal} {\bibinfo
  {journal} {Prog. Theor. Phys.}\ }\textbf {\bibinfo {volume} {99}},\ \bibinfo
  {pages} {489} (\bibinfo {year} {1998}{\natexlab{a}})}\BibitemShut {NoStop}%
\bibitem [{\citenamefont
  {Tasaki}(1998{\natexlab{b}})}]{Tasaki_1998_Review_Hubbard_rigorous}%
  \BibitemOpen
  \bibfield  {author} {\bibinfo {author} {\bibfnamefont {H.}~\bibnamefont
  {Tasaki}},\ }\bibfield  {title} {\bibinfo {title} {The {Hubbard} model - an
  introduction and selected rigorous results},\ }\href
  {https://doi.org/10.1088/0953-8984/10/20/004} {\bibfield  {journal} {\bibinfo
   {journal} {J. Phys. Condens. Matter}\ }\textbf {\bibinfo {volume} {10}},\
  \bibinfo {pages} {4353} (\bibinfo {year} {1998}{\natexlab{b}})}\BibitemShut
  {NoStop}%
\bibitem [{\citenamefont {Morris}\ \emph {et~al.}(2014)\citenamefont {Morris},
  \citenamefont {Nicholls}, \citenamefont {Pickard},\ and\ \citenamefont
  {Yates}}]{optados}%
  \BibitemOpen
  \bibfield  {author} {\bibinfo {author} {\bibfnamefont {A.~J.}\ \bibnamefont
  {Morris}}, \bibinfo {author} {\bibfnamefont {R.~J.}\ \bibnamefont
  {Nicholls}}, \bibinfo {author} {\bibfnamefont {C.~J.}\ \bibnamefont
  {Pickard}},\ and\ \bibinfo {author} {\bibfnamefont {J.~R.}\ \bibnamefont
  {Yates}},\ }\bibfield  {title} {\bibinfo {title} {{OptaDOS: A tool for
  obtaining density of states, core-level and optical spectra from electronic
  structure codes}},\ }\href
  {https://doi.org/https://doi.org/10.1016/j.cpc.2014.02.013} {\bibfield
  {journal} {\bibinfo  {journal} {Comp. Phys. Comm.}\ }\textbf {\bibinfo
  {volume} {185}},\ \bibinfo {pages} {1477} (\bibinfo {year}
  {2014})}\BibitemShut {NoStop}%
\bibitem [{\citenamefont {Segall}\ \emph {et~al.}(1996)\citenamefont {Segall},
  \citenamefont {Shah}, \citenamefont {Pickard},\ and\ \citenamefont
  {Payne}}]{Segall_Population_Analysis}%
  \BibitemOpen
  \bibfield  {author} {\bibinfo {author} {\bibfnamefont {M.~D.}\ \bibnamefont
  {Segall}}, \bibinfo {author} {\bibfnamefont {R.}~\bibnamefont {Shah}},
  \bibinfo {author} {\bibfnamefont {C.~J.}\ \bibnamefont {Pickard}},\ and\
  \bibinfo {author} {\bibfnamefont {M.~C.}\ \bibnamefont {Payne}},\ }\bibfield
  {title} {\bibinfo {title} {Population analysis of plane-wave electronic
  structure calculations of bulk materials},\ }\href
  {https://doi.org/10.1103/PhysRevB.54.16317} {\bibfield  {journal} {\bibinfo
  {journal} {Phys. Rev. B}\ }\textbf {\bibinfo {volume} {54}},\ \bibinfo
  {pages} {16317} (\bibinfo {year} {1996})}\BibitemShut {NoStop}%
\bibitem [{\citenamefont {Yates}\ \emph {et~al.}(2007)\citenamefont {Yates},
  \citenamefont {Wang}, \citenamefont {Vanderbilt},\ and\ \citenamefont
  {Souza}}]{Yates_Optados_Adaptive_Smearing}%
  \BibitemOpen
  \bibfield  {author} {\bibinfo {author} {\bibfnamefont {J.~R.}\ \bibnamefont
  {Yates}}, \bibinfo {author} {\bibfnamefont {X.}~\bibnamefont {Wang}},
  \bibinfo {author} {\bibfnamefont {D.}~\bibnamefont {Vanderbilt}},\ and\
  \bibinfo {author} {\bibfnamefont {I.}~\bibnamefont {Souza}},\ }\bibfield
  {title} {\bibinfo {title} {{Spectral and Fermi surface properties from
  Wannier interpolation}},\ }\href {https://doi.org/10.1103/PhysRevB.75.195121}
  {\bibfield  {journal} {\bibinfo  {journal} {Phys. Rev. B}\ }\textbf {\bibinfo
  {volume} {75}},\ \bibinfo {pages} {195121} (\bibinfo {year}
  {2007})}\BibitemShut {NoStop}%
\end{thebibliography}%

\end{document}

\title{Supplemental material for ``Local Exchange-Correlation Potentials by Density Inversion in Solids''}
\author{Visagan Ravindran}
\author{Nikitas I. Gidopoulos}
\author{Stewart J. Clark}
\affiliation{Department of Physics, Durham University, South Road, Durham, DH1 3LE, UK}
\email[Corresponding Author: ]{s.j.clark@durham.ac.uk}
\date{7th August 2025}

\begin{abstract}
    In this supplementary material, we present an inversion calculation of an LDA density, demonstrating the convergence of our density inversion algorithm when starting from a density whose local exchange-correlation potential (LXC) is known.
    We also present the projected density of states (PDOS) calculations demonstrating hybridisation effects in transition metal monoxides (TMOs) using LDA, LDA+$U$ and LXC-LDA+$U$.
\end{abstract}
\maketitle

\section*{Convergence of Inversion Algorithm: Inverting Local/Semi-local DFAs}\label{appendix:local_pot_inversion}
\begin{figure}[ht]
	\centering
	\includegraphics[width=0.8\columnwidth]{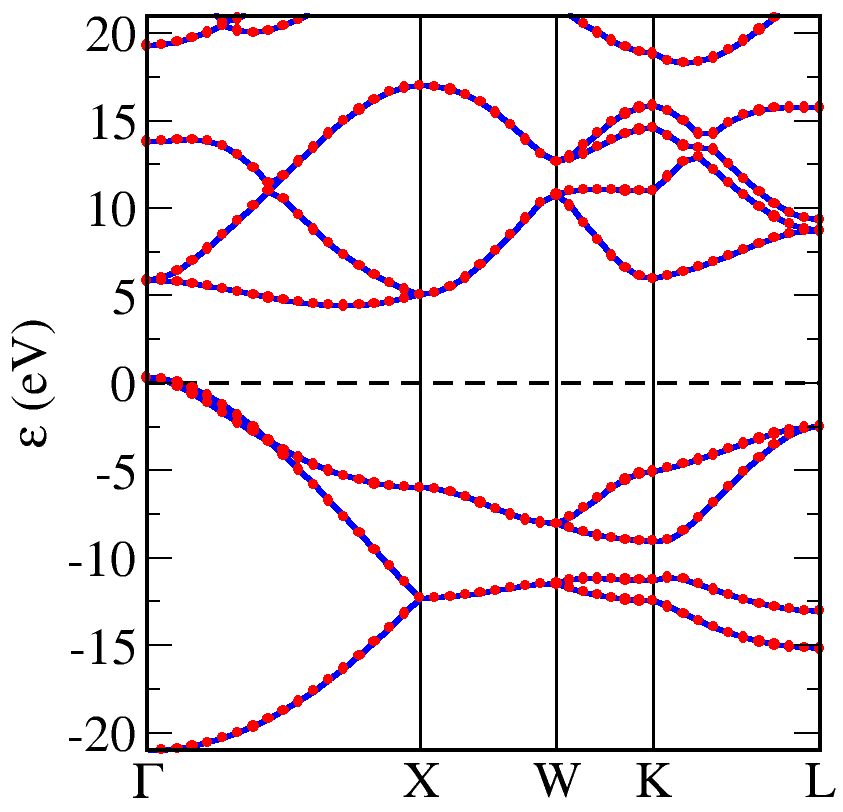}
	\caption{
		Computed band structures using LDA for diamond.
		The solid blue line indicates the band structure obtained using the self-consistent potential while the red dotted lines are obtained using the LDA-LXC potential via inversion of the LDA target density.
		Both band structures are indistinguishable from each other.}
	\label{fig:inversion_LDA}
\end{figure}

The algorithm to invert a given target density to find the Kohn-Sham (KS) potential with a local exchange-correlation potential (LXC) potential is detailed in the main text and in Ref. \cite{Hollins_LFX_2017}.
By definition, the LXC potential obtained from a target density of a local or semi-local DFA is identical to local potential of that DFA,
\begin{equation}
    v_\mathrm{xc}^\mathrm{DFA}(\vec{r}) = \frac{\delta E_\mathrm{xc}}{\delta \rho(\vec{r})}.
\end{equation}
To that end, it is useful to invert densities from the local density approximation\cite{Kohn-Sham_DFT,LDA_CASTEP} (LDA), whose potential has an explicit functional form in terms of the density to verify convergence of the inversion algorithm.
As discussed in the main text, we inverted LDA densities with the LXC potential initialised to the PBE potential calculated from the LDA density.
An example of this is given in Fig. \ref{fig:inversion_LDA} for diamond where we have plotted the LDA band structure and the band structure obtained via inversion of the LDA density, dubbed LXC-LDA.
One can see the two band structures are practically indistinguishable by eye.

\section*{Projected Density of States: LDA+$U$ vs LXC-LDA+$U$}\label{appendix:pdos_plots}
The non-locality in LDA+$U$ originates from the Hubbard-$U$ term\cite{Hubbard_Model_OG,Tasaki_1998_Review_Hubbard_intro,Tasaki_1998_Review_Hubbard_rigorous}.
By contrast, the LXC-LDA+$U$, by its very nature must act on all orbitals in the same manner since it is a local (multiplicative) potential.
The identification of the character of bands with regards to the angular momentum must thus be done with some care since it is not guaranteed that either method will preserve the character of bands that is obtained by the LDA.

\begin{figure*}[t]
	\subfloat[MnO LDA]{\includegraphics[width=0.35\linewidth]{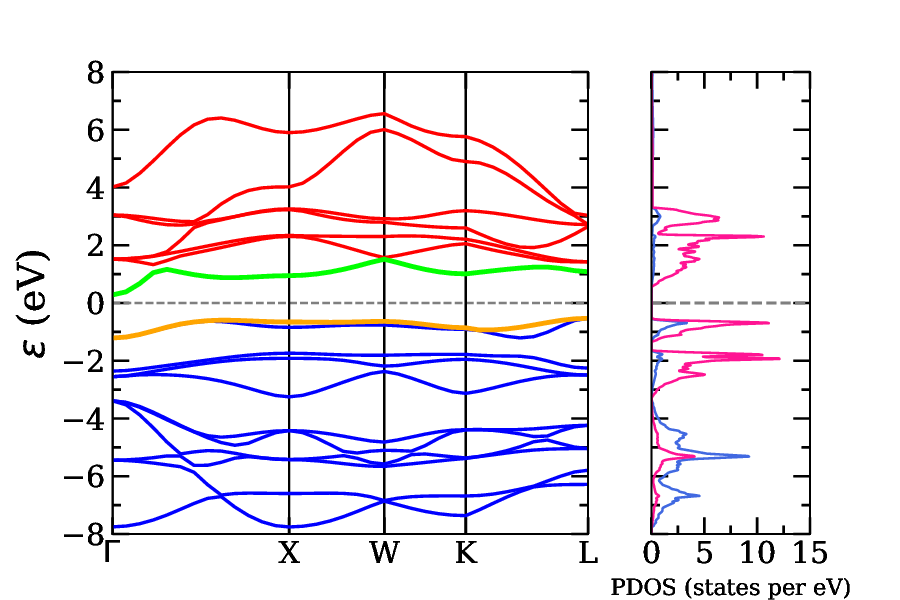}}
	\subfloat[MnO LDA+$U$]{\includegraphics[width=0.35\linewidth]{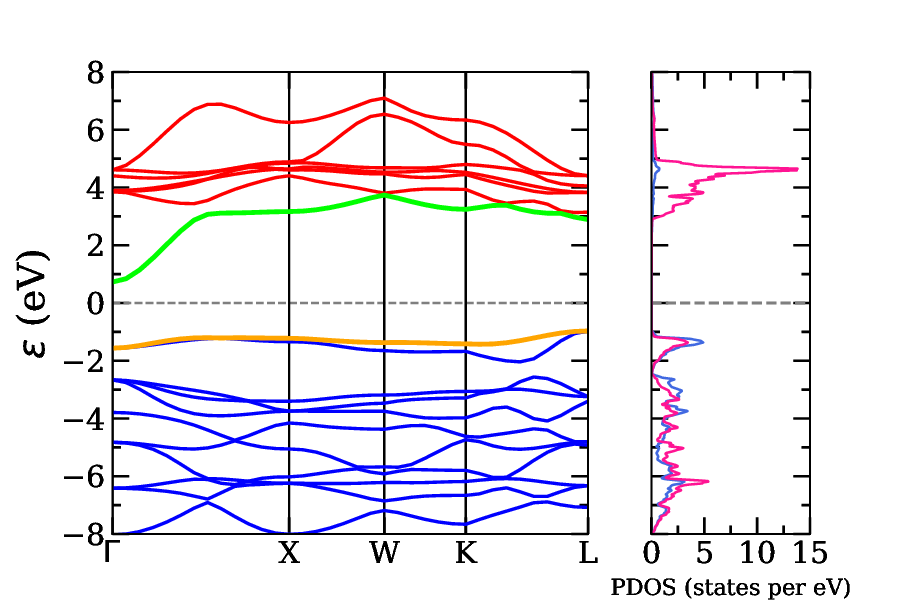}}
	\subfloat[MnO LXC-LDA+$U$]{\includegraphics[width=0.35\linewidth]{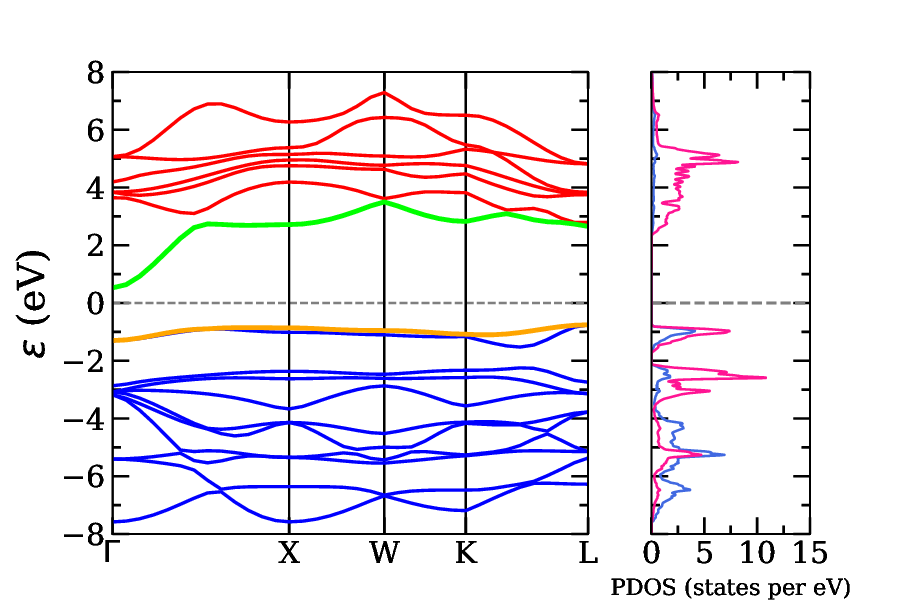}}
	\hfill
	\subfloat[NiO LDA]{\includegraphics[width=0.35\linewidth]{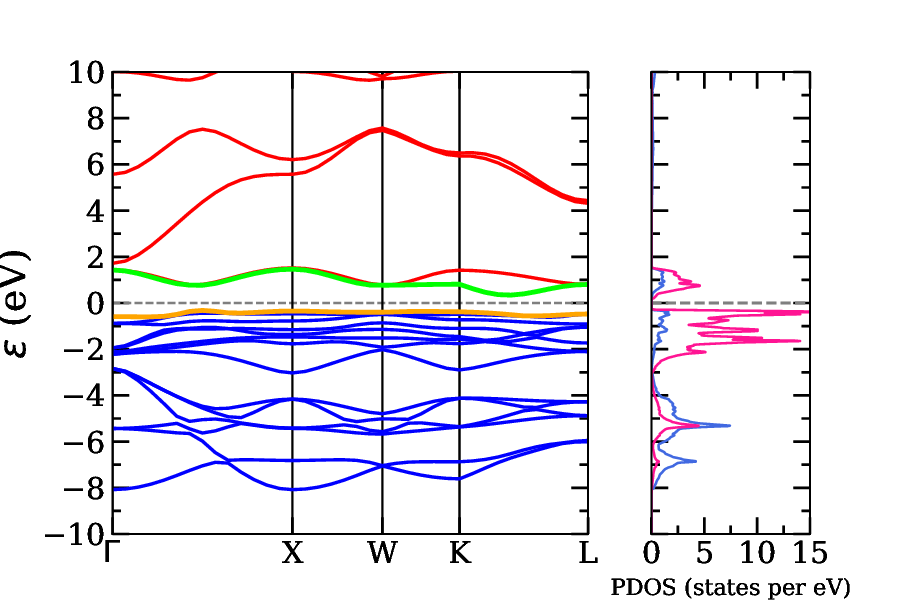}}
	\subfloat[NiO LDA+$U$]{\includegraphics[width=0.35\linewidth]{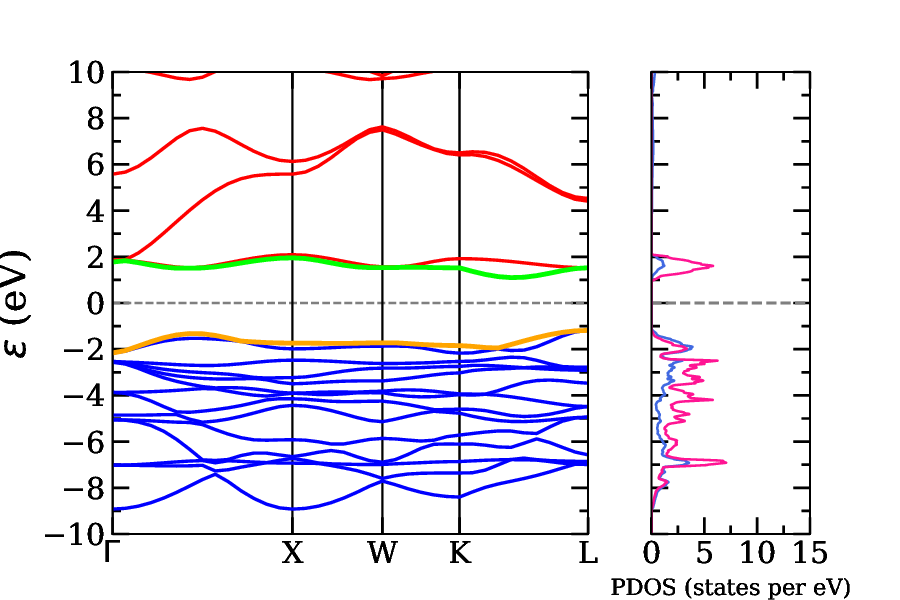}}
	\subfloat[NiO LXC-LDA+$U$]{\includegraphics[width=0.35\linewidth]{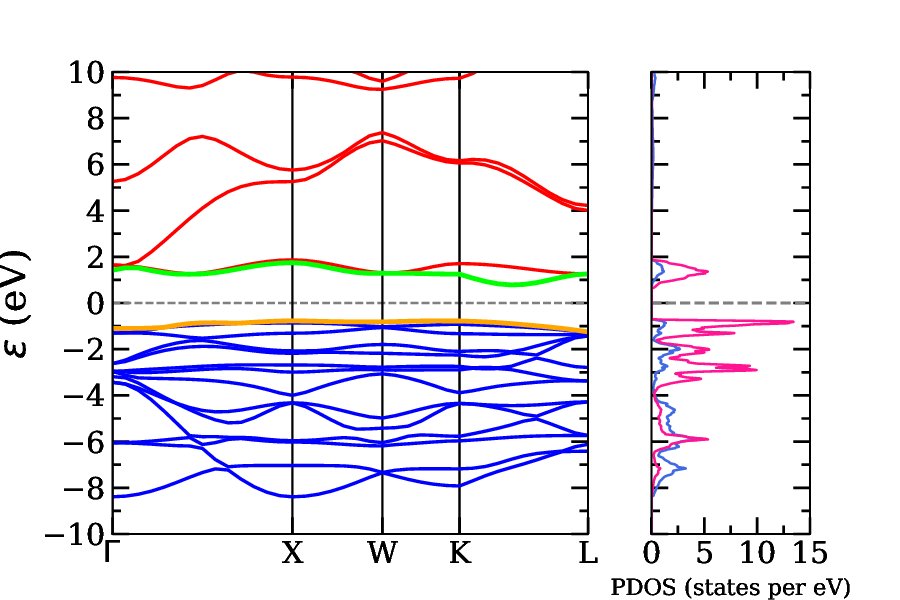}}
	\hfill
	\subfloat[CoO LDA]{\includegraphics[width=0.35\linewidth]{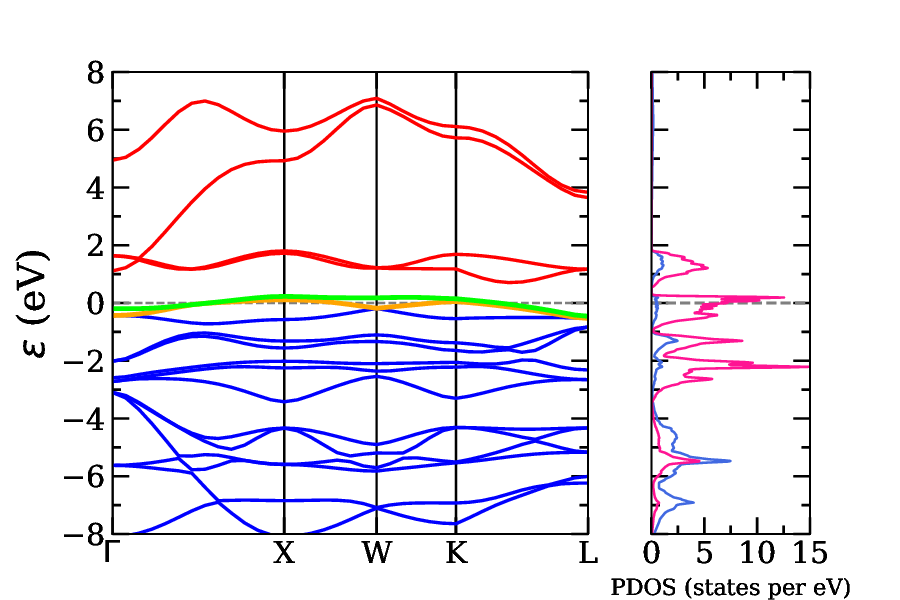}}
	\subfloat[CoO LDA+$U$]{\includegraphics[width=0.35\linewidth]{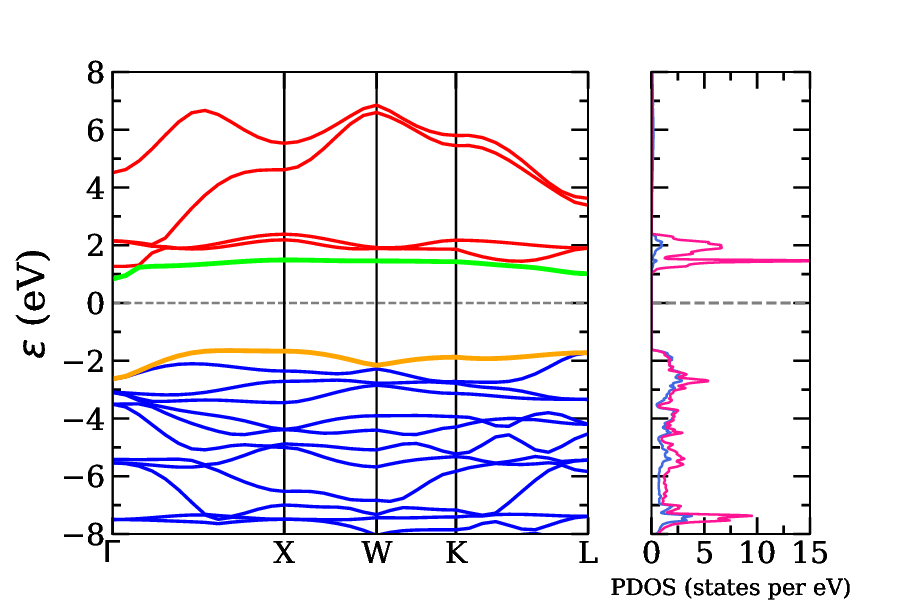}}
	\subfloat[CoO LXC-LDA+$U$]{\includegraphics[width=0.35\linewidth]{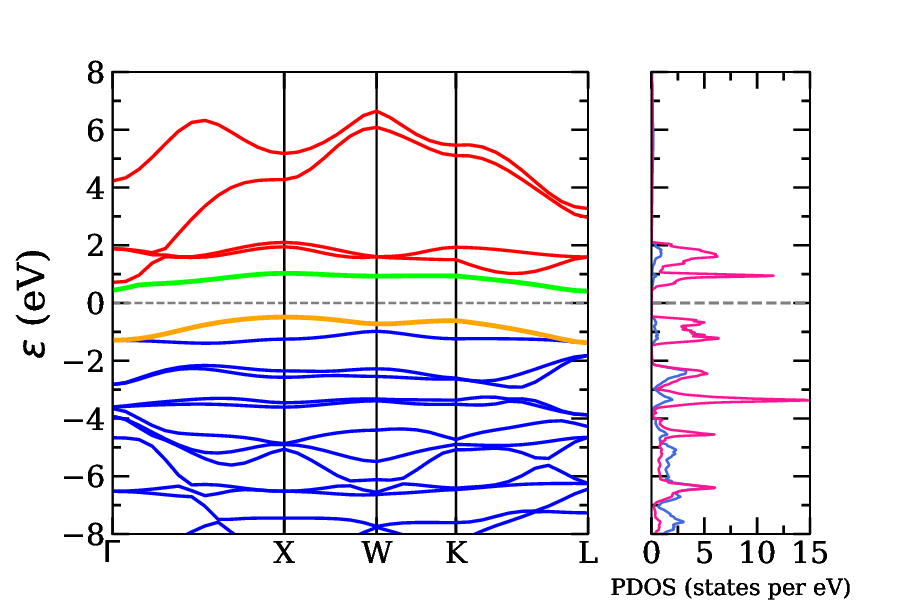}}
	\hfill
	\subfloat[FeO LDA]{\includegraphics[width=0.35\linewidth]{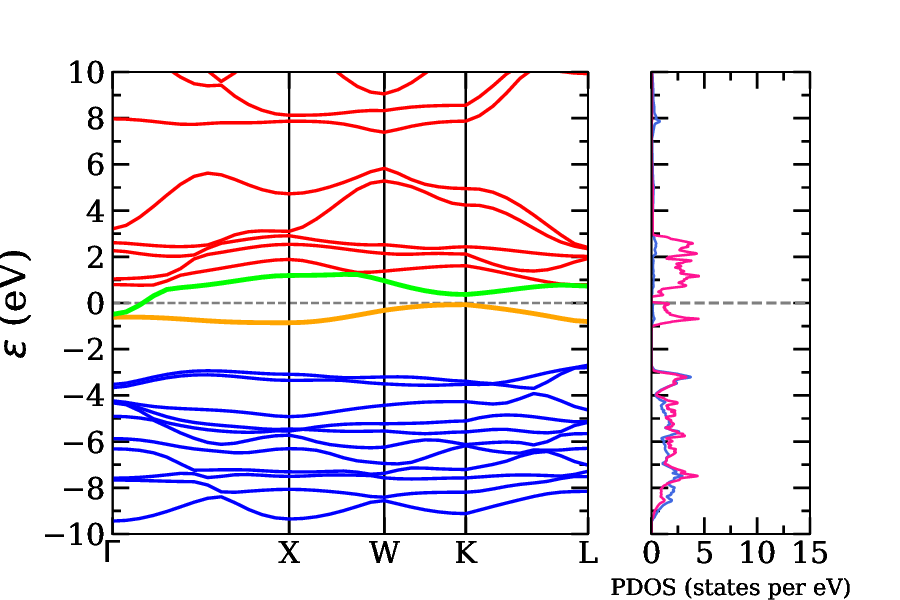}}
	\subfloat[FeO LDA+$U$]{\includegraphics[width=0.35\linewidth]{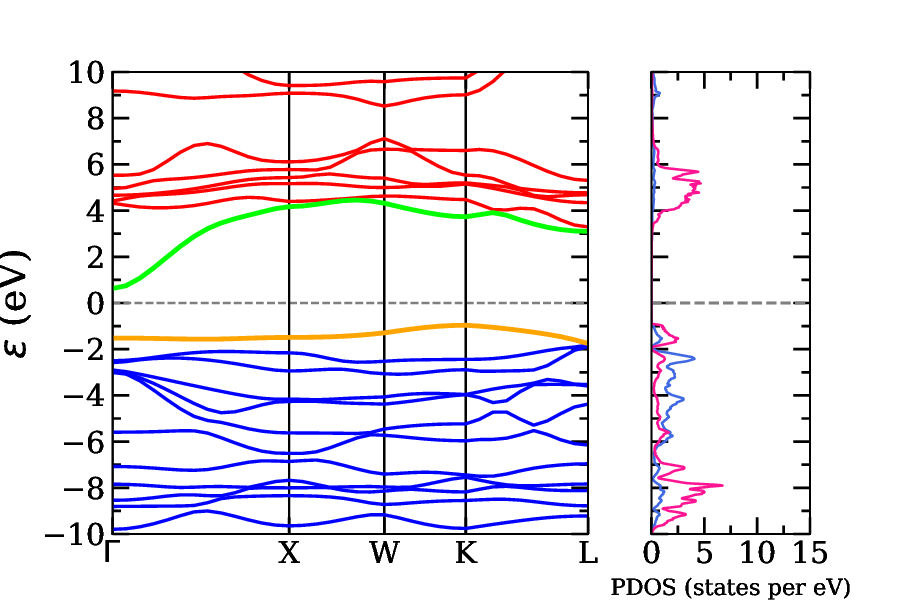}}
	\subfloat[FeO LXC-LDA+$U$]{\includegraphics[width=0.35\linewidth]{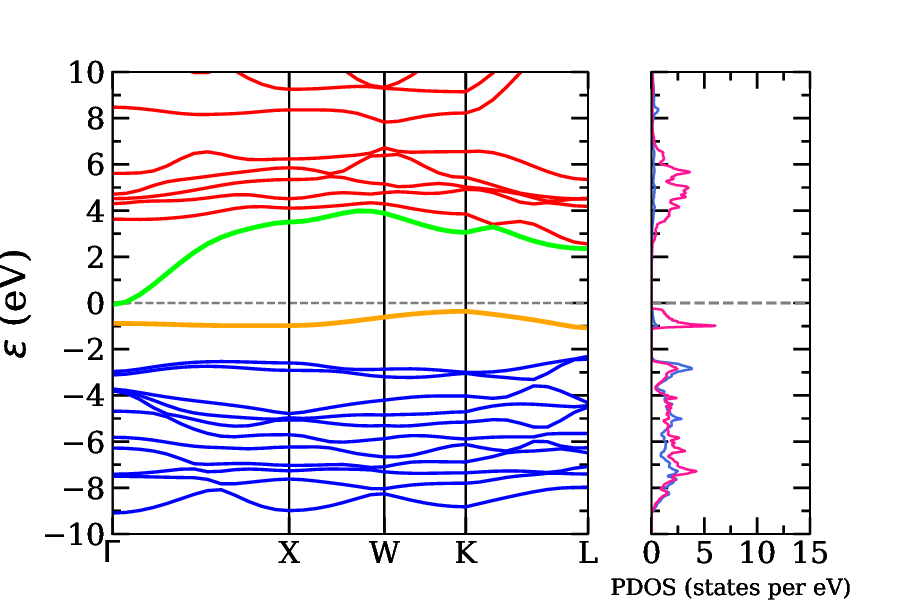}}
	\label{fig:pdos_bs_appendix}
	\caption{
		Projected density of states (PDOS) for MnO, NiO, CoO and FeO calculated using LDA, LDA+$U$ and LXC-LDA+$U$ methods using $U=5.0 \text{ eV}$ in the latter two cases.
		In the band structures, the blue and red lines indicate occupied and unoccupied bands respectively while the orange and green lines indicate the valence and conduction bands respectively.
		For the PDOS, we have only plotted a single spin channel for clarity since the second is degenerate as a result of the anti-ferromagnetic ordering. The pink line is for the transition metal cation's $3d$ electrons while the blue line indicates oxygen $2p$ states.
		The energy scale has been set such that the Fermi energy is at 0 eV in all plots.
	}
\end{figure*}

We performed a series of LDA+$U$ calculations with a Hubbard-$U$ applied to the transition metal cation's $d$ orbitals before inverting these densities to find the LXC-LDA+$U$ band structure.
We calculated the projected density of states (PDOS) using the OptaDOS code\cite{optados} which implements the population analysis methodology of Segall \textit{et al.}\cite{Segall_Population_Analysis} to determine the character of each band by projecting the Kohn-Sham orbital onto a linear combination of atomic orbitals (LCAO) with the broadening scheme of Yates \textit{et al.}\cite{Yates_Optados_Adaptive_Smearing} with a adaptive smearing ratio of 0.4.
A Monkhorst-Pack k-point grid with a spacing of $0.02 \text{ \AA}^{-1}$ was used for the PDOS calculation.
The PDOS is plotted alongside the associated band structure in Fig \ref{fig:pdos_bs_appendix} for each transition metal monoxide (TMO) for the LDA, LDA+$U$ and LXC-LDA+$U$ methods at $U=5$ eV.

In general, one can see that the inversion largely preserves the PDOS as the Hubbard-$U$ parameter is varied and the only substantial significant change one observes is in the calculated gap where it appears that the eigenvalues are merely shifted by a largely constant amount.
This behaviour also manifests itself in the insensitivity of the total band width $\Delta \tilde{E}$ bands with largely $d$-character with increasing $U$ and in particular, similar to the LDA band width.
This is not true however for the LDA+$U$ where one sees that as $U$ is applied, additional bands start acquiring $d$ character and that the valence band and other occupied states near it start to show hybridisation of the transition metal cation's $3d$ electrons with the oxygen $2p$ states.

Regarding material-specific differences, we point out that CoO has a larger difference in the LXC-LDA+$U$ PDOS compared to MnO and NiO and as pointed out in the main text, also does not recover the dispersion of LDA to the same extent as in MnO and NiO.
FeO proves once again to be somewhat of a special case amongst these class of materials where in the LDA case, almost all occupied bands shown have $d$ character and exhibit strong hybridisation with oxygen $p$ states, thus making it difficult to identify specific bands with $d$ character.
This contrasts with the behaviour with the other TMOs where the highest $n_d/2$ occupied bands where $n_d$ is the total number of $d$ electrons also have predominantly $d$-character.
With LDA+$U$, as we have previously stated, occupied bands near the valence band start to have oxygen $p$-character; the difference in FeO however is much more stark to the extent that these bands have \textit{predominantly} oxygen-$p$ character.

For calculations of the total band width for each TMO presented in the main text, we treated the LDA PDOS as a reference, obtaining the bands with pre-dominantly $d$ character.
We then calculated the total band width for these set of bands for LDA+$U$ and LXC-LDA+$U$.

\bibliography{refs}